\documentclass[11pt,letterpaper]{amsart}
\usepackage[T1]{fontenc}
\usepackage{lmodern}
\usepackage[english]{babel}
\usepackage{amsmath,amssymb,amsthm,mathtools,mathrsfs}
\usepackage[numbers,sort&compress]{natbib}
\usepackage{needspace}
\usepackage[colorlinks=true,linkcolor=blue,citecolor=blue,urlcolor=blue]{hyperref}

\numberwithin{equation}{section}

\theoremstyle{plain}

\newtheorem{theorem}{Theorem}

\newtheorem{proposition}{Proposition}[section]
\newtheorem{lemma}[proposition]{Lemma}
\newtheorem{corollary}[proposition]{Corollary}

\theoremstyle{definition}
\newtheorem{definition}{Definition}[section]

\theoremstyle{remark}
\newtheorem{remark}{Remark}[section]

\newcommand{\R}{\mathbb R}
\newcommand{\Q}{\mathbb Q}
\newcommand{\Z}{\mathbb Z}
\newcommand{\Hh}{\mathbb H}
\newcommand{\Prob}{\mathcal P}
\newcommand{\Leb}{\mathrm{Leb}}
\newcommand{\Lip}{\operatorname{Lip}}
\newcommand{\essosc}{\operatorname{ess\,osc}}
\newcommand{\SL}{\mathrm{SL}}
\newcommand{\PSL}{\mathrm{PSL}}
\newcommand{\tr}{\operatorname{tr}}

\newcommand{\Ereg}{\widetilde{\mathcal E}}
\newcommand{\GL}{\mathrm{GL}}

\makeatletter
\renewenvironment{abstract}{%
  \ifx\maketitle\relax
    \ClassWarning{\@classname}{Abstract should precede
      \protect\maketitle\space in AMS document classes; reported}%
  \fi
  \global\setbox\abstractbox=\vtop \bgroup
    \normalfont\Small
    \list{}{\labelwidth\z@
      \leftmargin3pc \rightmargin\leftmargin
      \listparindent\normalparindent \itemindent\z@
      \parsep\z@ \@plus\p@
      
    }%
    \item[]\hfil{\scshape\abstractname}\hfil\par\smallskip
}{%
  \endlist\egroup
  \ifx\@setabstract\relax \@setabstracta \fi
}
\makeatother

\begin{document}
\title[Finite-step projective Schr\"odinger dynamics]{Finite-step projective dynamics of Schr\"odinger cocycles: non-degeneracy, regularity, and rigidity}
\author{Mira Shamis}
\address{School of Mathematical Sciences, Queen Mary University of London,
Mile End Road, London E1 4NS, England}
\email{m.shamis@qmul.ac.uk}
\address{School of Mathematical Sciences, Holon Institute of Technology, 52 Golomb St.,
P.O.B. 305, Holon 5810201, Israel}
\email{miras@hit.ac.il}

\date{\today}
\begin{abstract}
We study the no deterministic images condition (NDI)
for Schr\"odinger cocycles, which excludes a common image
of a fixed probability measure under almost all
projective transfer transformations.
We prove two-step NDI at every energy for nonconstant
measurable potentials over mixing systems and ergodic
potentials with infinite essential range, and classify
persistence for finite-state locally constant potentials.
For bounded Markov and hidden Markov potentials
on general state spaces under a one-step $L^2$ norm gap
condition on the Markov operator, we prove H\"older bounds for solution-direction
distributions above an exponential cutoff and H\"older
conditional stationary measures.
Counterexamples show the essential sharpness of both assumptions.

For nonconstant real trigonometric potentials,
including the almost Mathieu operator, we prove NDI
and H\"older bounds for distributions of solution
directions at all steps $k\ge2$ and all frequencies
(rational or irrational). The exponent depends
on the step, while the constant is uniform in steps,
frequency and energy on compact intervals.
For every irrational frequency and every energy
with positive Lyapunov exponent, we prove non-atomicity
of both Oseledets direction distributions and convergence
to the unstable distribution from every phase-independent
initial probability.
This gives a new criterion for a zero Lyapunov exponent:
a subsequential limit with an atom for one fixed
initial direction suffices. All these results extend to an analytic class, 
with step-dependent regularity bounds.
\end{abstract}
\maketitle

\setcounter{tocdepth}{1}
\tableofcontents

\textit{Statement on the use of artificial intelligence.}
ChatGPT (6.1 Sol) was used for LaTeX transcription and editing,
mostly for the conversion of handwritten material into LaTeX,
for linguistic corrections, proofreading and bibliographic searches.
Its use for checking explicit calculations was limited to
the choice of small constants in \eqref{eq:zeta-choice}
and the author's handwritten expansion of the determinant
of a $4\times4$ tridiagonal matrix.
This calculation was the author's starting point for the
proof of \eqref{eq:continuant-coefficient-bound}.
It also assisted with preliminary drafts of parts of the
Introduction, subsequently fully rewritten by the author.
All the mathematical ideas and arguments are those of the author.

\section{Introduction}\label{sec:introduction}

We consider one-dimensional discrete Schr\"odinger operators
on $\ell^2(\Z)$,
\begin{equation}\label{eq:operator-introduction}
 (H_\omega u)_j
 =u_{j+1}+u_{j-1}+f(\sigma^j\omega)u_j,
 \qquad j\in\Z,
\end{equation}
where $\sigma$ is an invertible probability-preserving
transformation of a standard probability space
$(\Omega,\mathcal F,\mathbf P)$, and
$f:\Omega\to\R$ is measurable 
and not constant
$\mathbf P$-almost everywhere.
Any fixed nonzero coupling constant is included in $f$.
The potential is stationary.

The corresponding eigenvalue equation,
\begin{equation}\label{eq:schrodinger}
 u_{j+1}+u_{j-1}+f(\sigma^j\omega)u_j
 =\lambda u_j,
 \qquad \lambda\in\R,
\end{equation}
propagates consecutive values of a solution
through the transfer matrices
\begin{equation}\label{eq:scalar-matrix}
 A_\lambda(\omega)
 =\begin{pmatrix}\lambda-f(\omega)&-1\\1&0\end{pmatrix},
 \qquad
 \binom{u_{j+1}}{u_j}
 =A_\lambda(\sigma^j\omega)\binom{u_j}{u_{j-1}}.
\end{equation}
Their projective action describes the evolution
of the direction of this pair.
We study both the law of the transformation obtained
after a prescribed number of steps and the distribution
of the resulting direction.

We identify the projective line with
\begin{equation}\label{eq:prioj-space}
 M=\mathbb{RP}^1=\R/\pi\Z
\end{equation}
and use the angular distance
\begin{equation}\label{eq:angular-distance}
 d(\phi,\psi)=\min_{m\in\Z}|\phi-\psi+m\pi|.
\end{equation}
We write $\Prob(M)$ for the Borel probability measures
on $M$ and $[z]$ for the line spanned by a nonzero
vector $z\in\R^2$.
For $k\ge1$, put
\begin{equation}\label{eq:route-laws}
 \begin{aligned}
 F_{\lambda,\omega}^{(k)}([z])
 &=
 \left[
 A_\lambda(\sigma^{-(k-1)}\omega)^{-1}
 \cdots A_\lambda(\omega)^{-1}z
 \right],\\
 \mu_\lambda^{(k)}
 &=
 \operatorname{Law}_{\mathbf P}
       \bigl(F_{\lambda,\omega}^{(k)}\bigr) = \int_\Omega
       \delta_{F_{\lambda,\omega}^{(k)}}\,d\mathbf P(\omega),
 \end{aligned}
\end{equation}
where $\delta_F$
denotes the Dirac measure at $F$.
We regard $\mu_\lambda^{(k)}$ as a probability measure
on the space of smooth diffeomorphisms of $M$,
denoted by $\operatorname{Diff}^{\infty}(M)$, equipped with
the $C^\infty$ topology.

We first study whether almost every transformation
in \eqref{eq:route-laws} can send one fixed probability
measure on $M$ to a common image.
Excluding such deterministic images is a
non-degeneracy condition used in results on growth
of matrix products, regularity of projective
distributions and localisation
\cite{GoldsheidMarkov,GKNonstationary,GKM,DGKBlockFactors}.
We establish this condition under general dynamical
assumptions and determine when it persists
as the product length increases.

\subsection{Non-degeneracy of finite transfer products}
\label{subsec:main-results}

\begin{definition}\label{def:ndi}
A probability measure $\mu$ on $\operatorname{Diff}^1(M)$
satisfies the \emph{no deterministic images condition},
abbreviated NDI, if there are no
$\eta,\eta'\in\Prob(M)$ such that
\begin{equation}\label{eq:deterministic-image}
 F_*\eta=\eta'
 \qquad\text{for $\mu$-almost every }F.
\end{equation}
Here $F_*\eta$ denotes the push-forward of $\eta$ under $F$, 
namely, for every Borel set $E\subset M$,
\[
 (F_*\eta)(E)=\eta(F^{-1}(E)).
\]
\end{definition}

For the laws in \eqref{eq:route-laws}, NDI is equivalent
to the absence of an invariant probability measure
for the relative group
\begin{equation}\label{eq:group}
 G_\lambda^{(k)}
 =
 \overline{\left\langle
 F\circ\widetilde F^{-1}:
 F,\widetilde F\in\operatorname{supp}\mu_\lambda^{(k)}
 \right\rangle},
\end{equation}
where the closure is taken in the group of projective
transformations.
If $B$ and $C$ are transfer products representing
$F$ and $\widetilde F$, respectively, then
\[
 (F\circ\widetilde F^{-1})([z])=[BC^{-1}z].
\]
Indeed, a common-image relation extends to the support
by continuity, and its common image is invariant
under the relative group.
Conversely, if the group preserves $\eta'$, fixing
$F_0$ in the support and setting
$\eta=(F_0^{-1})_*\eta'$ gives
$F_*\eta=\eta'$ throughout the support.

We note that one step always fails NDI,
since every inverse matrix in \eqref{eq:scalar-matrix}
sends $[1:0]$ to $[0:1]$.
The first possible length is therefore two.

\begin{theorem}\label{thm:main-ndi}
Let $H_\omega$ be given by
\eqref{eq:operator-introduction}.
Assume that $\sigma$ is mixing in the sense of
\eqref{eq:mix-prob}--\eqref{eq:mix-test},
and that $f$ is measurable 
and not constant
$\mathbf P$-almost everywhere.
Then $\mu_\lambda^{(2)}$ satisfies NDI
for every $\lambda\in\R$.

If $f$ has finite essential range, then
$\mu_\lambda^{(k)}$ satisfies NDI for every
$\lambda\in\R$ and every $k\ge2$.
\end{theorem}

Quite surprisingly, when $f$ has infinite
essential range, mixing can be weakened
to ergodicity.
Here infinite essential range means that
\[
 \mathbf P\{f\in E\}<1
 \qquad
 \text{for every finite set }E\subset\R.
\]

\begin{corollary}\label{cor:ergodic-infinite-range}
Let $H_\omega$ be given by
\eqref{eq:operator-introduction}.
Assume that $\sigma$ is ergodic and that $f$
is measurable 
and has infinite essential range.
Then $\mu_\lambda^{(2)}$ satisfies NDI
for every $\lambda\in\R$.
\end{corollary}

In the case of Schr\"odinger cocycles, NDI has a direct geometric meaning:
a projective point represents a linear boundary condition, and
\eqref{eq:route-laws} describes its transfer across a finite interval
at a fixed energy.
NDI excludes fixed input and output probability distributions
related by \eqref{eq:deterministic-image} for almost every realisation
of the potential.
This condition, used in the random-product results discussed below,
thus expresses non-degeneracy of the family of transfer maps.
For the quasiperiodic potentials considered later,
we also relate finite-step phase distributions to distributions
of invariant directions, whenever such directions exist.

Schr\"odinger cocycles also become strictly monotone
with respect to the energy after two steps
\cite{AvilaKrikorianMonotonic}.
This is a different property, present even for
constant potentials, which concerns variation
with energy rather than non-degeneracy of
a transformation law at a fixed energy. Nevertheless, the coincidence 
of the two-step thresholds for these different properties is worth noting.

Theorem~\ref{thm:main-ndi} applies, in particular,
to the mixing subshifts of finite type and
hyperbolic systems considered in \cite{ADZ}.
Corollary~\ref{cor:ergodic-infinite-range}
gives the two-step conclusion for every ergodic
Schr\"odinger operator in that setting whenever
$f$ has infinite essential range.
Note that in these general NDI results we do not
assume any of the following hypotheses used
in \cite{ADZ}: H\"older continuity of $f$,
local product structure of the invariant
measure, or the existence of a fixed point.

Corollary~\ref{cor:ergodic-infinite-range} applies
to quasiperiodic potentials generated by any measurable
function with infinite essential range, for every
irrational frequency and every real spectral parameter.
For nonconstant real trigonometric polynomials,
including the almost Mathieu potential, and for the analytic
class of Subsection~\ref{subsec:analytic-extension}, we prove more:
NDI holds for every $k\ge2$ and every frequency,
including rational frequencies
(Theorems~\ref{prop:trigonometric-all-k-ndi}
and~\ref{thm:analytic-extension}).

Goldsheid \cite{GoldsheidMarkov} uses the relative-group
condition \eqref{eq:group} in the study of exponential growth
of Markov-dependent matrix products.
The work considers laws of consecutive products
and it is noted that an inclusion between relative groups,
valid for independent products, may fail under
Markov dependence.
Two-step non-degeneracy for Schr\"odinger matrices
is established under an additional support condition
on consecutive coefficients
\cite[Proposition~3.2, Lemma~3.3]{GoldsheidMarkov}.
Theorem~\ref{thm:main-ndi} and
Corollary~\ref{cor:ergodic-infinite-range}
give two-step conclusions without any additional
conditions on the joint distribution of consecutive
potential values.

The equivalent \emph{measures condition} is a principal
hypothesis in the non-stationary Furstenberg theorem
of Gorodetski--Kleptsyn \cite{GKNonstationary}
and the central limit theorem of
Gorodetski--Kleptsyn--Monakov \cite{GKMCLT}.
Gorodetski--Kleptsyn--Monakov \cite{GKM} use NDI
to obtain finite-step H\"older regularisation above
an exponentially decreasing scale, while
Monakov \cite{MonakovLogHolder} extends the regularity
theory to log-H\"older estimates.
The condition also enters the localisation results
of Damanik--Gorodetski--Kleptsyn \cite{DGKBlockFactors}.

The length of the product matters in these applications.
In the independent setting,
Gorodetski--Kleptsyn--Monakov \cite{GKM} prove that
absence of a common invariant probability implies
NDI for some convolution power.
Gorodetski--Kleptsyn explicitly discuss replacing
one-step assumptions by assumptions on longer
products \cite[Remark~1.8]{GKNonstationary}.
For correlated potentials, $\mu_\lambda^{(k)}$
is generally not a convolution power of the one-step law.
We therefore study persistence directly for
the laws in \eqref{eq:route-laws}.

For finite-state Markov chains, universal persistence
for locally constant potentials admits a complete
classification.

Let $\{X_j\}_{j\in\Z}$ be a stationary Markov chain
on a finite alphabet
$\mathcal A=\{1,\ldots,\ell\}$, $\ell\ge2$,
with irreducible transition matrix $P=\{p_{vw}\}$
and stationary distribution $\pi$.
Its Markov operator is
\begin{equation}\label{eq:finite-markov-operator}
 (Ph)(v)=\sum_{w\in\mathcal A}p_{vw}h(w).
\end{equation}
We allow periodic chains.
The matrix $P$ is primitive if there exists
an integer $r\ge1$ such that
\[
 (P^m)_{vw}>0
 \qquad\text{for every }m\ge r
 \text{ and all }v,w\in\mathcal A.
\]

The space of admissible two-sided trajectories is
\begin{equation}\label{eq:symbolic-space}
 \Omega_P
 =
 \left\{\omega\in\mathcal A^{\Z}:
 p_{\omega_j\omega_{j+1}}>0
 \text{ for every }j\in\Z\right\},
 \qquad X_j(\omega)=\omega_j.
\end{equation}
We equip $\Omega_P$ with its stationary Markov
measure $\mathbf P$, which has full support in the relative product topology.
The left shift is
\begin{equation}\label{eq:shift}
 (\sigma\omega)_j=\omega_{j+1},
\end{equation}
and the potential is
\begin{equation}\label{eq:first-potential}
 v_j(\omega)=f(\sigma^j\omega),
 \qquad j\in\Z.
\end{equation}
The function $f$ may depend on the entire trajectory.
It is locally constant if it depends on a fixed
finite set of coordinates.

\begin{theorem}\label{thm:main-classification}
For an irreducible finite-state transition matrix $P$,
the following conditions are equivalent.
\begin{enumerate}
\renewcommand{\labelenumi}{\textup{(\alph{enumi})}}
\item $P$ is primitive.

\item For some $r\ge1$, the corresponding Markov operator $P$ of \eqref{eq:finite-markov-operator} satisfies 
\[
 \left\|P^r\big|_{L^2_0(\pi)}\right\|<1,
 \qquad
 L^2_0(\pi)
 =
 \left\{h:\mathcal A\to\R:
 \sum_{v\in\mathcal A}\pi(v)h(v)=0\right\}.
\]

\item For every nonconstant locally constant function
$f:\Omega_P\to\R$, every $\lambda\in\R$ and every
$k\ge2$, the transformation law $\mu_\lambda^{(k)}$
satisfies NDI.
\end{enumerate}
If the chain has period $d>1$, there exists a
nonconstant potential depending only on the current
state whose $2d$-step projective transformation
is the identity at $\lambda=0$.
\end{theorem}

Thus all-length NDI holds for every nonconstant
locally constant potential on the mixing subshifts
considered in \cite{ADZ}.
For general mixing invariant measures this follows
from Theorem~\ref{thm:main-ndi}, since locally constant
functions have finite range.
Theorem~\ref{thm:main-classification} classifies the
universal property among irreducible finite-state
Markov chains and does not require a fixed point.

We next consider measurable functions of general-state
Markov chains.
Let $\{X_j\}_{j\in\Z}$ be stationary on a standard
Borel state space $S$, with stationary distribution
$F$, transition kernel $K(v,dw)$ and Markov operator
\begin{equation}\label{eq:markov-operator}
 (Ph)(v)=\int_S h(w)\,K(v,dw).
\end{equation}
Let $V:S\to\R$ be measurable, 
with
\begin{equation}\label{eq:essential-oscillation}
 \essosc_F V
 :=
 \operatorname*{ess\,sup}_F V
 -
 \operatorname*{ess\,inf}_F V>0.
\end{equation}
We take
\begin{equation}\label{eq:second-potential}
 v_j(\omega)=V(X_j(\omega)),
 \qquad j\in\Z,
\end{equation}
or $f=V\circ X_0$ on the stationary two-sided
trajectory space.
When $S\subseteq\R$ and $V(v)=v$, the potential
is the Markov chain itself.
This includes hidden Markov potentials,
for which the potential process need not
itself be Markov.
Both $S$ and the essential range of $V$ may be infinite.

Our assumption on the Markov operator is the strict
one-step condition, or one-step $L^2$ gap,
\begin{equation}\label{eq:one-step-gap}
 \rho
 :=
 \left\|P\big|_{L^2_0(F)}\right\|<1,
 \qquad
 L^2_0(F)
 =
 \left\{h\in L^2(F):\int_S h\,dF=0\right\}.
\end{equation}

\begin{theorem}\label{thm:main-markov-ndi}
Let $H_\omega$ be given by
\eqref{eq:operator-introduction}, with potential
\eqref{eq:second-potential}.
Under \eqref{eq:essential-oscillation} and
\eqref{eq:one-step-gap}, the transformation law
$\mu_\lambda^{(k)}$ satisfies NDI for every
$\lambda\in\R$ and every $k\ge2$.
\end{theorem}

\subsection{Uniform Wasserstein separation}

We now prove a positive lower bound for the averaged
Wasserstein distance between $F_*\eta$ and $\eta'$,
uniform over all probability measures $\eta,\eta'$
and over the spectral parameter on every compact
interval.

\begin{definition}\label{def:w1}
For $\eta,\eta'\in\Prob(M)$, let
$\operatorname{Cpl}(\eta,\eta')$ denote the probability
measures on $M\times M$ whose coordinate distributions are $\eta$ and $\eta'$.
Their $1$-Wasserstein distance is
\begin{equation}\label{eq:wasserstein-definition}
 W(\eta,\eta')
 =
 \inf_{\Gamma\in\operatorname{Cpl}(\eta,\eta')}
 \int_{M\times M}d(\phi,\psi)\,d\Gamma(\phi,\psi),
\end{equation}
where $d$ is the angular distance
\eqref{eq:angular-distance}.
\end{definition}

\begin{theorem}\label{thm:main-wasserstein}
Let $H_\omega$ be given by
\eqref{eq:operator-introduction}.
Fix $k\ge2$ and a compact interval $\Lambda\subset\R$.
Then $\mu_\lambda^{(k)}$ satisfies NDI for every
$\lambda\in\Lambda$ if and only if there exists $d_*>0$
such that
\begin{equation}\label{eq:main-wasserstein}
 \inf_{\lambda\in\Lambda}
 \inf_{\eta,\eta'\in\Prob(M)}
 \int_{\operatorname{Diff}^{\infty}(M)}
 W(F_*\eta,\eta')\,d\mu_\lambda^{(k)}(F)
 \ge d_*.
\end{equation}
\end{theorem}

Theorems~\ref{thm:main-ndi}--\ref{thm:main-markov-ndi}
and Corollary~\ref{cor:ergodic-infinite-range}
supply the corresponding NDI hypotheses.
The constant $d_*$ may depend on the stationary
system, the potential, $k$ and $\Lambda$.

To the best of our knowledge, this uniform
Wasserstein separation from deterministic images
has not previously been stated explicitly for any model.
It turns qualitative NDI into a quantitative
estimate for approximate common images, uniform
over both probability measures.
This formulation can be used beyond the present setting.
In particular, it simplifies part of the regularisation
argument of Gorodetski--Kleptsyn--Monakov \cite{GKM}
and is used in the proof of our Markov regularisation
theorem (Theorem~\ref{thm:main-regularity}).

Wasserstein distance arises naturally in spectral
theory, for example when estimating how close
the density-of-states measures are for nearby
distributions of independent potential values
\cite{HislopMarxI,HislopMarxII,ShamisIDS}.

\subsection{Regularisation of projective distributions}

We now consider the Markov and hidden Markov potentials
defined by \eqref{eq:markov-operator}--\eqref{eq:second-potential}.
For a fixed initial direction $\Phi_0\in M$, put
\[
 \Phi_N=F_{\lambda,\omega}^{(N)}\Phi_0,
 \qquad
 Y_N=X_{-N},
 \qquad N\ge0,
\]
where $F_{\lambda,\omega}^{(N)}$ is given by
\eqref{eq:route-laws} and
$F_{\lambda,\omega}^{(0)}=\operatorname{Id}_M$.
We write
\begin{equation}\label{eq:intro-joint-laws}
 \begin{aligned}
 \nu_{N,\lambda}
 &=\operatorname{Law}(\Phi_N)
 =\int_{\operatorname{Diff}^{\infty}(M)}
       F_*\delta_{\Phi_0}\,d\mu_\lambda^{(N)}(F),\quad
 \mathcal P_{N,\lambda}
 =\operatorname{Law}(\Phi_N,Y_N).
 \end{aligned}
\end{equation}
Here $\delta_{\Phi_0}$ is the Dirac measure at $\Phi_0$;
the integral representation is for $N\ge1$.
Thus $\mu_\lambda^{(N)}$ of \eqref{eq:route-laws} is the distribution of the
projective transformation, and $\nu_{N,\lambda}$
is the distribution of the resulting direction on $M$.

More generally, we allow any initial joint distribution
on $M\times S$ with state distribution $F$.
We evolve these distributions under \eqref{eq:skew-transition},
since the next transfer matrix depends on the current state.
A joint distribution is stationary if the transition preserves it.

Let $B_r(x)$ be the open ball of radius $r$
centred at $x\in M$ in the angular metric
\eqref{eq:angular-distance}.
We seek estimates of the form
\begin{equation}\label{eq:intro-holder-measure}
 \nu_{N,\lambda}(B_r(x))\le Cr^\alpha
\end{equation}
with constants independent of $N$ and $x$.
For finitely valued potentials and a fixed initial
direction, $\nu_{N,\lambda}$ has finite support
at every finite $N$.
The question is how many steps are needed for
a H\"older estimate to hold above a prescribed scale.
Under our assumptions, we obtain such an estimate
above an exponentially decreasing cutoff.

Finite-step H\"older estimates for iterated
averaged push-forwards are obtained in
\cite[Theorem~2.8]{GKM}, uniformly over the initial
probability and transformation laws in a compact
family satisfying NDI and a uniform moment condition.
For the potential \eqref{eq:second-potential},
these averaged iterations do not in general
describe the correlated evolution
\eqref{eq:intro-joint-laws}.
We use part of the method of \cite{GKM}
to obtain a substantial extension of finite-step
H\"older regularisation to Markov and hidden
Markov Schr\"odinger dynamics.

In addition to \eqref{eq:one-step-gap} and
\eqref{eq:essential-oscillation}, we assume
\begin{equation}\label{eq:bounded-observation}
 V\in L^\infty(F).
\end{equation}

\begin{theorem}
\label{thm:main-regularity}
Let $H_\omega$ be the operator
\eqref{eq:operator-introduction} with potential
\eqref{eq:second-potential}.
Assume \eqref{eq:one-step-gap},
\eqref{eq:essential-oscillation} and
\eqref{eq:bounded-observation}.
Fix a compact interval $\Lambda\subset\R$
of spectral parameters.
\begin{enumerate}
\item
There exist $\alpha>0$, $C<\infty$ and
$0<\kappa_0<1$ such that
\begin{equation}\label{eq:main-finite-holder}
 \nu_{N,\lambda}(B_r(x))\le Cr^\alpha
 \qquad\text{whenever}\qquad
 r>\kappa_0^{\lfloor N/3\rfloor}
\end{equation}
for every $N\ge0$, $\lambda\in\Lambda$ and $x\in M$.
The constants are uniform over all initial
joint distributions on $M\times S$
with state distribution $F$.

\item
Stationary joint distributions with state
distribution $F$ exist.
The distribution $\nu_\lambda$ of the direction
of every such stationary distribution satisfies
\begin{equation}\label{eq:main-stationary-holder}
 \nu_\lambda(B_r(x))\le Cr^\alpha,
 \qquad r>0,\quad x\in M.
\end{equation}
The same estimate holds for every weak limit
of the averages
\begin{equation}\label{eq:phase-law-averages}
 \overline\nu_{N,\lambda}
 =
 \frac1N\sum_{n=0}^{N-1}\nu_{n,\lambda},
 \qquad N\ge1.
\end{equation}

\item
For a stationary joint distribution, let
$\mu_{v,\lambda}$ be the conditional distribution
of the direction given that the state is $v$.
There exist $0<\gamma<1/2$ and $C_1<\infty$
such that
\begin{equation}\label{eq:main-averaged-fibre-energy}
 \int_S\iint_{M\times M}d(\phi,\psi)^{-\gamma}\,
 d\mu_{v,\lambda}(\phi)d\mu_{v,\lambda}(\psi)\,dF(v)
 \le C_1.
\end{equation}
In particular, for $F$-almost every $v$,
\begin{equation}\label{eq:main-conditional-holder}
 \mu_{v,\lambda}(B_r(x))
 \le C_1(v)r^{\gamma/2},
 \qquad r>0,\quad x\in M,
\end{equation}
where
\begin{equation}\label{eq:main-conditional-constant}
 \int_S C_1(v)^2\,dF(v)\le C_*<\infty.
\end{equation}
The constants in
\eqref{eq:main-stationary-holder},
\eqref{eq:main-averaged-fibre-energy}
and \eqref{eq:main-conditional-constant}
are uniform in $\lambda\in\Lambda$
and in the stationary joint distribution.
The function $C_1(v)$ may depend on both.
\end{enumerate}
\end{theorem}

\begin{remark}
\label{rem:cutoff-gap}
Let $d_*>0$ be the two-step separation constant
in \eqref{eq:main-wasserstein}, and let $\mathscr L$
bound the Lipschitz constants of the one-step
projective maps and their inverses, as in
\eqref{eq:uniform-lipschitz}.
The proof permits the choice
\begin{equation}\label{eq:cutoff-gap}
 \kappa_0=1-c(\mathscr L,d_*)(1-\rho^2)^2
\end{equation}
with $c(\mathscr L,d_*)>0$ sufficiently small.
Indeed, \eqref{eq:zeta-choice} allows
$\zeta=c(\mathscr L,d_*)(1-\rho^2)^2$,
and the subsequent choice of the cutoff allows
$\kappa_0=1-\zeta$.
The exponent $\alpha$ and the constant $C$
are chosen accordingly.
Thus
\[
 \kappa_0^{\lfloor N/3\rfloor}
 \le
 \exp\!\left(
 -c(\mathscr L,d_*)(1-\rho^2)^2
       \lfloor N/3\rfloor
 \right).
\]
We do not optimise or use this explicit choice
elsewhere in the paper.
It displays the quantitative contribution of
both the Markov gap and the Wasserstein separation.
The latter also depends on the chain.
Its positivity follows from
Theorem~\ref{thm:main-markov-ndi}
under \eqref{eq:one-step-gap}.
\end{remark}

The regularity of stationary projective measures
is a longstanding and important subject in the
theory of random matrix products.
These measures describe the asymptotic distribution
of directions and enter Furstenberg's formula
for the Lyapunov exponent
\cite{Furstenberg,FurstenbergKifer}.
For independent products, H\"older regularity
under strong irreducibility, proximality and
a positive moment assumption was established
in \cite{GuivarchRegularity} 
(see also \cite[Chapter~14]{BenoistQuintBook}).

For Markov cocycles, large deviation estimates
and continuity of Lyapunov exponents are studied
in \cite{DKRandom}.
H\"older dependence of the stationary projective
measure and the maximal Lyapunov exponent on
the cocycle and transition kernel is established
in \cite{CDKM}.

By the Frostman principle,
\eqref{eq:main-stationary-holder} implies that
$\nu_\lambda$ gives zero mass to every Borel set
of Hausdorff dimension smaller than $\alpha$.
An application of such estimates in spectral
theory is given in
\cite[Lemma~4.1 and Remark~4.2]{ALSZ}.

Theorem~\ref{thm:main-regularity} does not assume
positivity of the Lyapunov exponent.
The finite-step estimate is used to prove
uniform positivity of the exponent \cite{PSMarkov}.

The Markov condition \eqref{eq:one-step-gap}
enters the proof in two distinct ways.
It controls averaging over the states directly
and also gives the geometric NDI condition
through Theorem~\ref{thm:main-markov-ndi}.
The uniform Wasserstein lower bound connects 
these two parts of the proof.

\subsection{Trigonometric potentials and an analytic extension}
\label{subsec:amo-introduction}

We next consider deterministic quasiperiodic potentials
\begin{equation}\label{eq:trigonometric-potential}
 \begin{gathered}
 v_j(\theta)=V(\theta+2\pi\alpha j),
 \qquad
 V(\theta)=\sum_{k=-d}^{d}a_ke^{ik\theta},\\
 a_{-k}=\overline{a_k},
 \qquad a_d\ne0,
 \qquad d\ge1.
 \end{gathered}
\end{equation}
The base transformation is the circle rotation
$\sigma\theta=\theta+2\pi\alpha\pmod{2\pi}$,
with normalised Lebesgue probability
$dm(\theta)=d\theta/(2\pi)$ on $\R/2\pi\Z$.
We denote the transfer matrix
\eqref{eq:scalar-matrix} for this potential
by $A_{\lambda,V}(\theta)$.

Arguably the most famous and most extensively
studied representative of this class is the
almost Mathieu operator
\eqref{eq:operator-introduction} with potential
\begin{equation}\label{eq:amo-potential}
 v_j(\theta)=g\cos(2\pi\alpha j+\theta),
 \qquad j\in\Z,\qquad g\ne0.
\end{equation}


Even more surprisingly, NDI for trigonometric potentials does
not even require ergodicity: it holds for every $k\ge2$, every energy
and every frequency, including rational ones. The same is
true for the analytic class introduced in
Subsection~\ref{subsec:analytic-extension}.
For general real-analytic potentials, however, this need not
hold once ergodicity is dropped. We give an example of a
real-analytic potential and a rational rotation for which the
two-step transformations already have a common invariant Dirac
measure (see Remark~\ref{rmk:analytic}). This rotation is not ergodic with respect to Lebesgue
measure, so there is no contradiction with
Theorem~\ref{thm:main-ndi} or
Corollary~\ref{cor:ergodic-infinite-range}.

\begin{theorem}\label{prop:trigonometric-all-k-ndi}
Let $V$ be a nonconstant real trigonometric polynomial
as in \eqref{eq:trigonometric-potential}.
For every $\alpha\in\R$, $\lambda\in\R$ and
$k\ge2$, the transformation law $\mu_\lambda^{(k)}$
defined by \eqref{eq:route-laws}, with the initial
phase distributed according to normalised Lebesgue
measure on $\R/2\pi\Z$, satisfies NDI.
The same conclusion holds for the law of the
projective action of the forward product
\eqref{eq:trigonometric-forward-product}.

Moreover, for every fixed $k\ge2$ and compact
interval $I\subset\R$, there exists
$d_k=d_k(V,I)>0$ such that
\begin{equation}\label{eq:trigonometric-k-wasserstein}
 \inf_{\substack{\alpha\in\R/\Z,\ \lambda\in I\\
                 \eta,\eta'\in\Prob(M)}}
 \int_{\operatorname{Diff}^{\infty}(M)}
 W(F_*\eta,\eta')\,d\mu_\lambda^{(k)}(F)
 \ge d_k.
\end{equation}
\end{theorem}

For an initial probability $\eta\in\Prob(M)$
independent of $\theta$, define
\begin{equation}\label{eq:trigonometric-phase-law}
 \nu_{n,\lambda}^{\alpha,V,\eta}
 =
 \int_{\R/2\pi\Z}
 (F_{\lambda,\theta}^{(n)})_*\eta\,dm(\theta),
\end{equation}
where $F_{\lambda,\theta}^{(n)}$ is given by
\eqref{eq:route-laws} for
\eqref{eq:trigonometric-potential}.
For every Borel set $E\subset M$,
\[
 \nu_{n,\lambda}^{\alpha,V,\eta}(E)
 =
 \int_{\R/2\pi\Z}
 \eta\bigl(\{v\in M:F_{\lambda,\theta}^{(n)}v\in E\}\bigr)
 \,dm(\theta).
\]
For each fixed $\theta$, the integrand is the $\eta$-measure
of all initial directions whose images after $n$ steps lie in $E$.
For a fixed initial direction $v\in M$, take $\eta=\delta_v$,
where $\delta_v$ is the Dirac measure at $v$, and write
$\nu_{n,\lambda}^{\alpha,V,v}$.
Then, for every Borel set $E\subset M$,
\[
 \nu_{n,\lambda}^{\alpha,V,v}(E)
 =
 m\bigl(\{\theta\in\R/2\pi\Z:
             F_{\lambda,\theta}^{(n)}v\in E\}\bigr).
\]
Thus $\nu_{n,\lambda}^{\alpha,V,v}(E)$ is the measure of all
initial phases of the potential for which the direction
after $n$ steps lies in $E$.
For \eqref{eq:amo-potential}, the corresponding
notation is $\nu_{n,\lambda}^{\alpha,g,\eta}$.
These definitions include rational frequencies.

\begin{theorem}
\label{thm:trigonometric-regularity}
Let $V$ be given by \eqref{eq:trigonometric-potential}.
For every compact interval $I\subset\R$, there exists
$C=C(V,I)<\infty$ with the following properties.
\begin{enumerate}
\item
For every $n\ge2$, $\alpha\in\R$,
$\lambda\in I$, every phase-independent
$\eta\in\Prob(M)$, $w\in M$ and $0<r\le1$,
\begin{equation}\label{eq:trigonometric-holder}
 \nu_{n,\lambda}^{\alpha,V,\eta}(B_r(w))
 \le Cr^{1/(2dn)}.
\end{equation}
The ball is taken in the angular metric
\eqref{eq:angular-distance}.

\item
With the same constant, for every $a\ge0$,
\begin{equation}\label{eq:trigonometric-logarithmic-excess}
 \frac1n\int_M
 (-\log d(z,w)-an)_+\,
 d\nu_{n,\lambda}^{\alpha,V,\eta}(z)
 \le 2dC e^{-a/(2d)},
\end{equation}
uniformly over the parameters in part~\textup{(1)}.
Here $s_+=\max\{s,0\}$ and $d$ is the angular
metric \eqref{eq:angular-distance}.
\end{enumerate}
For fixed $d$ and $I$, the constant can be chosen
uniformly over coefficient families satisfying
\[
 \sum_{k=-d}^{d}|a_k|\le A_*,
 \qquad |a_d|\ge a_*>0,
\]
where $A_*<\infty$ and $a_*$ are fixed.
\end{theorem}
At the exponential scale $r=e^{-an}$,
\eqref{eq:trigonometric-holder} gives
\[
 \nu_{n,\lambda}^{\alpha,V,\eta}(B_{e^{-an}}(w))
 \le Ce^{-a/(2d)}.
\]

\begin{corollary}
\label{cor:trigonometric-logarithmic-ui}
Under the parameter restrictions of
Theorem~\ref{thm:trigonometric-regularity}, put
\[
 Z_{n,w}(z)=\frac1n\log^+\frac1{d(z,w)},
 \qquad
 \log^+s=\max\{\log s,0\},
\]
where $d$ is the angular metric
\eqref{eq:angular-distance}.
For every $a\ge0$,
\begin{align}
 \nu_{n,\lambda}^{\alpha,V,\eta}
       \{Z_{n,w}>a\}
 &\le Ce^{-a/(2d)},
       \label{eq:trigonometric-logarithmic-tail}\\
 \int_M Z_{n,w}\mathbf1_{\{Z_{n,w}>a\}}\,
       d\nu_{n,\lambda}^{\alpha,V,\eta}
 &\le C(a+2d)e^{-a/(2d)}.
       \label{eq:trigonometric-logarithmic-ui}
\end{align}
In particular, these normalised logarithms
are uniformly integrable over all the indicated
parameters, initial probabilities and directions.
\end{corollary}

Estimate~\eqref{eq:trigonometric-holder} bounds the measure of phases for
 which a solution satisfying one boundary condition nearly 
 satisfies the prescribed condition at the other endpoint.
Corollary~\ref{cor:trigonometric-logarithmic-ui}
controls the logarithmic singularity at
coincidence, with normalisation by the
interval length.

Small-value estimates for phase-dependent
finite-interval determinants play an important
role in quasiperiodic spectral theory.
Such estimates are used in
\cite{GoldsteinSchlagFine,TaoVodaIDS}
to study Dirichlet eigenvalues and regularity
of the integrated density of states under
positive Lyapunov exponent and arithmetic
assumptions.
Related questions involving periodic
approximants are studied in
\cite{ShamisConnections,JitomirskayaMarx}.

For every fixed $k\ge2$,
Theorems~\ref{prop:trigonometric-all-k-ndi}
and~\ref{thm:main-wasserstein} give a H\"older
concentration bound directly from Wasserstein separation.
Let $d_k>0$ be the separation constant, uniform
for $\alpha\in\R$ and $\lambda\in I$, and let
$B_k$ bound the Euclidean operator norms of the
matrices representing $F_{\lambda,\theta}^{(k)}$
over these parameters and all $\theta$.
Then
\begin{equation}\label{eq:intro-two-step-wasserstein-holder}
 \nu_{k,\lambda}^{\alpha,V,\eta}(B_r(w))
 \le
 C\left(\frac{B_k^2}{d_k}\right)^{1/(2dk)}
 r^{1/(2dk)},
 \qquad 0<r\le1.
\end{equation}
For the almost Mathieu potential
\eqref{eq:amo-potential}, the exponent is $1/(2k)$.

\subsection{Phase distributions of Oseledets directions}

For a bounded ergodic operator, the Lyapunov exponent is defined by
\begin{equation}\label{eq:doubling-lyapunov-definition}
 L(\lambda)
 =
 \lim_{N\to\infty}\frac1N
 \int
 \log\left\|
 A_\lambda(\sigma^{N-1}\omega)\cdots
 A_\lambda(\omega)
 \right\|\,d\mathbf P(\omega).
\end{equation}
For \eqref{eq:trigonometric-potential}, we write
$L_{\alpha,V}(\lambda)$.

\smallskip

We now assume $\alpha\in\R\setminus\Q$.

\smallskip

A measurable invariant projective direction
is a function $q:\R/2\pi\Z\to M$ satisfying
\[
 F_{\lambda,\theta}^{(1)}q(\theta)
 =q(\theta-2\pi\alpha)
\]
for almost every $\theta$.
Its graph carries the invariant probability
\begin{equation}\label{eq:intro-invariant-graph-measure}
 \mathcal P^q=(\operatorname{Id},q)_*m,
 \qquad
 \begin{aligned}
 (\operatorname{Id},q):
 \R/2\pi\Z&\longrightarrow(\R/2\pi\Z)\times M,\\
 \theta&\longmapsto(\theta,q(\theta)).
 \end{aligned}
\end{equation}
The vertical projection of this measure is
the push-forward measure
$q_*m$, which gives the distribution of the invariant
direction under phase averaging.
In particular,
\[
 (q_*m)(B)=m\{\theta:q(\theta)\in B\},
 \qquad B\subset M\ \text{Borel}.
\]
Although $\mathcal P^q$ is non-atomic because its projection
onto the circle is Lebesgue measure, the measure $q_*m$
can have atoms.

When $L_{\alpha,V}(\lambda)>0$, Oseledets'
theorem \cite{Oseledets} gives two such invariant
directions for the inverse transfer dynamics
\eqref{eq:route-laws}: the stable direction
$s(\theta)$ and the unstable one $u(\theta)$.
For almost every $\theta$, nonzero vectors
in these directions have exponential growth
rates $-L_{\alpha,V}(\lambda)$ and
$L_{\alpha,V}(\lambda)$, respectively.
In the following theorem we show that for
the trigonometric potentials
\eqref{eq:trigonometric-potential} both Oseledets
projections are non-atomic.

\begin{theorem}
\label{thm:oseledets-distributions}
Let $V$ be a nonconstant real trigonometric
polynomial and let $\alpha\in\R\setminus\Q$.
For every $\lambda\in\R$ with
$L_{\alpha,V}(\lambda)>0$, both
$s_*m$ and $u_*m$ are non-atomic.
Moreover,
\begin{equation}\label{eq:amo-positive-le-limit}
 \nu_{n,\lambda}^{\alpha,V,\eta_0}
 \xrightarrow{w^*}u_*m
\end{equation}
for every initial probability $\eta_0$
independent of the initial phase.
Consequently, for every such $\eta_0$,
\begin{equation}\label{eq:trigonometric-asymptotic-concentration}
 \lim_{r\downarrow0}\,
 \limsup_{n\to\infty}\,
 \sup_{w\in M}
 \nu_{n,\lambda}^{\alpha,V,\eta_0}(B_r(w))
 =0.
\end{equation}
\end{theorem}

More generally, the non-atomicity conclusion
holds for the phase distribution $q_*m$
of every measurable invariant projective
direction, whenever such a direction exists.

For independent non-stationary products, NDI
implies exponential decay of the largest atomic
mass of the evolving measure
\cite[Theorem~1.13]{GKNonstationary}.
Theorem~\ref{thm:oseledets-distributions}
concerns the distributions of invariant directions
for deterministic quasiperiodic products,
where the transfer matrix and the invariant
direction depend on the same phase.
Moreover, decay of finite-step atomic masses
alone does not imply non-atomicity of a weak limit.

Invariant graphs and projective dynamics
of quasiperiodic Schr\"odinger cocycles have
been studied extensively.
For the almost Mathieu operator,
\cite{HaroPuig} relates strange nonchaotic
attractors to spectral properties.
For large coupling, Diophantine frequency
and a class of $C^2$ potentials,
\cite{BjerklovDynamics} gives a detailed
description of the invariant graph measures
and minimality of the projective dynamics.
Further results for selected sets of
frequencies are obtained in
\cite{BjerklovFrequencies}.

Dimension results for measures supported
on invariant graphs are established in
\cite{GrogerJager}.
The extension of these results to the
projective dynamics of the almost Mathieu
operator is identified there as a further
problem.
For general measurable cocycles, the joint
distribution of the Oseledets subspaces
can have considerable freedom
\cite{BochiLessa}.
Questions about dimension of these
distributions without independence are
also raised in \cite{BochiLessa}.
For the Schr\"odinger cocycles considered here,
Theorem~\ref{thm:oseledets-distributions}
leads to the question of quantitative
regularity and dimension of the phase
distributions of the Oseledets directions.

The convergence
\eqref{eq:amo-positive-le-limit}
gives a criterion for vanishing of the
Lyapunov exponent involving only one
fixed initial direction.
Combined with analytic rotation conjugacy,
it gives the following almost-everywhere
characterisation.

\begin{corollary}
\label{cor:atomic-reducibility}
Let $V$ be a nonconstant real trigonometric
polynomial and let $\alpha\in\R\setminus\Q$.
There exists a Borel set
$\mathcal N_{\alpha,V}\subset\R$ of Lebesgue
measure zero such that, for every
$\lambda\in\R\setminus\mathcal N_{\alpha,V}$,
the following conditions are equivalent.
\begin{enumerate}
\item
$L_{\alpha,V}(\lambda)=0$.

\item
The Schr\"odinger cocycle is analytically
conjugate to a rotation-valued cocycle.

\item
There exists a sequence $n_j\to\infty$
such that
\[
 \nu_{n_j,\lambda}^{\alpha,V,v}
 \xrightarrow{w^*}\delta_v
 \qquad\text{for every }v\in M.
\]
The same sequence works for all $v$.

\item
For some fixed $v\in M$ and sequence
$n_j\to\infty$,
$\nu_{n_j,\lambda}^{\alpha,V,v}$
converges weakly to a probability with an atom.
\end{enumerate}
The implications
\[
 (2)\Longrightarrow(3)\Longrightarrow(4)
 \Longrightarrow(1)
\]
hold at every spectral parameter.
Only $(1)\Longrightarrow(2)$ requires
exclusion of a set of energies of
Lebesgue measure zero.
\end{corollary}

For the almost Mathieu potential
\eqref{eq:amo-potential}, write
$L_{g,\alpha}(\lambda)$ for the Lyapunov exponent.
Theorem~\ref{thm:trigonometric-regularity}
gives exponent $1/(2n)$, with the same
uniformity of the multiplicative constant.
For the subcritical almost Mathieu operator, the 
phase distributions can nevertheless return to 
the initial probability along a subsequence.

\begin{corollary}
\label{cor:amo-returns}
Fix $\alpha\in\R\setminus\Q$ and $0<|g|<2$.
For Lebesgue-almost every
$\lambda\in\Sigma_{g,\alpha}$, where
$\Sigma_{g,\alpha}$ is the spectrum of the
almost Mathieu operator, there exists
a sequence $N_j\to\infty$ such that
\begin{equation}\label{eq:amo-return-laws}
 \nu_{N_j,\lambda}^{\alpha,g,\eta}
 \xrightarrow{w^*}\eta
 \qquad\text{for every }\eta\in\Prob(M).
\end{equation}
For these spectral parameters and every
fixed $v\in M$, there are no constants
$C_0<\infty$, $\gamma>0$ and positive
sequence $r_n\to0$ such that
\begin{equation}\label{eq:amo-impossible-cutoff}
 \nu_{n,\lambda}^{\alpha,g,v}(B_r(w))
 \le C_0r^\gamma
 \qquad\text{for every }n\ge2,\quad
 w\in M,\quad r_n<r\le1.
\end{equation}
\end{corollary}

Thus finite-step H\"older regularity with
one multiplicative constant is compatible
with repeated concentration in the
zero-exponent regime.
The fixed positive exponent above an
exponentially decreasing cutoff in
Theorem~\ref{thm:main-regularity}
gives a stronger conclusion.  

\subsection{An extension to analytic potentials}
\label{subsec:analytic-extension}

Theorems~\ref{prop:trigonometric-all-k-ndi}
and~\ref{thm:oseledets-distributions} and
Corollary~\ref{cor:atomic-reducibility} extend to the following
analytic class. The estimates
\eqref{eq:trigonometric-holder}--\eqref{eq:trigonometric-logarithmic-ui}
extend with constants and exponents depending on the number of steps.

Let $\mathcal C$ consist of the real-valued functions $V$,
analytic on $\R/2\pi\Z$, with the following property.
For every $\alpha\in\R$ and $k\ge2$, the functions
$V(\theta+2\pi j\alpha)$, $0\le j<k$, admit simultaneous analytic
continuation from their original branches along a common path
starting at some real phase. We also assume that this path
contains a sequence $\theta_\ell$ at which all the continued
values are finite and
\begin{equation}\label{eq:analytic-simultaneous-growth}
 \min_{0\le j<k}|V(\theta_\ell+2\pi j\alpha)|
 \longrightarrow\infty\qquad \text{as}\quad \ell\to\infty.
\end{equation}
The path and the sequence may depend on $\alpha$ and $k$.

A sufficient condition is that, for almost every $x\in\R$, the
original branch of $V$ admits analytic continuation along the
ray $x-iy$, $y\ge0$, and
\[
 |V(x-iy)|\longrightarrow\infty\qquad \text{as}\quad  y\to+\infty.
\]
For any finite set of shifts, these continuations and limits
then hold simultaneously for almost every $x$.

The class $\mathcal C$ contains nonconstant real trigonometric
polynomials and functions $V(\theta)=R(e^{i\theta})$, where $R$
is rational, real on the unit circle, has no poles there and
has a pole at infinity. Examples include
\[
 V(\theta)=\cos\theta+\frac1{a-\cos\theta},\qquad a>1,
\]
\[
 V(\theta)=P(\theta)+e^{Q(\theta)},\qquad
 V(\theta)=e^{Q(\theta)}+e^{-Q(\theta)},
\]
where $P,Q$ are real trigonometric polynomials, with $P$
nonconstant in the second example and $Q$ nonconstant in the
third. Along almost every downward vertical ray, a polynomial
term or one of the exponentials supplies the growth.

The class also contains
\[
 V(\theta)=\sqrt{3+\cos\theta},\qquad
 V(\theta)=\log(3+\cos\theta),
\]
with the positive square root and the real logarithm on the real
circle. These branches continue along every downward vertical
ray except $x=\pi\pmod{2\pi}$, which meets a complex zero of
$3+\cos\theta$. Along the remaining rays,
\[
 \left|\sqrt{3+\cos(x-iy)}\right|\sim2^{-1/2}e^{y/2},\qquad
 \operatorname{Re}\log(3+\cos(x-iy))=y-\log2+o(1).
\]

\begin{theorem}\label{thm:analytic-extension}
Let $V\in\mathcal C$, and consider
$v_j(\theta)=V(\theta+2\pi j\alpha)$.
\begin{enumerate}
\item Theorem~\ref{prop:trigonometric-all-k-ndi} holds.

\item For every fixed $k\ge2$ and compact interval $I\subset\R$,
there exist $C_k=C_k(V,I)<\infty$ and $\beta_k=\beta_k(V,I)>0$
such that
\begin{equation}\label{eq:analytic-fixed-holder}
 \nu_{k,\lambda}^{\alpha,V,\eta}(B_r(w))\le C_kr^{\beta_k}
\end{equation}
for all $\alpha\in\R$, $\lambda\in I$, phase-independent
$\eta\in\Prob(M)$, $w\in M$ and $0<r\le1$, with the
notation \eqref{eq:trigonometric-phase-law}.

\item For any $\alpha\in\R\setminus\Q$,
Theorem~\ref{thm:oseledets-distributions}
and Corollary~\ref{cor:atomic-reducibility} hold.
\end{enumerate}
If instead $V$ is meromorphic near $\R/2\pi\Z$,
real-valued at real phases away from finitely many poles
of finite order, and satisfies the continuation and growth
assumptions defining $\mathcal C$, then parts~\textup{(1)}
and~\textup{(2)}, the logarithmic estimates, and
Theorem~\ref{thm:oseledets-distributions} hold.
For $\alpha\in\R\setminus\Q$, implication~\textup{(4)}$\Rightarrow$\textup{(1)}
of Corollary~\ref{cor:atomic-reducibility} holds at every energy.
\end{theorem}

For fixed $k$, the logarithmic estimates
\eqref{eq:trigonometric-logarithmic-excess}--\eqref{eq:trigonometric-logarithmic-ui}
hold with $n=k$, $C$ replaced by $C_k$ and $2d$ by $(k\beta_k)^{-1}$.

\begin{remark}
The case of poles on the unit circle is treated separately
in the proof of Theorem~\ref{thm:analytic-extension}.
\end{remark}


\subsection{Counterexamples to regularisation}

Theorem~\ref{thm:main-regularity} is essentially
sharp with respect to boundedness of the
potential and the strict one-step Markov
norm condition.
First, boundedness cannot be replaced
by finiteness of every positive moment.

\begin{proposition}
\label{prop:unbounded-obstruction}
There exists an operator $H_\omega$ of the form
\eqref{eq:operator-introduction} with potential
\eqref{eq:second-potential}, where
$\{X_j\}_{j\in\Z}$ is a stationary Markov chain
on a countable real state space $S$,
with stationary distribution $F$, and
$V:S\to\R$ is nonconstant, such that
\begin{equation}\label{eq:intro-unbounded-moments}
 \left\|P\big|_{L^2_0(F)}\right\|<1,
 \qquad
 \int_S|V(v)|^\delta\,dF(v)<\infty
 \quad\text{for every }\delta>0,
\end{equation}
but at $\lambda=0$ the distribution of the direction
under every stationary joint distribution
fails to be H\"older with any positive exponent.

For every fixed initial direction $\Phi_0\in M$,
there are no constants $C<\infty$, $\alpha>0$
and positive sequence $r_N\to0$ such that
\[
 \nu_{N,0}(B_r(x))\le Cr^\alpha
 \qquad\text{for}\qquad
 N\ge1,\quad x\in M,\quad r>r_N.
\]
At $\lambda=0$, every positive moment of
the Lipschitz constants of the one-step
projective transformations and their
inverses is finite.
\end{proposition}

We construct a ``sticky'' Markov chain: with a fixed probability, 
it stays in its current state, and otherwise chooses a 
state independently from the stationary distribution.
The potential takes widely separated
large values with rapidly decreasing
stationary probabilities.
During a run of retention steps, large
potential values produce arbitrarily strong
contraction towards an attracting direction.
The stationary weights can still be chosen
so that all positive moments are finite.
The probability of such a run is too large
relative to the scale of concentration for
a H\"older estimate with any positive exponent
and any cutoff tending to zero.

Our second example shows that contraction
of a power of the Markov operator cannot
replace \eqref{eq:one-step-gap}, even for
a bounded locally constant potential.
Consider the doubling map
\begin{equation}\label{eq:doubling-map}
 D(x)=2x\pmod1,\qquad 0\le x<1,
\end{equation}
with Lebesgue probability.
Its natural extension is the shift on
two-sided binary sequences $(b_n)_{n\in\Z}$
with Bernoulli product measure
$(\tfrac12\delta_0+\tfrac12\delta_1)^\Z$.
We take
\begin{equation}\label{eq:doubling-sampling}
 f(\omega)=b_0+b_1-1.
\end{equation}
On $[0,1)$, this is
\begin{equation}\label{eq:doubling-intervals}
 f(x)=
 \begin{cases}
 -1,&0\le x<1/4,\\
  0,&1/4\le x<3/4,\\
  1,&3/4\le x<1.
 \end{cases}
\end{equation}
The potential is also a function of the
four-state Markov chain
$X_n=(b_n,b_{n+1})$.

\begin{proposition}
\label{prop:doubling-resonance}
Let $H_\omega$ be the operator
\eqref{eq:operator-introduction} with potential
\eqref{eq:doubling-sampling}.
\begin{enumerate}
\item
At $\lambda=0$, for every fixed initial
direction, all distributions $\nu_{N,0}$
are supported on one fixed set of at most
twelve points, independently of $N$.
Consequently, there are no constants
$C<\infty$, $\alpha>0$ and positive
sequence $r_N\to0$ such that
\[
 \nu_{N,0}(B_r(x))\le Cr^\alpha
 \qquad\text{for}\qquad
 N\ge1,\quad x\in M,\quad r>r_N.
\]
Every weak limit of $\nu_{N,0}$ or of
the averages \eqref{eq:phase-law-averages}
is supported on that finite set.

\item
At $\lambda=0$, all scalar transfer products
belong to one fixed finite set of matrices.
In particular, their norms are uniformly
bounded over all trajectories and all
numbers of steps.
The Lyapunov exponent satisfies
\[
 L(\lambda)=0
 \quad\Longleftrightarrow\quad
 \lambda=0.
\]
\end{enumerate}
\end{proposition}

We choose projective coordinates depending
on the binary digit.
At $\lambda=0$, every transition in these
coordinates is, up to sign, a power of
one matrix of order three.
The intermediate changes of coordinates cancel,
leaving two choices for each endpoint digit
and three possible powers, hence at most
twelve directions in
Proposition~\ref{prop:doubling-resonance}.

At every step, at least one atom has mass
at least $1/12$, which excludes every
cutoff tending to zero.
The failure of regularisation is proved
independently of the zero Lyapunov exponent.

\begin{remark}
\label{rem:doubling-gap}
Let $\widehat P$ be the Markov operator
of $X_n=(b_n,b_{n+1})$, with stationary
distribution $\widetilde\pi(i,j)=1/4$.
Then
\[
 \left\|\widehat P
       \big|_{L^2_0(\widetilde\pi)}\right\|=1,
 \qquad
 \left\|\widehat P^2
       \big|_{L^2_0(\widetilde\pi)}\right\|=0.
\]
Thus a gap at the second step cannot replace
the strict one-step condition in
Theorem~\ref{thm:main-regularity}.

Both counterexamples satisfy NDI for
every $k\ge2$ and every spectral parameter,
by Theorems~\ref{thm:main-ndi}
and~\ref{thm:main-markov-ndi}.
For every fixed $k$, both have uniform
Wasserstein separation on compact intervals
by Theorem~\ref{thm:main-wasserstein}.
These properties do not suffice for
regularisation of the correlated evolution.
\end{remark}

\begin{remark}
\label{rem:doubling-dgk}
A seemingly similar potential is constructed
in \cite[Section~6]{DGKBlockFactors}.
That potential is the difference of consecutive
binary digits, whereas
\eqref{eq:doubling-sampling} is their sum minus one.
The processes have different laws, since our
potential can equal $-1$ at two consecutive
indices, while the difference potential cannot.

The example in \cite[Section~6]{DGKBlockFactors}
exhibits an exceptional energy for localisation.
Proposition~\ref{prop:doubling-resonance}
gives a fixed finite family of transfer products
and persistent finite support for the direction
distributions.
Its purpose is to exclude regularisation
with every cutoff tending to zero, despite
NDI and Wasserstein separation at every
fixed length.
\end{remark}

\begin{remark}
\label{rem:transparency}
At $\lambda=0$, state-dependent changes
of coordinates reduce the transfer dynamics
to a finite group, so all transfer products
remain uniformly bounded although the
potential is nonconstant.
This gives a resonance analogous to
transparency energies in random polymer
models \cite{MPS}.
The example has a primitive Markov
representation with fixed points and
exactly one zero of the Lyapunov exponent,
so an exceptional energy can occur under
the hypotheses considered in \cite{ADZ}.

On the other hand, in the regimes covered
by \cite[Theorem~5]{PS}, the expected
eigenfunction correlator decays exponentially.
Under the additional Fermi-energy conditions
of \cite[Theorem~4 and Criterion~2]{PS},
this gives the Area Law for the
entanglement entropy.
\end{remark}

The paper is organised as follows.
Section~\ref{sec:amo} contains the results for trigonometric
potentials and the almost Mathieu operator, followed by
the extension to the class introduced in
Subsection~\ref{subsec:analytic-extension}.
We place these proofs first to keep
the quasiperiodic results together before
turning to the Markov and projective
preliminaries in Section~\ref{sec:models}.
Two-step NDI is proved in
Section~\ref{sec:ndi}.
Section~\ref{sec:routes} treats longer
products and the finite-state classification.
The Wasserstein estimate is proved in
Section~\ref{sec:wasserstein}.
The analytic estimates in
Section~\ref{sec:energies} are used in
Section~\ref{sec:regularisation} to prove
Theorem~\ref{thm:main-regularity}.
Section~\ref{sec:sharpness} contains the
counterexamples.
Auxiliary proofs are collected in the appendix.

\section{Projective dynamics for quasiperiodic potentials}
\label{sec:amo}


We prove Theorems~\ref{prop:trigonometric-all-k-ndi}--\ref{thm:analytic-extension}
and Corollaries~\ref{cor:trigonometric-logarithmic-ui}--\ref{cor:amo-returns}.
We first establish NDI and the quantitative estimates for
trigonometric potentials, then relate atomic limits to the
Lyapunov exponent. We conclude with the analytic extension
and a counterexample at a rational frequency.

Throughout this section, $M=\mathbb{RP}^1$ has the
angular metric \eqref{eq:angular-distance}.
For unit representatives $\mathbf v,\mathbf w$ of
$v,w\in M$,
\[
 d(v,w)=\arccos|\langle\mathbf v,\mathbf w\rangle|
 \in[0,\pi/2].
\]
We write $dm(\theta)=d\theta/(2\pi)$ for normalised
Lebesgue measure on $\R/2\pi\Z$.

\subsection{Transfer products and non-degeneracy}

For the coefficient calculation, we use the forward product
\begin{equation}\label{eq:trigonometric-forward-product}
 \mathcal A^{(n)}(\theta)
 =A_{\lambda,V}(\theta+2\pi(n-1)\alpha)
       \cdots A_{\lambda,V}(\theta).
\end{equation}
The matrix representing our backward transformation
$F_{\lambda,\theta}^{(n)}$ is
\begin{equation}\label{eq:trigonometric-forward-backward}
 G^{(n)}(\theta)
 =\left[\mathcal A^{(n)}
       (\theta-2\pi(n-1)\alpha)\right]^{-1}.
\end{equation}
Indeed, inversion reverses the order of the factors
in the forward product starting at
$\theta-2\pi(n-1)\alpha$.
For every determinant-one matrix $B$,
\begin{equation}\label{eq:trigonometric-inverse-determinant}
 \det(B^{-1}\mathbf v,\mathbf w)
 =-\det(B\mathbf w,\mathbf v),\qquad
 \|B^{-1}\|=\|B\|.
\end{equation}
Thus a determinant estimate uniform in the two
unit vectors transfers to the inverse product by
interchanging them and translating the initial phase.

\begin{proof}[Proof of Theorem~\ref{prop:trigonometric-all-k-ndi}]
We first consider the forward product
\eqref{eq:trigonometric-forward-product}.
Suppose that there exist $\eta,\eta'\in\Prob(M)$
such that
\begin{equation}\label{eq:trigonometric-common-image}
 (\mathcal A^{(k)}(\theta))_*\eta=\eta'
\end{equation}
for $m$-almost every $\theta$.
By continuity, this equality holds for every $\theta$.

In the coordinate $z=x/y$, we identify
$M$ with $\R\cup\{\infty\}$.
The same fractional-linear formulas extend the
projective transformations to the upper half-plane
$\Hh=\{z\in\mathbb C:\operatorname{Im}z>0\}$.
We show that \eqref{eq:trigonometric-common-image}
implies the existence of directions
$[z],[w]\in M\cup\Hh$ independent of $\theta$, such that
\begin{equation}\label{eq:trigonometric-constant-image}
 \mathcal A^{(k)}(\theta)[z]=[w]
 \qquad\text{for every }\theta,
\end{equation}
where, as in the Introduction, $[z]$ and $[w]$
denote the projective lines spanned by nonzero
vectors $z,w\in\mathbb C^2$, respectively. In particular, 
$\mathcal A^{(k)}(\theta)[z]=[w]$ means that
$\mathcal A^{(k)}(\theta)z$ and $w$ are proportional.

In what follows, we suppress the brackets and
write $z,w$ for the corresponding projective points,
with the projective action understood throughout.

First suppose that $\eta$ has an atom $x$ of mass
$p>0$.
Since $\mathcal A^{(k)}(\theta)$ is invertible,
\[
 \begin{aligned}
 \eta'\bigl(\{\mathcal A^{(k)}(\theta)x\}\bigr)
 &=\eta\bigl((\mathcal A^{(k)}(\theta))^{-1}
              \{\mathcal A^{(k)}(\theta)x\}\bigr)
 =\eta(\{x\})=p.
 \end{aligned}
\]
Thus $\mathcal A^{(k)}(\theta)x$ belongs to the
set of atoms of $\eta'$ of mass $p$.
There are at most $1/p$ such atoms.
A continuous map from the circle to this finite
set is constant.
This proves \eqref{eq:trigonometric-constant-image}
with $z=x$ and some fixed $w\in M$.

Now suppose that $\eta$ is non-atomic and fix
$\theta_0$.
Equation~\eqref{eq:trigonometric-common-image} gives
\[
 \begin{aligned}
 \bigl(\mathcal A^{(k)}(\theta_0)^{-1}
       \mathcal A^{(k)}(\theta)\bigr)_*\eta
 &=\bigl(\mathcal A^{(k)}(\theta_0)^{-1}\bigr)_*\eta' 
 =\eta.
 \end{aligned}
\]
These transformations therefore belong to the
stabiliser of $\eta$ \eqref{eq:stabiliser}.
By Lemma~\ref{lem:stabiliser}, all these transformations
fix a common point $z\in\Hh$.
Thus,
\[
 \mathcal A^{(k)}(\theta)z
 =\mathcal A^{(k)}(\theta_0)z
 \qquad\text{for every }\theta,
\]
which proves \eqref{eq:trigonometric-constant-image}
with $w=\mathcal A^{(k)}(\theta_0)z\in\Hh$.

We now exclude \eqref{eq:trigonometric-constant-image}.
Put $t=e^{i\theta}$ and write
\[
 b_j(t)=\lambda-
 \sum_{\ell=-d}^{d}
 a_\ell e^{2\pi i\ell j\alpha}t^\ell.
\]
These expressions are defined for
$t\in\mathbb C\setminus\{0\}$.
Since $a_d\ne0$,
\begin{equation}\label{eq:trigonometric-coefficient-asymptotic}
 b_j(t)
 =-a_de^{2\pi i d j\alpha}t^d+O(t^{d-1})
 \qquad(t\to\infty).
\end{equation}
In the coordinate $z=x/y$, the transfer matrix
\eqref{eq:scalar-matrix} at $\theta+2\pi j\alpha$
acts by
\begin{equation}\label{eq:recurrence}
 z_{j+1}(t)=b_j(t)-\frac1{z_j(t)}.
\end{equation}

If $z_0=0$, then $z_1=\infty$ and $z_2=b_1(t)$.
If $z_0\ne0$, including $z_0=\infty$, then
\[
 z_1(t)=b_0(t)-\frac1{z_0}=b_0(t)+O(1),
\]
where $1/\infty=0$.

Since each $b_j(t)$ is a Laurent polynomial and
the initial point $z_0$ is independent of $t$,
the recurrence shows that $z_k(t)$ is a rational
 function of $t$, with poles interpreted as the projective
value $\infty$.
Using \eqref{eq:trigonometric-coefficient-asymptotic}
in the recurrence, we obtain, in both cases,
\begin{equation}\label{eq:trigonometric-riccati-asymptotic}
 z_k(t)=b_{k-1}(t)+O(t^{-d})
 =-a_de^{2\pi i d(k-1)\alpha}t^d+O(t^{d-1}),
 \qquad k\ge2.
\end{equation}
By \eqref{eq:trigonometric-constant-image},
this rational map is constant on the unit circle
and hence identically constant, contradicting
\eqref{eq:trigonometric-riccati-asymptotic}.



We have excluded
\eqref{eq:trigonometric-common-image}
and therefore proved NDI for the forward law.

For the backward law, the identity
$(F^{-1})_*\eta=\eta'$ is equivalent to
$F_*\eta'=\eta$.
Then, combining~\eqref{eq:trigonometric-forward-backward}
and translation invariance of $m$, we obtain
NDI for $\mu_\lambda^{(k)}$.

Theorem~\ref{thm:main-wasserstein} gives positive
Wasserstein separation for every fixed $\alpha$,
uniformly for $\lambda\in I$.
To obtain uniformity in $\alpha$, note that
\[
 (\alpha,\lambda,\eta,\eta')
 \longmapsto
 \int_{\R/2\pi\Z}
 W\bigl((F_{\lambda,\theta}^{(k)})_*\eta,\eta'\bigr)
 \,dm(\theta)
\]
is continuous.
Its infimum over the compact space
$I\times\Prob(M)\times\Prob(M)$ is therefore
continuous in $\alpha$.
This infimum is positive at every $\alpha$,
so its minimum over $\R/\Z$ is positive.
This proves \eqref{eq:trigonometric-k-wasserstein}.
\end{proof}







We use the following consequence of Nazarov's
inequality \cite{Nazarov}; see also
\cite[Section~2.3, equation~(2.2)]{ShamisConnections}.
If $Q$ is a nonzero trigonometric polynomial
of degree at most $d\ge1$, then
\begin{equation}\label{eq:trigonometric-nazarov}
 m\{\theta:|Q(\theta)|<\epsilon\}
 \le C
       \left(\frac{\epsilon}{\|Q\|_\infty}\right)^{1/(2d)},
 \qquad \epsilon>0,
\end{equation}
where $C\ge1$ is a constant independent of $Q$, $d$ and $\epsilon$.
Its Fourier expansion has at most $2d+1$ nonzero terms.
Applying Nazarov's inequality to the sublevel set and
rearranging gives \eqref{eq:trigonometric-nazarov} when
$\epsilon<\|Q\|_\infty$; for larger $\epsilon$, the
estimate follows from $m(\R/2\pi\Z)=1$.

To obtain one H\"older constant for all steps, we prove the following lower bound for
\[
 D_n(\theta)=\det(\mathcal A^{(n)}(\theta)
                            \mathbf v,\mathbf w).
\]

\subsection{Three Fourier coefficients}

\begin{lemma}\label{lem:trigonometric-three-coefficients}
Let $V$ be as in \eqref{eq:trigonometric-potential},
put $\varrho=|a_d|$, and fix a compact interval
$I\subset\R$. There exists $c=c(V,I)>0$ such that
\begin{equation}\label{eq:trigonometric-determinant-lower-bound}
 \|\det(\mathcal A^{(n)}(\cdot)
                     \mathbf v,\mathbf w)\|_\infty
 \ge c\varrho^{n-2}(1+n)^{-2d}
\end{equation}
for every $n\ge2$, $\alpha\in\R$, $\lambda\in I$
and all unit vectors $\mathbf v,\mathbf w\in\R^2$.
For fixed $d,I,A_*,a_*$, the constant is uniform over
$\sum_{k=-d}^{d}|a_k|\le A_*$ and $|a_d|\ge a_*>0$.
\end{lemma}

\begin{proof}
Fix $n\ge2$, $\lambda\in I$ and $\alpha$.
Write
\[
c_j(\theta)=\lambda-V(\theta+2\pi j\alpha),
\qquad
p_{[a,b]}(\theta)
=\det(\lambda-H_{\theta,[a,b]}),
\]
where $H_{\theta,[a,b]}$ is the Dirichlet
restriction to $[a,b]$. We use the convention
$p_{[a,a-1]}=1$ for an empty interval.

The standard determinant recurrence gives
\[
\mathcal A^{(n)}(\theta)=
\begin{pmatrix}
p_{[0,n-1]}&-p_{[1,n-1]}\\
p_{[0,n-2]}&-p_{[1,n-2]}
\end{pmatrix}.
\]

Choose unit representatives
$\mathbf v=(v_1,v_2)^{\mathsf T}$
and $\mathbf w=(w_1,w_2)^{\mathsf T}$.
The determinant of the two column vectors
$\mathcal A^{(n)}(\theta)\mathbf v$
and $\mathbf w$ is therefore
\[
\begin{aligned}
D_n(\theta)
&=\det\bigl(
\mathcal A^{(n)}(\theta)\mathbf v,
\mathbf w\bigr)
=Xp_{[0,n-1]}-Yp_{[1,n-1]}
+Zp_{[0,n-2]}-Wp_{[1,n-2]},
\end{aligned}
\]
where
\[
X=w_2v_1,\qquad Y=w_2v_2,\qquad
Z=-w_1v_1,\qquad W=-w_1v_2.
\]
Since both vectors are unit vectors,
\begin{equation}\label{eq:four-coefficient-identities}
XW=YZ,\qquad
X^2+Y^2+Z^2+W^2=1.
\end{equation}

Put
\[
z=e^{i\theta},\qquad
\varrho=|a_d|,\qquad
b_j=-a_de^{2\pi i dj\alpha},\qquad
\mathcal T_n=\prod_{j=0}^{n-1}b_j.
\]
Thus $b_j$ is the coefficient of $z^d$
in $c_j$, and
\[
|b_j|=\varrho,\qquad
|\mathcal T_n|=\varrho^n.
\]
We write $\widehat f(k)$ for the
coefficient of $z^k$ in a
trigonometric polynomial $f$.

We first record an elementary estimate
for the determinants $p_{[a,b]}$.
Let $J=[a,b]$ contain $m$ sites and put
$B_J=\prod_{j\in J}b_j$.
For $0\le\ell\le2d$, we claim that
\begin{equation}\label{eq:continuant-coefficient-bound}
\left|\widehat p_J(dm-\ell)\right|
\le C(1+m)^\ell |B_J|,
\end{equation}
where $C$ depends only on the coefficients
of $V$, $\varrho^{-1}$ and $I$.

Indeed, each diagonal entry can be written as
\[
c_j(\theta)=b_jz^d
\left(1+\sum_{r=1}^{2d}
\beta_{j,r}z^{-r}\right),
\qquad |\beta_{j,r}|\le C_0.
\]
In the determinant expansion of a
tridiagonal matrix, every term is obtained
by choosing some disjoint pairs of
neighbouring sites and multiplying
the diagonal entries at all remaining sites.
Each chosen pair contributes a factor $-1$
and removes two diagonal entries.

If no pair is chosen, lowering the degree
from $dm$ to $dm-\ell$ requires choosing
lower-degree terms at at most $\ell$ sites.
After division by $B_J$, their total
contribution is bounded by
\[
\sum_{k=0}^{\ell}
\binom mk(2dC_0)^k
\le C(1+m)^\ell.
\]
If one neighbouring pair is chosen,
the maximal degree drops by $2d$.
For $\ell\le2d$, such a contribution
is therefore possible only when
$\ell=2d$, and then all remaining
diagonal entries must contribute
their highest terms. There are at most
$m-1$ choices of this pair, and each
normalised contribution has modulus
$\varrho^{-2}$. Two or more pairs
lower the degree by at least $4d$.
This proves
\eqref{eq:continuant-coefficient-bound}.
The same argument includes the empty
interval, with determinant equal to $1$.

In particular, the highest coefficient
of a determinant comes only from the
product of all its diagonal entries:
\begin{equation}\label{eq:continuant-leading}
\widehat p_{[a,b]}(d(b-a+1))
=\prod_{j=a}^{b}b_j.
\end{equation}

We now use these two facts for the
four determinants appearing in $D_n$.

At degree $dn$, only $p_{[0,n-1]}$
contributes, and hence
\begin{equation}\label{eq:three-coeff-first}
\widehat D_n(dn)=\mathcal T_n X.
\end{equation}

At degree $d(n-1)$, the determinants
$p_{[1,n-1]}$ and $p_{[0,n-2]}$
contribute their highest coefficients,
whereas $p_{[1,n-2]}$ cannot contribute.
Thus
\[
\begin{aligned}
\widehat D_n(d(n-1))
=
X\widehat p_{[0,n-1]}(d(n-1))
-\frac{\mathcal T_n}{b_0}Y
+\frac{\mathcal T_n}{b_{n-1}}Z.
\end{aligned}
\]
Applying
\eqref{eq:continuant-coefficient-bound}
to the first term, with $\ell=d$, gives
\begin{equation}\label{eq:three-coeff-second}
\left|
\frac{\widehat D_n(d(n-1))}{\mathcal T_n}
+\frac{Y}{b_0}-\frac{Z}{b_{n-1}}
\right|
\le C(1+n)^d|X|.
\end{equation}

Finally, at degree $d(n-2)$, the
determinant $p_{[1,n-2]}$ contributes
its highest coefficient:
\[
\widehat p_{[1,n-2]}(d(n-2))
=\prod_{j=1}^{n-2}b_j
=\frac{\mathcal T_n}{b_0b_{n-1}}.
\]
For $n=2$, this is the determinant
of the empty interval, equal to $1$.

The contributions of the other three
determinants are bounded by
\eqref{eq:continuant-coefficient-bound},
with $\ell=2d$ for $p_{[0,n-1]}$
and $\ell=d$ for the two determinants
on $n-1$ sites. Consequently,
\begin{equation}\label{eq:three-coeff-third}
\left|
\frac{\widehat D_n(d(n-2))}{\mathcal T_n}
+\frac{W}{b_0b_{n-1}}
\right|
\le C\bigl((1+n)^{2d}|X|
          +(1+n)^d(|Y|+|Z|)\bigr).
\end{equation}
It remains to show that these three
coefficients cannot all be too small.

Set
\[
q=\frac{\|D_n\|_\infty}{\varrho^{n-2}}.
\]
Since every Fourier coefficient of $D_n$
has modulus at most $\|D_n\|_\infty$,
\eqref{eq:three-coeff-first} gives
\[
|X|\le\varrho^{-2}q.
\]
Next, multiplying
\eqref{eq:three-coeff-second} by
$|b_0|=\varrho$, we obtain
\[
|Y-\xi Z|
\le C(1+n)^dq,
\qquad
\xi=\frac{b_0}{b_{n-1}},\quad|\xi|=1.
\]
The identity $YZ=XW$, together with
$|W|\le1$, also gives
\[
|YZ|\le\varrho^{-2}q.
\]
Therefore
\[
\begin{aligned}
(|Y|+|Z|)^2
=(|Y|-|Z|)^2+4|YZ|
&
\le |Y-\xi Z|^2+4|YZ|\\
&
\le C(1+n)^{2d}q^2+Cq.
\end{aligned}
\]
Combining these estimates yields
\begin{equation}\label{eq:three-coeff-small}
|X|+|Y|+|Z|
\le C\bigl((1+n)^dq+\sqrt q\bigr).
\end{equation}

Suppose, towards a contradiction, that
\[
 q<c(1+n)^{-2d}
\]
for a sufficiently small $0<c\le1$.
By \eqref{eq:three-coeff-small},
\[
 |X|+|Y|+|Z|
 \le C(c+\sqrt c)(1+n)^{-d}
 \le C\sqrt c\,(1+n)^{-d}.
\]
Choosing $c$ small enough, the identity
$X^2+Y^2+Z^2+W^2=1$ implies $|W|\ge1/2$.
On the other hand, \eqref{eq:three-coeff-third},
together with $|X|\le\varrho^{-2}q$ and
\eqref{eq:three-coeff-small}, gives
\[
\begin{aligned}
 |W|
 &\le q+C\bigl((1+n)^{2d}|X|
              +(1+n)^d(|Y|+|Z|)\bigr)\\
&
 \le C\bigl((1+n)^{2d}q+(1+n)^d\sqrt q\bigr)
 \le C(c+\sqrt c)<\frac12
\end{aligned}
\]
after decreasing $c$ further.
This contradiction proves
\[
 \|D_n\|_\infty
 \ge c\varrho^{n-2}(1+n)^{-2d},
\]
as required.

The constants depend only on $d$, $I$,
upper bounds for the coefficients of $V$
and a positive lower bound for $|a_d|$,
giving the required uniformity over
coefficient families.
\end{proof}

\subsection{The H\"older and logarithmic estimates}

\begin{proof}[Proof of Theorem~\ref{thm:trigonometric-regularity}]
The polynomial $D_n$ has degree at most $dn$.
Also,
\begin{equation}\label{eq:trigonometric-product-norm}
 \|\mathcal A^{(n)}(\theta)\|\le C_{\mathrm{tr}}^n,
 \qquad C_{\mathrm{tr}}=\sup_{\lambda\in I}|\lambda|
                                    +\|V\|_\infty+2.
\end{equation}
For unit vectors $\mathbf v,\mathbf w$ and an
invertible matrix $B$, the angular metric gives
\begin{equation}\label{eq:trigonometric-angular-determinant}
 |\det(B\mathbf v,\mathbf w)|
 =\|B\mathbf v\|\sin d([B\mathbf v],w).
\end{equation}
Consequently,
\begin{equation}\label{eq:trigonometric-angular-inclusion}
 d([\mathcal A^{(n)}(\theta)\mathbf v],w)<r
 \quad\Longrightarrow\quad |D_n(\theta)|<C_{\mathrm{tr}}^n r.
\end{equation}
By \eqref{eq:trigonometric-nazarov} and
Lemma~\ref{lem:trigonometric-three-coefficients},
the measure of these phases is at most
\[
 C
 \left(\frac{C_{\mathrm{tr}}^n(1+n)^{2d}}
              {c\varrho^{n-2}}\right)^{1/(2dn)}
 r^{1/(2dn)}.
\]
The factor preceding $r^{1/(2dn)}$ equals
\[
 C
 \left(\frac{C_{\mathrm{tr}}}{\varrho}\right)^{1/(2d)}
 \left(\frac{\varrho^2}{c}\right)^{1/(2dn)}
 (1+n)^{1/n}.
\]
The first factor involving $\varrho$ is fixed.
Since $n\ge2$, the next is bounded by
\[
 \max\left\{1,\left(\frac{\varrho^2}{c}\right)^{1/(4d)}\right\}.
\]
Finally, $1+n\le2^n$, so the last factor is at most $2$.
One constant therefore works for all
$n\ge2$ and all frequencies. The same reasoning
is uniform over the coefficient families in the
statement of the theorem.

Equations~\eqref{eq:trigonometric-forward-backward}
and~\eqref{eq:trigonometric-inverse-determinant}
give the same bound for the backward products.

For an arbitrary phase-independent initial probability $\eta$,
since the fixed-direction estimate is uniform in $v$
and $\eta$ is a probability measure, we conclude
\[
\begin{aligned}
\nu_{n,\lambda}^{\alpha,V,\eta}(B_r(w))
&=\int_M
\nu_{n,\lambda}^{\alpha,V,v}(B_r(w))
\,d\eta(v)
\le\int_M Cr^{1/(2dn)}\,d\eta(v)
=Cr^{1/(2dn)}.
\end{aligned}
\]
This proves \eqref{eq:trigonometric-holder}.

For part~\textup{(2)}, we note that for each $z\in M$, the elementary identity (the layer-cake formula)
\[
(-\log d(z,w)-an)_+
=\int_{an}^{\infty}
\mathbf 1_{\{d(z,w)<e^{-t}\}}\,dt
\]
holds, with both sides interpreted as
$+\infty$ when $z=w$.
Integrating and applying Tonelli's theorem, we obtain
\[
 \begin{aligned}
& \frac1n\int_M(-\log d(z,w)-an)_+\,
                   d\nu_{n,\lambda}^{\alpha,V,\eta}(z)
 =\frac1n\int_{an}^{\infty}\int_M
       \mathbf1_{\{d(z,w)<e^{-t}\}}\,
                   d\nu_{n,\lambda}^{\alpha,V,\eta}(z)\,dt\\
 &=\frac1n\int_{an}^{\infty}
       \nu_{n,\lambda}^{\alpha,V,\eta}(B_{e^{-t}}(w))\,dt
 \le\frac Cn\int_{an}^{\infty}e^{-t/(2dn)}\,dt
 =2dCe^{-a/(2d)}.
 \end{aligned}
\]
This proves \eqref{eq:trigonometric-logarithmic-excess}.
\end{proof}

\begin{proof}[Proof of Corollary~\ref{cor:trigonometric-logarithmic-ui}]
For $a\ge0$, $\{Z_{n,w}>a\}=B_{e^{-an}}(w)$,
so \eqref{eq:trigonometric-holder} gives
\eqref{eq:trigonometric-logarithmic-tail}.
For every nonnegative measurable function $Z$,
\[
 \int Z\mathbf1_{\{Z>a\}}\,d\nu
 =a\nu\{Z>a\}+\int_a^{\infty}\nu\{Z>t\}\,dt.
\]
Applying this identity to $Z_{n,w}$ yields
\[
 \begin{aligned}
 \int_M Z_{n,w}\mathbf1_{\{Z_{n,w}>a\}}\,
                    d\nu_{n,\lambda}^{\alpha,V,\eta}
 &\le Ca e^{-a/(2d)}+C\int_a^{\infty}e^{-t/(2d)}\,dt\\
 &=C(a+2d)e^{-a/(2d)}.
 \end{aligned}
\]
For fixed $d$, the bound $C(a+2d)e^{-a/(2d)}$
tends to zero as $a\to\infty$.
Since the bound is independent of $n$, $\alpha$, $\lambda$,
$w$ and $\eta$ under the assumptions of the theorem,
the normalised logarithms are uniformly integrable.
\end{proof}

\begin{remark}\label{rem:amo-wasserstein}
For each fixed $k\ge2$, we can also obtain
a H\"older estimate for the phase distributions
by combining Theorem~\ref{prop:trigonometric-all-k-ndi}
with Nazarov's inequality
\eqref{eq:trigonometric-nazarov}.
Let $G^{(k)}(\theta)$ be the matrix in
\eqref{eq:trigonometric-forward-backward} and put
\[
 B_k=
 \sup_{\substack{\alpha\in\R/\Z,\ \lambda\in I\\
                 \theta\in\R/2\pi\Z}}
       \|G^{(k)}(\theta)\|.
\]
For unit representatives $\mathbf v,\mathbf w$,
determinant one gives
$\|G^{(k)}(\theta)\mathbf v\|\ge B_k^{-1}$.
Hence, using that $\sin t\ge2t/\pi$ for $0\le t\le\pi/2$,
we obtain
\[
 \begin{aligned}
 \|\det(G^{(k)}(\cdot)\mathbf v,\mathbf w)\|_\infty
 &\ge\int|\det(G^{(k)}(\theta)\mathbf v,\mathbf w)|
                     \,dm(\theta)\\
 &\ge\frac{2}{\pi B_k}
       \int d(F_{\lambda,\theta}^{(k)}v,w)\,dm(\theta)
 \ge\frac{2d_k}{\pi B_k}.
 \end{aligned}
\]
The last inequality follows from
\eqref{eq:trigonometric-k-wasserstein}
with $\eta=\delta_v$ and $\eta'=\delta_w$.

By \eqref{eq:trigonometric-inverse-determinant},
this determinant is a trigonometric polynomial
of degree at most $dk$.
Moreover,
\[
 d(F_{\lambda,\theta}^{(k)}v,w)<r
 \quad\Longrightarrow\quad
 |\det(G^{(k)}(\theta)\mathbf v,\mathbf w)|<B_kr.
\]
Applying \eqref{eq:trigonometric-nazarov}, we obtain
\[
 \nu_{k,\lambda}^{\alpha,V,v}(B_r(w))
 \le
 C\left(\frac{B_k^2}{d_k}\right)^{1/(2dk)}
 r^{1/(2dk)}.
\]
Averaging in $v$ gives the same estimate for every
phase-independent initial probability $\eta$.

Thus NDI and Wasserstein separation give the
exponent $1/(2dk)$, with a multiplicative constant
that can depend on $k$.
Lemma~\ref{lem:trigonometric-three-coefficients}
provides the quantitative bound needed for
one constant valid for all $k\ge2$ in
Theorem~\ref{thm:trigonometric-regularity}.
\end{remark}

\subsection{Returns of rotation-valued cocycles}

For $t\in\R$, write
\[
 R_t=
 \begin{pmatrix}
  \cos t&-\sin t\\
  \sin t&\cos t
 \end{pmatrix}.
\]

\begin{lemma}\label{lem:analytic-rotation-return}
Let $\alpha\in\mathbb R\setminus\mathbb Q$.
Suppose that an analytic matrix function
$A:\R/2\pi\Z\to\SL(2,\R)$ has the representation
\begin{equation}\label{eq:amo-rotation-conjugacy}
 A(\theta)
 =B(\theta+2\pi\alpha)
       R_{\varphi(\theta)}B(\theta)^{-1},
\end{equation}
where $B:\R/2\pi\Z\to\SL(2,\R)$ and
$\varphi:\R/2\pi\Z\to\R$ are analytic.
Then there is a sequence $N_j\to\infty$ such that
\begin{equation}\label{eq:amo-uniform-matrix-return}
 \sup_\theta\|A^{(N_j)}(\theta)-I\|
 \longrightarrow0,
\end{equation}
where
$A^{(N)}(\theta)=A(\theta+2\pi(N-1)\alpha)
\cdots A(\theta)$.
\end{lemma}

\begin{proof}
Put
\[
 \bar\varphi=\int\varphi\,dm,\qquad
 S_N\varphi(\theta)
 =\sum_{\ell=0}^{N-1}
       \varphi(\theta+2\pi\ell\alpha).
\]
The intermediate factors $B^{-1}B$ cancel
in the product, while rotations commute.
Therefore,
\begin{equation}\label{eq:amo-rotation-product}
 A^{(N)}(\theta)
 =B(\theta+2\pi N\alpha)
       R_{S_N\varphi(\theta)}B(\theta)^{-1}.
\end{equation}
We shall find a sequence $N_j\to\infty$
such that both the base translation and
the rotation on the right approach
the identity uniformly in $\theta$.

By  \cite[Proposition~3.1]{AFK},
for the analytic function $\varphi$
there exist a sequence $Q_j\to\infty$
of denominators of continued-fraction
convergents of $\alpha$ and a constant
$C<\infty$ such that
\[
\begin{aligned}
\|S_{Q_j}\varphi-Q_j\bar\varphi\|_\infty
&\le C\left(Q_j^{-2}+Q_j^{-1/2}\right)
\le C_1Q_j^{-1/2}.
\end{aligned}
\]
Here we use the supremum norm on the
real circle; the analytic norm estimate
in \cite[Proposition~3.1]{AFK} implies
this bound. Since $Q_j$ are denominators
of convergents, we also have
\[
 \|Q_j\alpha\|_{\R/\Z}\le\frac1{Q_j}.
\]

For all sufficiently large $j$, put
\[
 H_j=\lfloor Q_j^{1/4}\rfloor,
 \qquad
 \omega_j=\frac{Q_j\bar\varphi}{2\pi}.
\]
Consider the $H_j+1$ fractional parts
\[
 \{r\omega_j\},\qquad 0\le r\le H_j.
\]
Divide $[0,1)$ into $H_j$ intervals
of length $1/H_j$.
Two of these fractional parts belong
to the same interval.
Subtracting their indices gives
an integer $1\le m_j\le H_j$ such that
\[
 \|m_j\omega_j\|_{\R/\Z}
 \le\frac1{H_j}.
\]
Hence,
\begin{equation}\label{eq:mj}
 \operatorname{dist}
 (m_jQ_j\bar\varphi,2\pi\Z)
 \le\frac{2\pi}{H_j}\longrightarrow0.
\end{equation}

Set $N_j=m_jQ_j$. By the definition
of the Birkhoff sums,
\[
 S_{N_j}\varphi(\theta)
 =
 \sum_{r=0}^{m_j-1}
 S_{Q_j}\varphi
       (\theta+2\pi rQ_j\alpha).
\]
Thence, 
\[
 \|S_{N_j}\varphi
       -m_jQ_j\bar\varphi\|_\infty
 \le C_1m_jQ_j^{-1/2}
 \le C_1H_jQ_j^{-1/2}
 \longrightarrow0.
\]
Together with the estimate \eqref{eq:mj}
on $m_jQ_j\bar\varphi$, this shows that
\[
 \sup_\theta
 \operatorname{dist}
 (S_{N_j}\varphi(\theta),2\pi\Z)
 \longrightarrow0.
\]
Thus
\[
 R_{S_{N_j}\varphi(\theta)}
 \longrightarrow I
\]
uniformly in $\theta$.

The base translation also returns,
since
\[
 \|N_j\alpha\|_{\R/\Z}
 \le m_j\|Q_j\alpha\|_{\R/\Z}
 \le\frac{H_j}{Q_j}
 \longrightarrow0.
\]
Moreover, $N_j\ge Q_j\to\infty$.
Uniform continuity and periodicity
of $B$, together with
\eqref{eq:amo-rotation-product},
now give
\[
 \sup_\theta\|A^{(N_j)}(\theta)-I\|
 \longrightarrow0.
\]
\end{proof}

\subsection{Phase distributions and atomic limits}

\begin{proof}[Proof of Theorem~\ref{thm:oseledets-distributions}]
Put
\[
 T\theta=\theta-2\pi\alpha,
 \qquad G(\theta)=A_{\lambda,V}(\theta)^{-1},
 \qquad
 G^{(n)}(\theta)=G(T^{n-1}\theta)\cdots G(\theta).
\]
The products of $G$ over $T$ represent
$F_{\lambda,\theta}^{(n)}$.
Their Lyapunov exponent equals $L_{\alpha,V}(\lambda)$.

We first prove that $q_*m$ is non-atomic for every
measurable invariant projective direction $q$.
Suppose that, for some $v\in M$, the set
\[
 E=\{\theta:q(\theta)=v\}
\]
has positive measure.
Recurrence for $T^2$ gives an even integer $k\ge2$ such that
\[
 m(E\cap T^{-k}E)>0.
\]
Indeed, otherwise the sets $T^{-2j}E$, $j\ge0$,
would be pairwise disjoint modulo null sets, giving
$(N+1)m(E)\le1$ for every $N\ge0$.
On this intersection, invariance of $q$ gives
\[
 F_{\lambda,\theta}^{(k)}v=q(T^k\theta)=v.
\]
Hence
\[
 \nu_{k,\lambda}^{\alpha,V,v}(\{v\})
 =m\{\theta:F_{\lambda,\theta}^{(k)}v=v\}
 \ge m(E\cap T^{-k}E)>0.
\]
But Theorem~\ref{thm:trigonometric-regularity} implies
\[
 \nu_{k,\lambda}^{\alpha,V,v}(\{v\})
 \le \lim_{r\downarrow0}Cr^{1/(2dk)}=0,
\]
a contradiction.

Now suppose that $L=L_{\alpha,V}(\lambda)>0$.
Oseledets' theorem gives stable and unstable directions
$s(\theta),u(\theta)$ with
\[
 G(\theta)q(\theta)=q(T\theta),\qquad q=s,u,
\]
projectively, almost everywhere.
Thus we proved that both $s_*m$ and $u_*m$
are non-atomic.

Fix $v\in M$ and choose a unit representative $\mathbf v$.
Since $s_*m$ is non-atomic, $v\ne s(\theta)$ for almost every $\theta$.
For a unit representative $\mathbf u(\theta)$ of $u(\theta)$,
Oseledets' theorem gives
\[
 \|G^{(n)}(\theta)\mathbf v\|=e^{nL+o(n)},
 \qquad
 \|G^{(n)}(\theta)\mathbf u(\theta)\|=e^{nL+o(n)}
\]
almost everywhere.
Since $\det G^{(n)}(\theta)=1$, the angular metric
\eqref{eq:angular-distance} gives
\[
 \begin{aligned}
 &\sin d\bigl(F_{\lambda,\theta}^{(n)}v,u(T^n\theta)\bigr)
=
 \frac{|\det(\mathbf v,\mathbf u(\theta))|}
 {\|G^{(n)}(\theta)\mathbf v\|\,
  \|G^{(n)}(\theta)\mathbf u(\theta)\|}
 \le e^{-2nL+o(n)}\longrightarrow0.
 \end{aligned}
\]
The distance takes values in $[0,\pi/2]$, so it also tends to zero.

For every $h\in C(M)$, uniform continuity and dominated convergence imply
\[
 \int h(F_{\lambda,\theta}^{(n)}v)\,dm(\theta)
 -\int h(u(T^n\theta))\,dm(\theta)\longrightarrow0.
\]
By invariance of $m$ under $T$,
\[
 \int h(u(T^n\theta))\,dm(\theta)
 =\int h(u(\theta))\,dm(\theta).
\]
Thus $\nu_{n,\lambda}^{\alpha,V,v}\xrightarrow{w^*}u_*m$
for every fixed $v$.
For any phase-independent initial probability $\eta_0$,
another application of dominated convergence gives
\[
 \begin{aligned}
 \int_M h\,d\nu_{n,\lambda}^{\alpha,V,\eta_0}
 &=\int_M\left(\int h(F_{\lambda,\theta}^{(n)}v)
                  \,dm(\theta)\right)d\eta_0(v)
 \longrightarrow\int h(u(\theta))\,dm(\theta).
 \end{aligned}
\]
This proves \eqref{eq:amo-positive-le-limit}.

To prove \eqref{eq:trigonometric-asymptotic-concentration}, fix $\eta_0$.
If the conclusion failed, there would be $\varepsilon>0$,
$r_j\downarrow0$, $n_j\to\infty$ and $w_j\in M$ such that
\[
 \nu_{n_j,\lambda}^{\alpha,V,\eta_0}(B_{r_j}(w_j))
 \ge\varepsilon.
\]
By compactness, after passing to a subsequence, $w_j\to w\in M$.
For every $r>0$, eventually
$B_{r_j}(w_j)\subset\overline B_r(w)$.
Weak convergence and the closed-set inequality therefore give
\[
 \varepsilon
 \le\limsup_{j\to\infty}
       \nu_{n_j,\lambda}^{\alpha,V,\eta_0}(\overline B_r(w))
 \le (u_*m)(\overline B_r(w)).
\]
Letting $r\downarrow0$ contradicts the non-atomicity of $u_*m$.
\end{proof}

\begin{proof}[Proof of Corollary~\ref{cor:atomic-reducibility}]
Suppose that condition~\textup{(4)} holds,
and denote its subsequential limit by $\nu$.
If $L_{\alpha,V}(\lambda)>0$,
Theorem~\ref{thm:oseledets-distributions}
and uniqueness of weak limits give $\nu=u_*m$,
contradicting the existence of an atom in $\nu$.
Hence $L_{\alpha,V}(\lambda)=0$, proving
$\textup{(4)}\Longrightarrow\textup{(1)}$
at every energy.

Suppose next that condition~\textup{(2)} holds.
In the convention used here, there is an analytic
map $B:\R/2\pi\Z\to\SL(2,\R)$ such that
\[
 A_{\lambda,V}(\theta)
 =
 B(\theta+2\pi\alpha)Q(\theta)B(\theta)^{-1},
 \qquad Q(\theta)\in \mathrm{SO}(2,\R).
\]
Write the rotation-valued matrix function as
\[
 Q(\theta)=
 \begin{pmatrix}
 a(\theta)&-b(\theta)\\
 b(\theta)&a(\theta)
 \end{pmatrix},
 \qquad a(\theta)^2+b(\theta)^2=1.
\]
An angle $\varphi$ can be chosen continuously
on $[0,2\pi]$ so that
$a(\theta)+ib(\theta)=e^{i\varphi(\theta)}$.
Since the endpoints represent the same point
of the circle,
\[
 \varphi(2\pi)-\varphi(0)=2\pi k
\]
for an integer $k$.
We need $k=0$ in order to choose the angle
periodically.

For the Schr\"odinger cocycle this follows
by continuously deforming the matrix function
$Q$ to the identity.
With the convention
\[
 Q(\theta)
 =B(\theta+2\pi\alpha)^{-1}
       A_{\lambda,V}(\theta)B(\theta),
\]
first replace $V$ by $tV$ and let $t$ decrease from one to zero.
Next replace $B(\theta+2\pi\alpha)$ continuously
by $B(\theta)$.
The resulting matrix function is
$B(\theta)^{-1}A_{\lambda,0}B(\theta)$.
Finally, deform the constant matrix
$A_{\lambda,0}$ to $I$.
An explicit path from $I$ to $A_{\lambda,0}$ is
\[
 \begin{pmatrix}1&s\lambda\\0&1\end{pmatrix}
 R_{s\pi/2},
 \qquad 0\le s\le1.
\]

Throughout these deformations the matrices are
invertible.
Their first columns, divided by their lengths,
therefore give a continuous deformation of
the circle-valued function $(a,b)$ to a constant.
The integer $k$, which counts its turns around
the circle, cannot change during such a deformation.
It is zero for the constant function and hence
also for $(a,b)$.

Thus $\varphi(2\pi)=\varphi(0)$, and the chosen angle
extends to a $2\pi$-periodic function on $\R$.
It is real analytic, as also follows locally
from
\[
 \varphi'(\theta)
 =a(\theta)b'(\theta)-b(\theta)a'(\theta).
\]

Lemma~\ref{lem:analytic-rotation-return}
now provides a sequence $n_j\to\infty$ such that
\[
 \sup_\theta
 \|\mathcal A^{(n_j)}(\theta)-I\|
 \longrightarrow0.
\]
Using
\[
 G^{(n_j)}(\theta)
 =
 \left[
 \mathcal A^{(n_j)}
       (\theta-2\pi(n_j-1)\alpha)
 \right]^{-1},
\]
we obtain
\[
 \sup_\theta\|G^{(n_j)}(\theta)-I\|
 \longrightarrow0.
\]
Thus
\[
 \sup_{\theta,v}
 d(F_{\lambda,\theta}^{(n_j)}v,v)
 \longrightarrow0,
\]
which proves condition~\textup{(3)}
with one sequence for every direction.
The implication
$\textup{(3)}\Longrightarrow\textup{(4)}$
is immediate.

Finally, by \cite[Theorem~1.2]{AFK},
outside a Borel set $\mathcal N_{\alpha,V}$
of Lebesgue measure zero, the cocycle
either has a positive Lyapunov exponent
or is analytically conjugate to a
rotation-valued cocycle.
Under condition~\textup{(1)}, the first
alternative is excluded.
This proves
$\textup{(1)}\Longrightarrow\textup{(2)}$
outside $\mathcal N_{\alpha,V}$.
Here conjugacy to a rotation-valued cocycle uses an
analytic $\SL(2,\R)$-valued map, as in the definition
in \cite[Section~1]{AFK}.
\end{proof}

\begin{proof}[Proof of Corollary~\ref{cor:amo-returns}]
For $V(\theta)=g\cos\theta$ and irrational $\alpha$,
\cite[Corollary~2]{BourgainJitomirskaya} gives
\[
 L_{g,\alpha}(\lambda)
 =\max\left\{0,\log\frac{|g|}{2}\right\},
 \qquad\lambda\in\Sigma_{g,\alpha}.
\]
For $0<|g|<2$, the exponent therefore vanishes
throughout the spectrum.
Corollary~\ref{cor:atomic-reducibility} yields, for
Lebesgue-almost every such $\lambda$, one sequence
$N_j\to\infty$ for which
$\nu_{N_j,\lambda}^{\alpha,g,v}\xrightarrow{w^*}\delta_v$
for every $v$.
Integration against any initial probability $\eta$,
using dominated convergence for continuous test functions,
proves \eqref{eq:amo-return-laws} with the same sequence.

If \eqref{eq:amo-impossible-cutoff} held for a fixed $v$,
choose $0<r<1$ with $C_0r^\gamma<1$.
For all sufficiently large $j$, $r_{N_j}<r$.
Since $B_r(v)$ is open and contains the support of
$\delta_v$, weak convergence gives
\[
 1\le\liminf_{j\to\infty}
       \nu_{N_j,\lambda}^{\alpha,g,v}(B_r(v))
 \le C_0r^\gamma<1,
\]
a contradiction.
\end{proof}

\subsection{The analytic extension}
\label{subsec:analytic-extension-proof}

\begin{proof}[Proof of Theorem~\ref{thm:analytic-extension}]
Suppose that the law of the forward $k$-step product
\eqref{eq:trigonometric-forward-product} has a common image.
Using the reduction to \eqref{eq:trigonometric-constant-image}, 
we obtain fixed  directions $z,w\in M\cup\Hh$ such that $\mathcal A^{(k)}(\theta)z=w$.
We continue \eqref{eq:trigonometric-constant-image} along
the path in the definition of $\mathcal C$ and set
$b_j(\theta)=\lambda-V(\theta+2\pi j\alpha)$.
Since all $|b_j(\theta_\ell)|\to\infty$, the recurrence
\eqref{eq:recurrence} yields
\begin{equation}\label{eq:analytic-iterate-asymptotic}
 z_k(\theta_\ell)=b_{k-1}(\theta_\ell)+o(1),
 \qquad\ell\to\infty.
\end{equation}
If $z_0=0$, we first use $z_1=\infty$, $z_2=b_1$.
Thus $z_k(\theta_\ell)$ is finite for large $\ell$ and unbounded,
contradicting \eqref{eq:trigonometric-constant-image}.
The conclusions for the backward law and uniform Wasserstein
separation follow as in the proof of
Theorem~\ref{prop:trigonometric-all-k-ndi}.

We next prove \eqref{eq:analytic-fixed-holder}. For fixed $k$ and $I$, let $G^{(k)}$ be given by
\eqref{eq:trigonometric-forward-backward} and consider
\begin{equation}\label{eq:analytic-determinant-family}
 D_k(\theta)=\det(G^{(k)}(\theta)\mathbf v,\mathbf w),
 \qquad\|\mathbf v\|=\|\mathbf w\|=1.
\end{equation}
For each $k\ge 2$, these determinants form a compact family $\mathscr D_k$
in the analytic norm
$\|D\|_\rho=\sup_{|\operatorname{Im}\theta|\le\rho}|D(\theta)|$
on a fixed strip: they depend continuously on $\alpha$, $\lambda\in I$
and the two unit vectors and are $1$-periodic in $\alpha$.
If $D_k\equiv0$, the projective image $G^{(k)}(\theta)v$
would equal the fixed point $w$ for every $\theta$.
By \eqref{eq:trigonometric-forward-backward}, this would give
a constant forward image \eqref{eq:trigonometric-constant-image},
contradicting \eqref{eq:analytic-iterate-asymptotic}.

For each $\widetilde D\in\mathscr D_k$,
by \cite[Lemma~2.1]{DKAnalytic} there exist
\[
\delta_0,S_0,b_0>0,\quad \text{such that}\quad m\{|D|<t\}\le S_0t^{b_0},\quad  t>0,
\]
for every $D\in\mathscr D_k$ satisfying
$\|D-\widetilde D\|_\rho<\delta_0$.
These neighbourhoods cover the determinant family. Choose a
finite subcover with constants $S_i,b_i$, $1\le i\le N$, and take
$S_k=\max\{1,S_1,\ldots,S_N\}$,
$\beta_k=\min_{1\le i\le N}b_i>0$. Then
\begin{equation}\label{eq:analytic-uniform-sublevel}
 m\{\theta:|D_k(\theta)|<t\}\le S_kt^{\beta_k},
 \qquad t>0,
\end{equation}
uniformly in all the parameters in \eqref{eq:analytic-determinant-family}.

Since the estimate \eqref{eq:trigonometric-product-norm}
requires only boundedness of $V$, it applies here with $C_{\mathrm{tr}}$ defined in the same way.
Combining \eqref{eq:trigonometric-product-norm} with
\eqref{eq:trigonometric-forward-backward}--\eqref{eq:trigonometric-inverse-determinant},
we obtain $\|G^{(k)}(\theta)\|\le C_{\mathrm{tr}}^k$.
By \eqref{eq:trigonometric-angular-determinant}, applying
\eqref{eq:analytic-uniform-sublevel} with $t=C_{\mathrm{tr}}^kr$
and integrating against $\eta$ gives
\eqref{eq:analytic-fixed-holder}, with
$C_k=S_kC_{\mathrm{tr}}^{k\beta_k}$.

For fixed $k$, the logarithmic estimates
\eqref{eq:trigonometric-logarithmic-excess}--\eqref{eq:trigonometric-logarithmic-ui},
with $n=k$, $C$ replaced by $C_k$ and $2d$ by $(k\beta_k)^{-1}$,
follow from \eqref{eq:analytic-fixed-holder} as in the proofs
of Theorem~\ref{thm:trigonometric-regularity}(2) and
Corollary~\ref{cor:trigonometric-logarithmic-ui}.

Note that the polynomial structure entered through the asymptotic formula
\eqref{eq:trigonometric-riccati-asymptotic}, the degree bound for
the determinant, and the lower bound
\eqref{eq:trigonometric-determinant-lower-bound} used with
\eqref{eq:trigonometric-nazarov}.
We have replaced these by
\eqref{eq:analytic-iterate-asymptotic} and
\eqref{eq:analytic-uniform-sublevel}.
The remaining arguments do not use polynomial structure.
In particular, \eqref{eq:analytic-fixed-holder} excludes atoms
for every fixed $k$, and the proof of
Theorem~\ref{thm:oseledets-distributions} gives non-atomicity of
$q_*m$ for every measurable invariant direction $q$, together with
\eqref{eq:amo-positive-le-limit}--\eqref{eq:trigonometric-asymptotic-concentration}.
The proof of Corollary~\ref{cor:atomic-reducibility} then applies
unchanged, using \cite[Theorem~1.2]{AFK} and
Lemma~\ref{lem:analytic-rotation-return}. This finishes the proof when $V$ has no real poles.

When $V$ has real poles, choose an open arc on which all
$V(\theta+2\pi j\alpha)$, $0\le j<k$, are finite.
A common image would give fixed $z,w$ with $z_k(\theta)=w$
throughout this arc, by the reduction to \eqref{eq:trigonometric-constant-image}.
This equality persists under analytic continuation along the
path defining $\mathcal C$, contradicting
\eqref{eq:analytic-iterate-asymptotic}.
The Wasserstein integral is continuous by bounded convergence:
at each limiting frequency only finitely many phases meet a pole.

For the regularity estimates, we first remove the poles
from the matrix entries as follows.
Choose a real-analytic $2\pi$-periodic function $h\not\equiv0$
such that $hV$ is real analytic on $\R/2\pi\Z$.
With $G(\theta)=A_{\lambda,V}(\theta)^{-1}$, put
\begin{equation}\label{eq:cleared-transfer-matrix}
 \widetilde G(\theta)=h(\theta)G(\theta)
 =\begin{pmatrix}
 0&h(\theta)\\-h(\theta)&h(\theta)(\lambda-V(\theta))
 \end{pmatrix}.
\end{equation}
With $T\theta=\theta-2\pi\alpha$ and
$H_k(\theta)=\prod_{j=0}^{k-1}h(T^j\theta)$, we have
\begin{equation}\label{eq:cleared-transfer-product}
 \widetilde G^{(k)}=H_kG^{(k)},\qquad
 \widetilde D_k:=\det(\widetilde G^{(k)}\mathbf v,\mathbf w)=H_kD_k,
\end{equation}
whenever $V(\theta-2\pi j\alpha)$ is finite for $0\le j<k$.
The determinants $\widetilde D_k$ are analytic and not identically
zero. Whenever $H_k\ne0$, the scalar factor preserves the projective
action, and \eqref{eq:trigonometric-angular-determinant} gives
\[
 d(F_{\lambda,\theta}^{(k)}v,w)<r
 \quad\Longrightarrow\quad
 |\widetilde D_k(\theta)|<\widetilde B^kr,
 \qquad \widetilde B=\sup_{\lambda\in I,\theta}\|\widetilde G(\theta)\|.
\]
We now apply the finite-cover argument leading to
\eqref{eq:analytic-uniform-sublevel} to $\widetilde D_k$
and obtain \eqref{eq:analytic-fixed-holder}.
Analyticity of $H_k$, combined with $H_k\not\equiv0$,
implies that there are finitely many zeros in one period,
which do not affect phase averaging.

It remains to check the integrability required in Oseledets' theorem.
We have $\|G(\theta)\| \le(1+|\lambda|)(1+|V(\theta)|)$, and near a pole $p$ of order $q$,
\[
 \log(1+|V(\theta)|)\le C+q\bigl|\log|\theta-p|\bigr|,
 \qquad 0<|\theta-p|<\varepsilon<1.
\]
Since $\int_0^\varepsilon|\log t|\,dt<\infty$,
\begin{equation}\label{eq:meromorphic-log-integrability}
 \int\log^+\|G(\theta)\|\,dm(\theta)
 \le C+\int\log(1+|V(\theta)|)\,dm(\theta)<\infty.
\end{equation}
Oseledets' theorem and the proof of
Theorem~\ref{thm:oseledets-distributions} thus give, for
$L_{\alpha,V}(\lambda)>0$,
\[
 \nu_{n,\lambda}^{\alpha,V,\eta}\xrightarrow{w^*}u_*m,
 \qquad (u_*m)(\{w\})=0,\quad \text{for every}\quad w\in M,
\]
contradicting the existence of an atomic subsequential limit for
one fixed initial direction, proving \textup{(4)}$\Rightarrow$\textup{(1)}
of Corollary~\ref{cor:atomic-reducibility}.
The matrix $\widetilde G$ is analytic, but its determinant
$h^2$ vanishes at the zeros of $h$. Thus it is not an analytic
$\SL(2,\R)$-valued cocycle, as required by
\cite[Theorem~1.2]{AFK}. Note that \cite[Theorem~1.2]{AFK} cannot be applied to $G$
either, since division by $h$ restores the original poles.
\end{proof}

\begin{remark}\label{rmk:analytic}
Consider
\[
 V(\theta+2\pi j\alpha)
 =\frac{a+\cos(\theta+2\pi j\alpha)}
        {a-\cos(\theta+2\pi j\alpha)},
 \qquad a>1,\quad\alpha=\frac12,\quad\lambda=1.
\]
Then $V$ is real analytic on the circle and
$V(\theta+\pi)=1/V(\theta)$. For \eqref{eq:recurrence} with $z_0=1$,
\[
 z_1(\theta)=-V(\theta),\qquad
 z_2(\theta)=1-V(\theta+\pi)+\frac1{V(\theta)}=1.
\]
Thus $(\mathcal A^{(2)}(\theta))_*\delta_1=\delta_1$ for every
$\theta$, and NDI fails at two steps. The identity
$V(\theta)V(\theta+\pi)=1$ prevents
\eqref{eq:analytic-simultaneous-growth}, so $V\notin\mathcal C$.
Since rotation by $1/2$ is not ergodic with respect to Lebesgue measure,
there is no contradiction with Theorem~\ref{thm:main-ndi} or
Corollary~\ref{cor:ergodic-infinite-range}.
For irrational frequencies, two-step NDI for this potential follows
from Corollary~\ref{cor:ergodic-infinite-range}.
\end{remark}

\section{Markov and projective preliminaries}\label{sec:models}

\subsection{Markov chains, reversal
and hidden potentials}
\label{subsec:symbolic-preliminaries}
\label{subsec:general-preliminaries}

We use the two classes of potentials
\eqref{eq:first-potential} and
\eqref{eq:second-potential}, with the
probability spaces and Markov operators
defined in the Introduction.
Here we give the additional facts needed
in the proofs.

For a measurable function $f$ on
$\Omega_P$ of \eqref{eq:symbolic-space}, nonconstancy almost everywhere
is equivalent to
\begin{equation}\label{eq:observation-nondegeneracy}
 \essosc_{\mathbf P}f
 :=\operatorname*{ess\,sup}_{\mathbf P}f
   -\operatorname*{ess\,inf}_{\mathbf P}f>0,
\end{equation}
or, equivalently, to $f_*\mathbf P$
not being a point mass.
For a locally constant function, we can
choose $r,s\ge0$ and write
\[
 f(\omega)=F_0(\omega_{-r},\ldots,\omega_s).
\]
For continuous functions on $\Omega_P$, nonconstancy is
equivalent to nonconstancy almost everywhere,
since $\mathbf P$ has full support on $\Omega_P$.

We use mixing in the measure-theoretic sense.
A probability-preserving transformation $\sigma$ of
$(\Omega,\mathbf P)$ is mixing if, for every pair
of measurable sets $A,B$,
\begin{equation}\label{eq:mix-prob}
 \mathbf P(A\cap\sigma^{-n}B)
 \longrightarrow\mathbf P(A)\mathbf P(B),
 \qquad n\to\infty.
\end{equation}
Equivalently, for bounded measurable functions $h_1,h_2$,
\begin{equation}\label{eq:mix-test}
 \int_\Omega h_1(h_2\circ\sigma^n)\,d\mathbf P
 \longrightarrow
 \left(\int_\Omega h_1\,d\mathbf P\right)
 \left(\int_\Omega h_2\,d\mathbf P\right).
\end{equation}

For a primitive finite-state transition matrix,
the stationary shift is mixing; see
\cite[Proposition~4.4.4, p.~79]{BrinStuck}.

\medskip
\noindent\textbf{Two-sided extensions
and reversal.}

\smallskip

We use the kernel notation for both
classes of chains.
For the finite-state chain,
$K(a,\{b\})=p_{ab}$ and $F(\{a\})=\pi(a)$.
For the general-state chain, $K$ and $F$
are those of \eqref{eq:markov-operator}.


By the Kolmogorov extension theorem, a stationary Markov
chain on a standard Borel state space, initially defined
at nonnegative indices, admits a two-sided stationary
extension with the same stationary distribution $F$
and transition kernel $K$.
Its restriction to nonnegative indices has the original
distribution.
Here we work on this two-sided trajectory space,
where the left shift \eqref{eq:shift} is invertible.

For a non-invertible probability-preserving
transformation, we use the natural extension
and lift the potential by the factor
projection.
This preserves all joint distributions
of potential values at nonnegative indices.
 
For the backward evolution, we write
\begin{equation}\label{eq:tau}
\tau=\sigma^{-1}.
\end{equation}
 
The inverse of an invertible mixing
probability-preserving transformation
is mixing. For the left shift $\sigma$ of
\eqref{eq:shift}, invariance yields
\[
 \mathbf P(A\cap\sigma^n B)
 =\mathbf P(\tau^{n}A\cap B)
 \longrightarrow\mathbf P(A)\mathbf P(B).
\]
Reversing a symbolic trajectory preserves
finite-coordinate dependence.
For a finite-state chain, reversal
preserves irreducibility and period:
admissible paths are reversed, while
closed paths retain their lengths.
In particular, reversal preserves
primitivity.

\medskip
\noindent\textbf{The reverse transition
and its operator.}

\smallskip

For the second potential
\eqref{eq:second-potential}, we use
the specified stationary measure $F$.
We do not require $F$ to be the unique
stationary measure on the whole state space.
The norm condition \eqref{eq:one-step-gap}
refers to this measure.

Stationarity means that the transition kernel $K(v, dy)$ in \eqref{eq:markov-operator} satisfies
\[
 \int_S K(v,D)\,dF(v)=F(D)
\]
for every Borel set $D\subseteq S$.
Jensen's inequality therefore gives
\begin{equation}\label{eq:markov-contraction}
 \|Ph\|_{L^2(F)}^2
 \le\int_S\int_S|h(w)|^2K(v,dw)\,dF(v)
 =\|h\|_{L^2(F)}^2.
\end{equation}
The Markov operator $P$ of \eqref{eq:markov-operator}
preserves constants and means, and hence
preserves $L^2_0(F)$.

Since $S$ is a standard Borel space, we can
condition the stationary pair
$(X_0,X_1)$ on its second coordinate.
We obtain a reverse transition kernel
$K^\leftarrow$ satisfying
\begin{equation}\label{eq:reversed-kernel}
 dF(v)K(v,dw)
 =dF(w)K^\leftarrow(w,dv).
\end{equation}
The kernel $K^\leftarrow$ is also
stationary with respect to $F$.
By \eqref{eq:reversed-kernel}, its operator satisfies
\begin{equation}\label{eq:reversed-operator}
 P^\leftarrow=P^*,\qquad
 (P^\leftarrow)^*=P,\qquad
 \left\|P^\leftarrow\big|_{L^2_0(F)}\right\|
 =\left\|P\big|_{L^2_0(F)}\right\|.
\end{equation}
The restrictions of $P^\leftarrow$
and $P$ to the invariant subspace
$L^2_0(F)$ are also adjoint operators,
so their norms agree.
For the finite-state chain, we similarly have
\[
 p^\leftarrow_{ab}
 =\frac{\pi(b)p_{ba}}{\pi(a)}.
\]
We do not assume that $P=P^*$.

Potentials taking two distinct values are
allowed, provided that both values occur
with positive $F$-probability.
In particular, a $\{-1,1\}$-valued potential
is nonconstant almost everywhere whenever
both signs have positive probability,
regardless of its mean.

\medskip

\noindent\textbf{Hidden potentials defined
by an additional probability kernel.}

\smallskip

We can also allow the potential value
at a state to be chosen according to
a probability kernel $E(v,dt)$ from $S$
to $\R$.
We include this case in
\eqref{eq:second-potential} by enlarging
the state space to $S\times\R$ and defining
\begin{equation}\label{eq:emission-model}
 \widetilde F(dv,dt)=F(dv)E(v,dt),\qquad
 \widetilde K((v,t),dw,ds)=K(v,dw)E(w,ds).
\end{equation}
Under $\widetilde F$, the first coordinate
has distribution $F$, and the conditional
distribution of the second one
given the first coordinate $v$ is $E(v,dt)$.
Stationarity of $F$ implies stationarity
of $\widetilde F$.
We take the potential to be
$\widetilde V(v,t)=t$.

For $h\in L^2(\widetilde F)$, we define the average with respect to $E(v,dt)$ by
\[
 \bar h(v)=\int_\R h(v,t)\,E(v,dt).
\]
Jensen's inequality gives
\[
 |\bar h(v)|^2
 \le\int_\R|h(v,t)|^2\,E(v,dt),
\]
and integrating with respect to $F$ yields
\[
 \|\bar h\|_{L^2(F)}^2
 \le\int_S\int_{\R}|h(v,t)|^2\,E(v,dt)\,dF(v)
 =\|h\|_{L^2(\widetilde F)}^2.
\]
The Markov operator of the chain on
$S\times\mathbb R$ with transition kernel
\eqref{eq:emission-model} is
\[
 (\widetilde P h)(v,t)
 =\int_S K(v,dw)\int_{\mathbb R}
       h(w,s)\,E(w,ds)
 =(P\bar h)(v).
\]
If $h$ has mean zero with respect to $\widetilde F$, then
\[
 \int_S\bar h(v)\,dF(v)
   =\int_S\int_{\R}h(v,t)\,E(v,dt)\,dF(v)
   =\int_{S\times\R}h\,d\widetilde F
   =0.
\]
Under \eqref{eq:one-step-gap}, we therefore obtain
\[
 \|\widetilde P h\|_{L^2(\widetilde F)}
 =\|P\bar h\|_{L^2(F)}
 \le\rho\|\bar h\|_{L^2(F)}
 \le\rho\|h\|_{L^2(\widetilde F)}.
\]
Conversely, taking $h(v,t)=h_0(v)$ with
$h_0\in L^2_0(F)$ gives
\begin{equation}\label{eq:augmented-gap}
 \left\|\widetilde P
       \big|_{L^2_0(\widetilde F)}\right\|
 =\left\|P\big|_{L^2_0(F)}\right\|=\rho.
\end{equation}
Thus Theorems~\ref{thm:main-markov-ndi}
and~\ref{thm:main-wasserstein}
apply whenever $t$ is nonconstant $\widetilde F$-almost everywhere.
If $t$ is also bounded
$\widetilde F$-almost everywhere,
Theorem~\ref{thm:main-regularity} applies.
The hidden state space may be infinite.

\subsection{Polar coordinates and backward Cauchy evolution}
\label{subsec:cauchy-preliminaries}

We work on the projective space $M$ \eqref{eq:prioj-space} with the angular
metric $d$ of 
\eqref{eq:angular-distance}.
We now introduce the coordinates used in the proofs
and derive the evolution corresponding to the
transfer matrices \eqref{eq:scalar-matrix}.

For the eigenvalue equation \eqref{eq:schrodinger}, we use
the Cauchy boundary conditions
\begin{equation}\label{eq:cauchy-data}
 u_1=\cos\varphi,\qquad u_0=\sin\varphi,
 \qquad -\frac\pi2\le\varphi<\frac\pi2.
\end{equation}
We write the transfer matrix
\eqref{eq:scalar-matrix} as
\begin{equation}\label{eq:scalar-matrix-notation}
 A_\lambda(\omega)=B(\lambda-f(\omega)),
 \qquad
 B(t)=\begin{pmatrix}t&-1\\1&0\end{pmatrix}.
\end{equation}

We use a polar representation of solutions.
For the standard Pr\"ufer representation and
its relation to the discrete analogue of the Riccati equation,
see \cite[Section~4.1]{Teschl}.

A nonzero solution cannot have two consecutive
zero values.
Thus $[u_j:u_{j+1}]$ is defined for every $j$,
and we use the projective ratio
$\xi_j=u_j/u_{j+1}\in\widehat\R
=\R\cup\{\infty\}$.

Fix $\beta\in(0,\pi)$ and write
\begin{equation}\label{eq:beta-constants}
 C=\cos\beta,\qquad S_\beta=\sin\beta>0.
\end{equation}
We define $\phi_j\in M$ by
\begin{equation}\label{eq:phase-coordinate}
 \xi_j=C-S_\beta\tan\phi_j
       =\frac{\cos(\phi_j+\beta)}{\cos\phi_j}.
\end{equation}
Thus the initial phase
$\Phi_0^{\rm C}=\phi_0$ is determined by
$u_0/u_1$ in \eqref{eq:cauchy-data}.
The matrix
\begin{equation}\label{eq:phase-coordinate-matrix}
 J_\beta=
 \begin{pmatrix}C&-S_\beta\\1&0\end{pmatrix}
\end{equation}
sends $[\cos\phi_j:\sin\phi_j]$ to
$[u_j:u_{j+1}]$. Where $\tan\phi$ and $\xi$ are finite,
we may equivalently write these lines as
$[1:\tan\phi]$ and $[\xi:1]$.
The homogeneous-coordinate identity
\eqref{eq:phase-coordinate} holds on the
whole projective line, including at infinity.

The equation \eqref{eq:schrodinger}
at site $j+1$ gives
\begin{equation}\label{eq:backward-ratio}
 \xi_j=\lambda-f(\sigma^{j+1}\omega)
                  -\xi_{j+1}^{-1}.
\end{equation}
By \eqref{eq:phase-coordinate},
\[
 \xi_{j+1}^{-1}
 =\frac{\cos\phi_{j+1}}
        {\cos(\phi_{j+1}+\beta)}.
\]
The identity
\[
 \cos\phi_{j+1}
 =C\cos(\phi_{j+1}+\beta)
  +S_\beta\sin(\phi_{j+1}+\beta)
\]
therefore yields
\[
 \xi_{j+1}^{-1}
 =C+S_\beta\tan(\phi_{j+1}+\beta).
\]
Substituting in \eqref{eq:backward-ratio},
we obtain
\[
 C-S_\beta\tan\phi_j
 =\lambda-f(\sigma^{j+1}\omega)-C
       -S_\beta\tan(\phi_{j+1}+\beta).
\]
We rearrange this equality and divide by
$S_\beta>0$ to get
\begin{equation}\label{eq:phase-backward-step}
 \tan\phi_j
 =\tan(\phi_{j+1}+\beta)
       +c_\lambda(\sigma^{j+1}\omega),
\end{equation}
where
\begin{equation}\label{eq:coefficient}
 c_\lambda(\omega)
 =\frac{f(\omega)-\lambda+2C}{S_\beta}.
\end{equation}

For the second potential
\eqref{eq:second-potential}, we also write
\[
 c_\lambda(v)
 =\frac{V(v)-\lambda+2C}{S_\beta},
 \qquad v\in S.
\]

We express \eqref{eq:phase-backward-step}
through the projective map
\begin{equation}\label{eq:riccati-map}
 T_c(\phi)
 =\arctan\bigl(\tan(\phi+\beta)+c\bigr)
       \pmod\pi.
\end{equation}
The map \eqref{eq:riccati-map} extends
smoothly through the poles.
Thus \eqref{eq:phase-backward-step}
means
\[
 \phi_j
 =T_{c_\lambda(\sigma^{j+1}\omega)}
       (\phi_{j+1})
\]
on all of $M$.

With $\tau=\sigma^{-1}$ as in \eqref{eq:tau},
we write
\begin{equation}\label{eq:evolution-shift}
 \Phi_N=\phi_{-N},\qquad Y_N=X_{-N},
 \qquad N\ge0.
\end{equation}
Taking $j=-N-1$ in
\eqref{eq:phase-backward-step}, we obtain
the backward evolution with the Cauchy
conditions \eqref{eq:cauchy-data}
\begin{equation}\label{eq:actual-phase-recursion}
 \Phi_{N+1}
 =T_{c_\lambda(\tau^N\omega)}\Phi_N,
 \qquad \Phi_0=\Phi_0^{\rm C}.
\end{equation}
The coefficient in this step is evaluated
at the site $-N$.

We write $R_\beta(\phi)=\phi+\beta$
and let $L_c$ be translation by $c$
in the coordinate $z=\tan\phi$.
Then $T_c=L_c\circ R_\beta$.
All translations fix $q=\pi/2$.
Consequently, with $p=\pi/2-\beta$,
\begin{equation}\label{eq:one-step-deterministic}
 T_c(p)=q,\qquad
 (T_c)_*\delta_p=\delta_q
 \qquad(c\in\R),
\end{equation}
where $\delta_p$ and $\delta_q$ denote the Dirac
probability measures at $p$ and $q$, respectively.
This is the one-step deterministic image
from the Introduction in the coordinates
\eqref{eq:phase-coordinate}.

In the coordinate $z=y/x$, we represent
these maps by the matrices
\begin{equation}\label{eq:matrix-factorisation}
 L_c=\begin{pmatrix}1&0\\c&1\end{pmatrix},
 \qquad
 R_\beta=\begin{pmatrix}C&-S_\beta\\S_\beta&C\end{pmatrix},
 \qquad
 A(c)=L_cR_\beta.
\end{equation}
The matrix $A(c)$ induces a smooth projective map on $M$,
and its inverse is induced by $A(c)^{-1}$.
Thus each $T_c$ is a smooth projective diffeomorphism of $M$.
For the matrices $J_\beta$, $A(c)$ and $B(t)$
in \eqref{eq:phase-coordinate-matrix},
\eqref{eq:matrix-factorisation} and
\eqref{eq:scalar-matrix-notation}, respectively,
direct multiplication gives
\begin{equation}\label{eq:schrodinger-conjugacy}
 J_\beta A(c)=B(2C-S_\beta c)J_\beta,
 \qquad
 2C-S_\beta c_\lambda(\omega)
       =\lambda-f(\omega),
\end{equation}
where the second equality follows from
\eqref{eq:beta-constants} and \eqref{eq:coefficient}.
Thus, the phase step is conjugate
to the scalar transfer step
\eqref{eq:scalar-matrix}
\begin{equation}\label{eq:phase-scalar-conjugacy}
 J_\beta\circ T_{c_\lambda(\omega)}
 =A_\lambda(\omega)\circ J_\beta.
\end{equation}
The matrices in this identity act on
projective lines.

The coordinate order in
\eqref{eq:phase-coordinate-matrix}
is $(u_j,u_{j+1})$.
To compare with the order $(u_{j+1},u_j)$
in \eqref{eq:scalar-matrix}, set
\begin{equation}\label{eq:phase-transfer-coordinate}
 H_\beta=
 \begin{pmatrix}0&1\\1&0\end{pmatrix}J_\beta
 =\begin{pmatrix}1&0\\C&-S_\beta\end{pmatrix}.
\end{equation}
Then \eqref{eq:schrodinger-conjugacy}
also gives
\begin{equation}\label{eq:backward-phase-conjugacy}
 H_\beta A(c_\lambda(\omega))
 =A_\lambda(\omega)^{-1}H_\beta.
\end{equation}
Hence \eqref{eq:actual-phase-recursion}
is the projective evolution
\eqref{eq:route-laws} expressed in the
fixed coordinates \eqref{eq:phase-coordinate}.
We use the same notation for transformations
and distributions in these coordinates.
The fixed bi-Lipschitz change of coordinates
is independent of $\lambda$ and the trajectory
and preserves NDI, regularity and Wasserstein
separation, with possibly different constants.

The coefficient is nonconstant at every
spectral parameter whenever the potential
is nonconstant. Indeed, for the second
potential \eqref{eq:second-potential},
\begin{equation}\label{eq:energy-oscillation}
 c_\lambda(v)-c_\lambda(w)
 =\frac{V(v)-V(w)}{S_\beta},
 \qquad
 \essosc_F c_\lambda
 =\frac{1}{S_\beta}\essosc_F V.
\end{equation}
The same identities hold for $f$ on
the space of trajectories, with
$\mathbf P$ in place of $F$.

\subsection{Finite-step transformations
and joint distributions}
\label{subsec:phase-law-preliminaries}

In the coordinates
\eqref{eq:phase-coordinate}, we write
the transformations \eqref{eq:route-laws} with $\tau$ of \eqref{eq:tau} as
\begin{equation}\label{eq:route-laws-sec2}
 F^{(k)}_{\lambda,\omega}
 =T_{c_\lambda(\tau^{k-1}\omega)}
       \circ\cdots\circ T_{c_\lambda(\omega)},
 \qquad k\ge1.
\end{equation}
We continue to denote the distribution
of \eqref{eq:route-laws-sec2} by
$\mu_\lambda^{(k)}$.
In the Introduction we allowed an arbitrary initial
direction. For the Cauchy conditions
\eqref{eq:cauchy-data}, we have
\begin{equation}\label{eq:cauchy-phase-laws}
 \Phi_N=F^{(N)}_{\lambda,\omega}\Phi_0^{\rm C},
 \qquad
 \nu_{N,\lambda}=\mathbf E_{\mathbf P}\{\delta_{\Phi_N}\},
\end{equation}
where $F^{(0)}_{\lambda,\omega}$ is the identity.
We continue to use \eqref{eq:phase-law-averages}
for the averages of these distributions.

For the second potential \eqref{eq:second-potential},
the chain $Y_N=X_{-N}$ in
\eqref{eq:evolution-shift} has transition kernel
\[
 K^\leftarrow(v,D)
 =\mathbf P\{X_{-1}\in D\mid X_0=v\},
 \qquad F\text{-almost every }v.
\]
We write $T_{\lambda,v}=T_{c_\lambda(v)}$.
We define the joint evolution for an
arbitrary initial distribution
$\mathcal P_{0,\lambda}$ on $M\times S$
with state distribution $F$ by
\begin{equation}\label{eq:skew-transition}
 (\phi,v)\longmapsto(T_{\lambda,v}\phi,w),
 \qquad w\sim K^\leftarrow(v,dw),
\end{equation}
where $K^\leftarrow(v,dw)$ is the reversed
transition kernel defined by
\eqref{eq:reversed-kernel}.
More explicitly, for every bounded
measurable function $H$ on $M\times S$,
\begin{equation}\label{eq:joint-transition-integral}
 \int H\,d\mathcal P_{N+1,\lambda}
 =\int_{M\times S}\mathcal P_{N,\lambda}(d\phi,dv)
       \int_S H(T_{\lambda,v}\phi,w)
                        K^\leftarrow(v,dw).
\end{equation}
The state distribution remains $F$
by stationarity of $K^\leftarrow$.
We denote the distribution of the phase
by $\nu_{N,\lambda}$, consistently with
\eqref{eq:intro-joint-laws}.

For \eqref{eq:cauchy-data}, we take
\[
 \mathcal P_{0,\lambda}(d\phi,dv)
 =\delta_{\Phi_0^{\rm C}}(d\phi)\,dF(v).
\]
Here $\delta_{\Phi_0^{\rm C}}$ is the Dirac
probability measure at the initial phase
determined by \eqref{eq:cauchy-data}
and \eqref{eq:phase-coordinate}.
Then \eqref{eq:joint-transition-integral}
gives exactly the distribution in
\eqref{eq:cauchy-phase-laws}:
we apply all transformations along
the same Markov trajectory.

\subsection{Arclength and uniform
projective bounds}

We use the angular metric
\eqref{eq:angular-distance} for balls,
the Wasserstein distance
\eqref{eq:wasserstein-definition}
and the kernels $d(\phi,\psi)^{-\gamma}$
in \eqref{eq:main-averaged-fibre-energy}.
We denote arclength measure by
$d\Leb=d\phi$, so that $\Leb(M)=\pi$, and write
\begin{equation}\label{eq:metric-diameter}
 D=\operatorname{diam}(M)=\pi/2.
\end{equation}
For a diffeomorphism $F$ of $M$, we define
\begin{equation}\label{eq:lipschitz-size}
 \mathfrak L(F)
 =\max\{\Lip F,\Lip F^{-1}\}.
\end{equation}
Here $\Lip F$ and $\Lip F^{-1}$ are the
Lipschitz constants of $F$ and its inverse
with respect to the angular metric
\eqref{eq:angular-distance}; explicitly,
\[
 \Lip F
 =\sup_{\phi\ne\psi}
   \frac{d(F\phi,F\psi)}{d(\phi,\psi)}.
\]

If the potential is bounded and
$\lambda$ belongs to a fixed compact
interval $\Lambda$, then
$c_\lambda$ in \eqref{eq:coefficient}
ranges over a bounded set.
The matrices \eqref{eq:matrix-factorisation}
and their inverses depend smoothly on $c$.
Hence there exists $L\ge1$ such that
\begin{equation}\label{eq:uniform-lipschitz}
 \mathfrak L(T_{c_\lambda})\le L,
 \qquad \lambda\in\Lambda
\end{equation}
almost surely.
The constant $L$ is uniform over the states
and over $\lambda\in\Lambda$.

\subsection{Hyperbolic and parabolic projective transformations}
\label{subsec:projective-classification}

We record the elementary classification of real hyperbolic and parabolic projective
transformations of $M$ in the coordinates
\eqref{eq:prioj-space}--\eqref{eq:angular-distance}.

Recall that, for $v\in\R^2\setminus\{0\}$, $[v]$ denotes the
one-dimensional subspace $\R v$.  Under the identification
$M=\R/\pi\Z$ in \eqref{eq:prioj-space}, the line
$[(\cos\phi,\sin\phi)]$ corresponds to the angle $\phi$ modulo $\pi$.
In particular, $\phi$ and $\phi+\pi$ represent the same point of $M$.

For $A\in \GL(2,\R)$, let $\psi_A$ denote its projective action,
\begin{equation}\label{eq:projective-matrix-action}
 \psi_A[v]=[Av],
 \qquad [v]\in M.
\end{equation}
Multiplying $A$ by a nonzero scalar does not change
$\psi_A$.  The sign of $\det A$ determines whether $\psi_A$
preserves or reverses orientation.

Suppose that $\psi=\psi_A$ preserves orientation.  After multiplying
$A$ by a scalar, we may take $A\in\SL(2,\mathbb R)$.  The two matrices
$A$ and $-A$ define the same projective transformation.  Since
\begin{equation}\label{eq:projective-characteristic-polynomial}
 t^2-(\tr A)t+1
\end{equation}
is the characteristic polynomial of $A$,
its discriminant distinguishes the
hyperbolic and parabolic cases.

\smallskip

\noindent
\emph{Hyperbolic transformations.}
We call $\psi$ hyperbolic if
\begin{equation}\label{eq:hyperbolic-trace}
 |\tr A|>2.
\end{equation}
Replacing $A$ by $-A$ if necessary, we may assume
$\tr A>2$.  Then $A$ has two distinct positive eigenvalues
$\lambda$ and $\lambda^{-1}$, with $\lambda>1$, and for some
$B\in\GL(2,\R)$,
\begin{equation}\label{eq:hyperbolic-normal-form}
 A
 =
 B
 \begin{pmatrix}
  \lambda&0\\
  0&\lambda^{-1}
 \end{pmatrix}
 B^{-1}.
\end{equation}
Thus $\psi$ has exactly two fixed points, the two eigendirections
of $A$.  Every point different from these two fixed points converges
under forward iteration to the eigendirection corresponding to
$\lambda$, and under backward iteration to the eigendirection
corresponding to $\lambda^{-1}$.

\smallskip
\noindent
\emph{Parabolic transformations.}
We call $\psi$ parabolic if
\begin{equation}\label{eq:parabolic-trace}
 |\tr A|=2
\end{equation}
and $A\ne\pm I$.  Replacing $A$ by $-A$ if necessary, we may
assume $\tr A=2$.  Then $A$ has the repeated eigenvalue $1$ and
a one-dimensional eigenspace.  Hence, for some
$B\in\GL(2,\R)$ and some $s\ne0$,
\begin{equation}\label{eq:parabolic-normal-form}
 A
 =
 B
 \begin{pmatrix}
  1&s\\
  0&1
 \end{pmatrix}
 B^{-1}.
\end{equation}
In particular, a nontrivial parabolic transformation has exactly
one fixed point $p\in M$, namely the projective eigendirection of
$A$.  Moreover,
\begin{equation}\label{eq:parabolic-iterates}
 \begin{pmatrix}
  1&s\\
  0&1
 \end{pmatrix}^{n}
 =
 \begin{pmatrix}
  1&ns\\
  0&1
 \end{pmatrix},
 \qquad n\ge1.
\end{equation}
It follows from \eqref{eq:parabolic-iterates} that every point
different from $p$ has an infinite orbit and converges under
forward iteration to $p$.

\begin{lemma}
\label{lem:parabolic-finite-set}
Let $\psi$ be a nontrivial parabolic projective transformation
with fixed point $p$.  If a nonempty finite set $E\subset M$
satisfies $\psi(E)=E$, then $E=\{p\}$.
\end{lemma}

\begin{proof}
Suppose that $x\in E$ and $x\ne p$.  Since $\psi(E)=E$,
all forward iterates $\psi^n(x)$ belong to $E$.  On the other hand,
\eqref{eq:parabolic-iterates} shows that these iterates are all
distinct, contradicting the finiteness of $E$.  Hence every point
of $E$ equals $p$, and since $E$ is nonempty, $E=\{p\}$.
\end{proof}

\section{Projective rigidity and deterministic images}\label{sec:ndi}

The points $p,q$ are those in \eqref{eq:one-step-deterministic}.
That calculation excludes NDI at one step and fixes the exceptional
directions $p$ and $q$.

\subsection{Invariant measures and two-step transformations}

Comparing two one-step transformations with
different coefficients produces a translation
in the affine coordinate $z=\tan\phi$.
We first identify the probability measures
that such a translation can preserve.

\begin{lemma}\label{lem:translation}
If $t\ne0$ and a probability on $\widehat\R$ is invariant under
$z\mapsto z+t$, then it is $\delta_\infty$.
\end{lemma}

\begin{proof}
The disjoint half-open intervals
$[j|t|,(j+1)|t|)$, $j\in\Z$, partition $\R$ and have the same mass
by invariance.  Finite total mass forces their common mass to be zero.
Thus the whole mass is at infinity.
\end{proof}

We apply this fact to two one-step
transformations having the same image
of a probability measure.

\begin{corollary}\label{cor:equal-one-step}
If $a\ne b$ and $(T_a)_*\eta=(T_b)_*\eta$, then $\eta=\delta_p$.
Equivalently, a probability preserved by $T_bT_a^{-1}$ with $a\ne b$
must be $\delta_q$.
\end{corollary}

\begin{proof}
From \eqref{eq:matrix-factorisation},
$T_bT_a^{-1}=L_{b-a}$.  Applying Lemma~\ref{lem:translation} to the
common image gives $\delta_q$, whose inverse image under any $T_c$
is $\delta_p$.
\end{proof}

For two-step transformations, we shall again
compare maps having the same image of a
probability measure.
Fixing one such map $F_0$, we obtain
\[
 F_*\eta=(F_0)_*\eta
 \quad\Longrightarrow\quad
 (F_0^{-1}F)_*\eta=\eta.
\]
We therefore study the projective transformations
preserving $\eta$.

We first assume that $\eta$ is non-atomic.
The next lemma gives a common fixed point
in the upper half-plane for all projective
transformations preserving that measure.

\smallskip

Recall that by $\Prob(M)$ we denote the Borel probability measures on
$M$, and by $\Hh$ the upper half-plane.

\smallskip

\begin{lemma}\label{lem:stabiliser}
Let \(G\) be either \(\SL(2,\R)\) or \(\PSL(2,\R)\), acting on
\(M=\mathbb P^1(\R)\), and for \(\eta\in\Prob(M)\) define the
stabiliser of $\eta$
\begin{equation}\label{eq:stabiliser}
    \mathcal G_\eta
    =
    \{g\in G:g_*\eta=\eta\}.
\end{equation}
If \(\eta\) is non-atomic, then \(\mathcal G_\eta\) is compact.
Moreover, all elements of \(\mathcal G_\eta\) fix one common point
of \(\Hh\).
\end{lemma}

\begin{proof}
For \(G=\SL(2,\R)\), compactness is Furstenberg's lemma,
see \cite[Lemma~3.6]{AvilaBochi}.

For \(G=\PSL(2,\R)\), let
\[
    \pi:\SL(2,\R)\longrightarrow\PSL(2,\R)
\]
be the natural projection.  Recall that
\[
    \PSL(2,\R)=\SL(2,\R)/\{\pm I\};
\]
thus two matrices \(A\) and \(-A\) represent the same element of
\(\PSL(2,\R)\).  Since \(A\) and \(-A\) induce the same
transformation of \(\mathbb P^1(\R)\), the full preimage
\[
    \pi^{-1}(\mathcal G_\eta)
\]
is precisely the stabiliser of \(\eta\) in \(\SL(2,\R)\).
It is therefore compact by the preceding case, and hence its
continuous image
\[
    \mathcal G_\eta
    =
    \pi\bigl(\pi^{-1}(\mathcal G_\eta)\bigr)
\]
is compact as well.

Finally, both groups act isometrically on \(\Hh\), and a compact
group of isometries of \(\Hh\) has a common fixed point by the
Cartan--Bruhat--Tits fixed-point theorem,
see \cite[Corollary~II.2.8, p.~179]{BridsonHaefliger}.
\end{proof}


The following lemma bounds the number of coefficient pairs with 
prescribed images and describes the solution set for one prescribed real image.

Recall $C,S_\beta$ from \eqref{eq:beta-constants}, the matrix
$R_\beta$ from \eqref{eq:matrix-factorisation} and the projective
map $T_c$ from \eqref{eq:riccati-map}.
In the affine coordinate $z=y/x$ used for that matrix action, write
\[
 \mathcal R_\beta(z)=\frac{S_\beta+Cz}{C-S_\beta z},\qquad
 T_a(z)=a+\mathcal R_\beta(z),\qquad F_{a,b}=T_bT_a.
\]

\begin{lemma}\label{lem:two-parameter}
The following statements hold.
\begin{enumerate}
\item For fixed $z,z'\in\Hh$, there are at most two real pairs $(a,b)$
      satisfying
      \[
      F_{a,b}(z)=z'.
      \]

\item For fixed distinct $x_1,x_2\in M$ and fixed distinct
      $y_1,y_2\in M$, there are at most two real pairs $(a,b)$
      satisfying
      \[
      F_{a,b}(x_i)=y_i,
      \qquad i=1,2.
      \]

\item For fixed $x,y\in M$, the set of real solutions $(a,b)$ of
      \[
      F_{a,b}(x)=y
      \]
      is one of the following:
      \begin{enumerate}
      \item the empty set;
      \item a horizontal line $b=b_0$;
      \item a vertical line $a=a_0$;
      \item the graph in the $(a,b)$-plane of an orientation-reversing real projective transformation
           \[
            b=\psi(a),
            \]
            with the pole of $\psi$ omitted.
      \end{enumerate}
\end{enumerate}
\end{lemma}

The proof consists of elementary calculations
with fractional linear transformations;
we give the full details in Appendix~\ref{app:proofs}.

\subsection{The finite-state two-step law}

The next proposition treats a coefficient \eqref{eq:coefficient} that is a function
of a finite-state chain.
In the proof of Theorem~\ref{thm:main-ndi},
projective rigidity will reduce a possible deterministic-image
obstruction to a finite coefficient range, allowing us to
apply this finite-state result.

\begin{proposition}\label{prop:finite-two-step}
Let $Q$ be a primitive finite-state transition matrix with stationary
probability $\widetilde\pi$, and let $c(\cdot)$ be a nonconstant real function
of its states.  Then the law
\[
 \sum_{u,v}\widetilde\pi(u)q_{uv}\,
             \delta_{T_{c(v)}T_{c(u)}}
\]
satisfies NDI.
\end{proposition}

\begin{proof}
Suppose $(T_{c(v)}T_{c(u)})_*\eta=\eta'$ for every edge $u\to v$.  
If $\eta'=\delta_q$, fix a state $u$
and an allowed successor $v$, namely $q_{uv}>0$.
By \eqref{eq:one-step-deterministic},
we have $T_{c(v)}^{-1}q=p$.
Thus the relation for this edge gives
\[
 (T_{c(u)})_*\eta
    =(T_{c(v)}^{-1})_*\eta'
    =\delta_p.
\]
Since $u$ was arbitrary, the equality holds for every
state $u$.
Since $c(\cdot)$ takes at least two distinct values,
Corollary~\ref{cor:equal-one-step} implies $\eta=\delta_p$.
By \eqref{eq:one-step-deterministic}, its image under
every $T_c$ is $\delta_q$, contradicting $p\ne q$.

Hence $\eta'\ne\delta_q$. Suppose $q_{uv}>0$ and $q_{uw}>0$, so both $v$ and $w$ can
follow the same state $u$.
Put $\xi_u=(T_{c(u)})_*\eta$.
The relations for these two edges give
\[
 (T_{c(v)})_*\xi_u=\eta',
 \qquad
 (T_{c(w)})_*\xi_u=\eta'.
\]
Applying $T_{c(v)}^{-1}$ to the first equality and then
$T_{c(w)}$ yields
\[
 (T_{c(w)}T_{c(v)}^{-1})_*\eta'
    =(T_{c(w)})_*\xi_u=\eta'.
\]
But $T_{c(w)}T_{c(v)}^{-1}=L_{c(w)-c(v)}$.
If $c(w)\ne c(v)$, Lemma~\ref{lem:translation} would therefore
imply $\eta'=\delta_q$, which has already been excluded.
Hence $c(v)=c(w)$. 
If $c(u)=c(u')$, then
$\xi=(T_{c(u)})_*\eta=(T_{c(u')})_*\eta$.
For successors $v$ of $u$ and $w$ of $u'$, the edge relations give
$(T_{c(v)})_*\xi=(T_{c(w)})_*\xi=\eta'$.
Distinct successor coefficients would force this common image to be
$\delta_q$, again by Lemma~\ref{lem:translation}.
Since $\eta'\ne\delta_q$, the successor coefficients agree.  

Let $\mathcal B$ denote the state space of $Q$.
For each $a\in c(\mathcal B)$, choose a state $u$ with $c(u)=a$
and an allowed next state $v$, and define $\psi(a)=c(v)$.
The preceding argument shows that the result depends neither
on the choice of $u$ with coefficient $a$ nor on the choice
of the allowed next state $v$.
Thus the coefficient of the next state is determined by the
coefficient of the current state, through the map
\[
 \psi\colon c(\mathcal B)\longrightarrow c(\mathcal B),
 \qquad
 c(v)=\psi(c(u))\quad\text{whenever }q_{uv}>0.
\]
For every state $v$, there is a state $u$
with $q_{uv}>0$, so every
coefficient value $c(v)$ occurs as $\psi(c(u))$.
Hence $\psi$ is surjective.
Since its domain and codomain are the same finite set,
it is a permutation.
Irreducibility of $Q$ makes this
permutation a single cycle: a path between any two states joins their
coefficient values by iterates of $\psi$.

Let the cycle length be $d$.  Every return path of length $n$ at a
state $u$ satisfies $\psi^n(c(u))=c(u)$, so $d$ divides every
return length.  Primitivity makes their greatest common divisor one.
Thus $d=1$, and $c$ is constant, a contradiction.
\end{proof}

\subsection{Two steps on a mixing base}

Two further facts are needed for the passage
from finite-state chains to an arbitrary
mixing base.
First, a finite-valued mixing process has
a primitive matrix of possible consecutive
values, even when the process itself
is not Markov.

\begin{lemma}\label{lem:factor-graph}
Let $\{Y_n\}_{n\in\mathbb Z}$ be a stationary mixing process
\eqref{eq:mix-prob}--\eqref{eq:mix-test} whose essential range is a
finite set $\mathcal C$.  Define a matrix $D=(d_{ab})_{a,b\in\mathcal C}$ by
\[
d_{ab}
=
\begin{cases}
1, & \mathbf P\{Y_0=a,Y_1=b\}>0,\\
0, & \mathbf P\{Y_0=a,Y_1=b\}=0.
\end{cases}
\]
Then $D$ is primitive: there exists an integer $N\ge1$ such that,
for every $n\ge N$, every entry of $D^n$ is positive.
\end{lemma}

\begin{proof}
We show that, for all sufficiently large $n$, any two values
$a,b\in\mathcal C$ can be joined by a sequence of $n$ transitions,
each having positive probability.

Let $\vartheta$ be the left shift on the stationary
trajectory space of the process, so that
$Y_n=Y_0\circ\vartheta^n$.
Fix $a,b\in\mathcal C$ and apply
\eqref{eq:mix-prob} to
\[
 A=\{Y_0=a\},
 \qquad B=\{Y_0=b\}.
\]
Since
\[
 \vartheta^{-n}B=\{Y_n=b\},
\]
we obtain
\[
\mathbf P\{Y_0=a,Y_n=b\}
\longrightarrow
\mathbf P\{Y_0=a\}\mathbf P\{Y_0=b\}>0.
\]
Thus there exists an integer $N_{a,b}$ such that
\[
\mathbf P\{Y_0=a,Y_n=b\}>0
\qquad \text{for all}\quad n\ge N_{a,b}.
\]
For each such $n$, summing over the possible intermediate values gives
\[
\mathbf P\{Y_0=a,Y_n=b\}
=
\sum_{a_1,\ldots,a_{n-1}\in\mathcal C}
\mathbf P\{Y_0=a,Y_1=a_1,\ldots,
Y_{n-1}=a_{n-1},Y_n=b\}.
\]
The sum is finite and positive, so at least one summand is positive.
Choose the corresponding sequence
\[
a_0=a,\quad a_1,\ldots,a_{n-1},\quad a_n=b.
\]
For every $j=0,\ldots,n-1$, stationarity gives
\[
\begin{aligned}
\mathbf P\{Y_0=a_j,Y_1=a_{j+1}\}
&=
\mathbf P\{Y_j=a_j,Y_{j+1}=a_{j+1}\}\\
&\ge
\mathbf P\{Y_0=a_0,\ldots,Y_n=a_n\}
>0.
\end{aligned}
\]
Hence $d_{a_j a_{j+1}}=1$ for every $j$, and therefore
\[
(D^n)_{ab}
\ge
d_{a_0a_1}d_{a_1a_2}\cdots d_{a_{n-1}a_n}
=1.
\]
Finally, since $\mathcal C$ is finite, we may take
\[
N=\max_{a,b\in\mathcal C}N_{a,b}.
\]
Then every entry of $D^n$ is positive whenever $n\ge N$.
Thus $D$ is primitive.
\end{proof}

The projective transformation in part~(3)
of Lemma~\ref{lem:two-parameter} reverses
orientation.
We now describe its possible ergodic
invariant probabilities.
The mixing conclusion will be used in the
main proof, and the ergodic conclusion in
Corollary~\ref{cor:ergodic-infinite-range}.

\begin{lemma}
\label{lem:projective-mixing}
Let $\psi$ be an orientation-reversing real
projective transformation preserving
$m\in\Prob(M)$.
If $(M,m,\psi)$ is ergodic, then $m$ is
supported on at most two points.
If it is mixing, then $m$ is a point mass.
\end{lemma}

\begin{proof}
Choose a real matrix $Q=(q_{ij})$ representing
$\psi$.
In the affine coordinate $t=y/x$, its action
and derivative are
\[
 \psi(t)=\frac{q_{21}+q_{22}t}{q_{11}+q_{12}t},
 \qquad
 \psi'(t)=\frac{\det Q}{(q_{11}+q_{12}t)^2},
\]
away from the pole.
Since $\psi$ reverses orientation, this
derivative is negative, and hence $\det Q<0$.
The characteristic polynomial of $Q$ has
discriminant
\[
 (\tr Q)^2-4\det Q>0.
\]
Thus $Q$ has two distinct real eigenvalues
$\lambda_1,\lambda_2$, and
$\lambda_1\lambda_2=\det Q<0$.

Choose corresponding eigenvectors $e_1,e_2$
and use the projective coordinate
$[e_1+t e_2]$, with $t=\infty$ representing
$[e_2]$.
In this coordinate,
\[
 \psi(t)=\frac{\lambda_2}{\lambda_1}t=-s t,
 \qquad s>0.
\]
We retain the notation $\psi,m$ after this
change of coordinates.
Invariance, ergodicity and mixing are
unchanged by this conjugacy.

Assume first that $(M,m,\psi)$ is ergodic.
We consider separately $s\ne1$ and $s=1$.

\smallskip
\noindent
\emph{Case 1: $s\ne1$.}
The points $0$ and $\infty$ are fixed by
$\psi$, and
\[
 \psi^{2n}(t)=s^{2n}t.
\]
For every $t\notin\{0,\infty\}$, these iterates
converge to $0$ if $s<1$, and to $\infty$
if $s>1$.
In either case the limit differs from $t$.
A sufficiently small neighbourhood of $t$
is therefore visited only finitely
often, so $t$ is not recurrent under $\psi^2$.

Since $\psi^2$ preserves $m$, namely
\[
 m(\psi^{-2}E)=m(E),
\]
Poincar\'e
recurrence
\cite[Proposition~4.2.2]{BrinStuck}
implies that $m$-almost every point is
recurrent under $\psi^2$.
Consequently,
\[
 m(\{0,\infty\})=1.
\]
Each of these two points is fixed by $\psi$,
so its singleton is an invariant set:
\[
 \psi^{-1}\{0\}=\{0\},
 \qquad
 \psi^{-1}\{\infty\}=\{\infty\}.
\]
Ergodicity gives
$m(\{0\}),m(\{\infty\})\in\{0,1\}$.
Their sum is one, so one of the two points
has mass one.

\smallskip
\noindent
\emph{Case 2: $s=1$.}
Now $\psi(t)=-t$, and hence
\[
 |\psi(t)|=|t|,
\]
where $|\infty|=\infty$.
We show that $|t|$ is constant
$m$-almost everywhere.
Otherwise, for some finite $r\ge0$,
\[
 0<m([-r,r])<1.
\]
But $[-r,r]$ is invariant under
$t\mapsto-t$, contrary to ergodicity.

Thus $|t|=r_0$ almost everywhere for some
$r_0\in[0,\infty]$.
If $r_0=0$ or $r_0=\infty$, then $m$ is
a point mass.
If $0<r_0<\infty$, then $m$ is supported
on $\{-r_0,r_0\}$.
This proves the ergodic conclusion.

For an invariant Borel set $E$, mixing gives
$m(E)=m(E)^2$, so $(M,m,\psi)$ is ergodic
and $m$ is supported on at most two points.

The only remaining possibility other than
a point mass is a two-point orbit
$\{x,\psi(x)\}$ with $\psi(x)\ne x$.
Invariance assigns mass $1/2$ to each point,
and $\psi^2(x)=x$.
Hence, for every $n\ge1$,
\[
 m\bigl(\{x\}\cap\psi^{-2n}\{x\}\bigr)
 =\frac12,
\]
contradicting mixing, which requires the
limit $m(\{x\})^2=1/4$.
\end{proof}

We now prove the two-step conclusion of
Theorem~\ref{thm:main-ndi}.
The conclusion for functions with finite essential
range will be proved in Section~\ref{sec:routes}.

\begin{proof}[{Proof of the two-step conclusion
of Theorem~\ref{thm:main-ndi}}]
Fix $\lambda\in\R$.
We use the backward transformation
$\tau=\sigma^{-1}$ of \eqref{eq:tau},
which is also mixing, and write
\[
 a(\omega)=c_\lambda(\omega),
 \qquad
 b(\omega)=a(\tau\omega),
\]
where $c_\lambda$ is defined in
\eqref{eq:coefficient}.
Since $\tau$ preserves $\mathbf P$,
\begin{equation}\label{eq:ndi-coefficient-distributions}
 \operatorname{Law}(b)=\operatorname{Law}(a).
\end{equation}
By \eqref{eq:energy-oscillation},
$a$ is not almost surely constant.

Suppose, contrary to NDI, that there are
$\eta,\eta'\in\Prob(M)$ such that
\begin{equation}\label{eq:mixing-common-image}
 (F_{a(\omega),b(\omega)})_*\eta=\eta'
 \qquad\text{for $\mathbf P$-almost every }\omega,
\end{equation}
where $F_{u,v}=T_v\circ T_u$ is the two-step
transformation used in
Lemma~\ref{lem:two-parameter}.

We first exclude
\eqref{eq:mixing-common-image} when $a$ has
finite essential range.
We then distinguish three possibilities
for $\eta$: it is non-atomic, it has at least
two atoms of maximal mass, or it has exactly
one atom of maximal mass.
In the first two cases we obtain a finite
essential range for $a$.
In the last case we obtain a mixing projective
transformation and apply
Lemma~\ref{lem:projective-mixing}.

\medskip
\noindent
\emph{A finite essential range for $a$.}
Suppose that the essential range of $a$ is
a finite set $\mathcal C$.
Since $a$ is not almost surely constant,
$\mathcal C$ contains at least two values.

We use these coefficient values as the states
of an auxiliary finite Markov chain.
We choose its transition probabilities so that
the stationary distribution of two consecutive
states agrees with the distribution of $(a,b)$.
This will allow us to apply
Proposition~\ref{prop:finite-two-step}.

For $u,v\in\mathcal C$, define
\begin{equation}\label{eq:ndi-auxiliary-chain}
 \pi(u)=\mathbf P\{a=u\},
 \qquad
 q_{uv}
 =\frac{\mathbf P\{a=u,\ b=v\}}{\pi(u)}.
\end{equation}
Every $\pi(u)$ is positive, and
\[
 \sum_{v\in\mathcal C}q_{uv}
 =\frac{\mathbf P\{a=u\}}{\pi(u)}
 =1.
\]
Moreover, using
\eqref{eq:ndi-coefficient-distributions},
we obtain
\[
 \sum_{u\in\mathcal C}\pi(u)q_{uv}
 =\mathbf P\{b=v\}
 =\mathbf P\{a=v\}
 =\pi(v).
\]
Thus $Q=(q_{uv})$ is a stochastic matrix
with stationary probability $\pi$.
The original coefficient process need not
be Markov; we use only the agreement of
the consecutive-pair distributions
in \eqref{eq:ndi-auxiliary-chain}.

The process $\{a\circ\tau^n\}_{n\in\Z}$
is stationary and mixing.
Indeed, events defined by this process
are events on the mixing probability
space $(\Omega,\mathbf P,\tau)$.
Applying Lemma~\ref{lem:factor-graph},
we find that the matrix
\[
 D=(d_{uv})_{u,v\in\mathcal C},
 \qquad
 d_{uv}=\mathbf1_{\{q_{uv}>0\}},
\]
is primitive.
The matrices $D$ and $Q$ have positive
entries in exactly the same positions.
Every product contributing a positive term
to $(D^n)_{uv}$ therefore contributes a
positive term to $(Q^n)_{uv}$.
Hence $Q$ is primitive.

If $q_{uv}>0$, then
$\mathbf P\{a=u,b=v\}>0$.
On this event $F_{a,b}$ is the fixed
transformation $T_v\circ T_u$.
The almost-sure relation
\eqref{eq:mixing-common-image} therefore gives
\begin{equation}\label{eq:ndi-auxiliary-common-image}
 (T_v\circ T_u)_*\eta=\eta'
 \qquad\text{whenever }q_{uv}>0.
\end{equation}
We now apply
Proposition~\ref{prop:finite-two-step}
to the primitive matrix $Q$ on $\mathcal C$
and the coefficient function $c(u)=u$.
This function is nonconstant because
$\mathcal C$ contains at least two values.
We conclude that the probability law
\[
 \sum_{u,v\in\mathcal C}
       \pi(u)q_{uv}\,\delta_{T_v\circ T_u}
\]
satisfies NDI.
But \eqref{eq:ndi-auxiliary-common-image}
gives a deterministic image for this law,
a contradiction.

Thus, under
\eqref{eq:mixing-common-image},
$a$ cannot have finite essential range.

\medskip
\noindent
\emph{Case 1: $\eta$ is non-atomic.}
Choose $\omega_0$ from the set of full
probability on which
\eqref{eq:mixing-common-image} holds,
and denote the corresponding transformation by
\[
 F_0=F_{a(\omega_0),b(\omega_0)}.
\]
Then $(F_0)_*\eta=\eta'$.
For almost every $\omega$, we have
\[
 (F_0^{-1}F_{a(\omega),b(\omega)})_*\eta
 =(F_0^{-1})_*\eta'
 =\eta.
\]
Hence every such relative transformation
belongs to the stabiliser $\mathcal G_\eta$
defined in Lemma~\ref{lem:stabiliser}.

Applying Lemma~\ref{lem:stabiliser}, we obtain a point
$z\in\Hh$ fixed by every element of
$\mathcal G_\eta$.
Consequently,
\[
 F_0^{-1}F_{a(\omega),b(\omega)}z=z
 \qquad\text{for almost every }\omega.
\]
With $z'=F_0z$, this becomes
\[
 F_{a(\omega),b(\omega)}z=z'
 \qquad\text{for almost every }\omega.
\]
The points $z,z'\in\Hh$ are independent of $\omega$.
By 
Lemma~\ref{lem:two-parameter}~(1),
there are at most two real pairs $(u,v)$
satisfying $F_{u,v}z=z'$.
It follows that $(a,b)$ belongs almost surely
to a fixed finite set.
In particular, $a$ has finite essential range,
which we have already excluded.

\medskip
\noindent
\emph{Case 2: $\eta$ has at least two atoms
of maximal mass.}
If $m_*>0$ is the supremum of the atom masses
of $\eta$, it is attained among the finitely
many atoms of mass at least $m_*/2$.

Let $E$ and $E'$ be the sets of atoms
of maximal mass of $\eta$ and $\eta'$,
respectively.
For any bijective transformation $F$,
\[
 (F_*\eta)(\{F(x)\})=\eta(\{x\}).
\]
Thus a push-forward by $F$ preserves
the masses of atoms.
By \eqref{eq:mixing-common-image},
$E$ and $E'$ are finite nonempty sets,
and
\begin{equation}\label{eq:ndi-maximal-atom-images}
 F_{a(\omega),b(\omega)}(E)=E'
 \qquad\text{for almost every }\omega.
\end{equation}
The restriction of each transformation
to $E$ is a bijection onto $E'$.

Suppose that $E$ contains distinct points
$x_1,x_2$.
For almost every $\omega$, the ordered pair
\[
 \bigl(
 F_{a(\omega),b(\omega)}x_1,\,
 F_{a(\omega),b(\omega)}x_2
 \bigr)
\]
is one of the finitely many ordered pairs
$(y_1,y_2)$ of distinct points of $E'$.
For each such pair, Lemma~\ref{lem:two-parameter}~(2) 
gives at most two real pairs $(u,v)$ satisfying
\[
 F_{u,v}x_1=y_1,
 \qquad
 F_{u,v}x_2=y_2.
\]
Taking the union over the finitely many
image pairs $(y_1,y_2)$ gives finite essential
range for $a$, contradicting the first
part of the proof.

\medskip
\noindent
\emph{Case 3: $\eta$ has a unique atom
of maximal mass.}
It remains to consider
\[
 E=\{x\},
 \qquad E'=\{y\}.
\]
We do not assume that $\eta$ or $\eta'$
is a point mass.
However, by
\eqref{eq:ndi-maximal-atom-images},
every transformation in
\eqref{eq:mixing-common-image}
must send the unique maximal atom $x$
to the unique maximal atom $y$.
Hence
\[
 F_{a(\omega),b(\omega)}x=y
 \qquad\text{for almost every }\omega.
\]

We apply 
Lemma~\ref{lem:two-parameter}~(3)
to these fixed points $x,y$.
The set of real solutions $(u,v)$ of
$F_{u,v}x=y$ cannot be empty.
A vertical or horizontal solution line makes
$a$ or $b$, respectively, almost surely constant.
Since $a$ and $b$ have the same distribution,
both alternatives contradict nonconstancy.

We are therefore left with one
orientation-reversing real projective
transformation $\psi$ such that
\[
 b=\psi(a)
 \qquad\text{almost surely}.
\]
By the definition of $b$, this is
\begin{equation}\label{eq:ndi-projective-factor}
 a\circ\tau=\psi\circ a
 \qquad\text{almost surely}.
\end{equation}
We show that $\psi$ preserves the
probability measure
\[
 m=a_*\mathbf P
\]
on $M$ and is mixing with respect to $m$.
Here we regard the real coefficient values
as points of the projective line in
the coordinate used for $\psi$.

For every Borel set $B\subset M$,
\eqref{eq:ndi-projective-factor}
and invariance of $\mathbf P$ under $\tau$
give
\[
 \begin{aligned}
 m(\psi^{-1}B)
 &=\mathbf P\{\psi(a)\in B\}
 =\mathbf P\{a\circ\tau\in B\}
 =\mathbf P\{a\in B\}
 =m(B).
 \end{aligned}
\]
Thus $\psi_*m=m$.

Iterating
\eqref{eq:ndi-projective-factor},
we obtain, for every $n\ge1$,
\[
 a\circ\tau^n=\psi^n\circ a
 \qquad\text{almost surely}.
\]
For Borel sets $A,B\subset M$,
mixing of $\tau$ now gives
\[
 \begin{aligned}
 m(A\cap\psi^{-n}B)
 &=\mathbf P\{a\in A,\ \psi^n(a)\in B\}
 =\mathbf P\bigl(
       a^{-1}(A)\cap\tau^{-n}(a^{-1}(B))
                  \bigr)\\
 &\longrightarrow
   \mathbf P\{a\in A\}\mathbf P\{a\in B\}
 =m(A)m(B).
 \end{aligned}
\]
Hence $(M,m,\psi)$ is mixing.
Applying Lemma~\ref{lem:projective-mixing},
we conclude that $m$ is a point mass.
Since $m=a_*\mathbf P$, $a$ is almost
surely constant, contradicting
\eqref{eq:energy-oscillation}.
\end{proof}

\begin{remark}
The two-step conclusion of Theorem~\ref{thm:main-ndi} requires 
only mixing of the underlying transformation
and
measurability and nonconstancy almost
everywhere of the potential function.
We use no $L^2$ norm condition.

For the first potential
\eqref{eq:first-potential},
primitivity of the finite-state transition
matrix gives the required mixing even
when the potential depends on the whole
trajectory.
For the second potential
\eqref{eq:second-potential},
the strict norm condition
\eqref{eq:one-step-gap} implies mixing
and hence also gives the two-step conclusion.

The strict norm condition has an additional
role in the NDI result for every $k\ge2$,
which we prove in the next section,
Section~\ref{sec:routes},
and in the regularisation theorem,
which we prove in
Section~\ref{sec:regularisation}.
\end{remark}

\begin{proof}[Proof of
Corollary~\ref{cor:ergodic-infinite-range}]
Fix $\lambda\in\R$ and write
\[
 a=c_\lambda,\qquad
 b=a\circ\tau,\qquad
 \tau=\sigma^{-1},
\]
using \eqref{eq:coefficient} and \eqref{eq:tau}.
The transformation $\tau$ preserves
$\mathbf P$ and is ergodic.
Since $c_\lambda$ is an affine nonconstant
function of $f$, the function $a$ has
infinite essential range.
Moreover, $a$ and $b$ have the same
distribution.

Suppose, contrary to NDI, that
$\eta,\eta'\in\Prob(M)$ satisfy
\[
 (F_{a(\omega),b(\omega)})_*\eta=\eta'
 \qquad\text{for $\mathbf P$-almost every }\omega.
\]
We use the three cases from the proof of
Theorem~\ref{thm:main-ndi}.

If $\eta$ is non-atomic, the stabiliser
argument and 
Lemma~\ref{lem:two-parameter}~(1)  give a finite
set containing $(a,b)$ almost surely.
If $\eta$ has at least two atoms of maximal
mass, we apply Lemma~\ref{lem:two-parameter}~(2) to two
such atoms and their finitely many possible
ordered pairs of images.
Again $(a,b)$ belongs almost surely to
a finite set.
Neither reduction uses mixing.
Both contradict the infinite essential
range of $a$.

It remains to consider a unique atom $x$
of maximal mass of $\eta$.
A bijective push-forward preserves the
masses of atoms.
Thus $\eta'$ also has a unique atom $y$
of maximal mass, and
\[
 F_{a(\omega),b(\omega)}x=y
 \qquad\text{for almost every }\omega.
\]
We apply 
Lemma~\ref{lem:two-parameter}~(3).
The horizontal and vertical alternatives
would make $b$ or $a$ almost surely constant.
Since these functions have the same
distribution, neither alternative is
possible.
We therefore obtain an orientation-reversing
real projective transformation $\psi$ with
\[
 a\circ\tau=\psi\circ a
 \qquad\text{almost surely}.
\]

Put $m=a_*\mathbf P$, regarding the real
coefficient values as points of $M$.
For every Borel set $B\subset M$,
\[
 \begin{aligned}
 m(\psi^{-1}B)
 &=\mathbf P\{\psi(a)\in B\}
 =\mathbf P\{a\circ\tau\in B\}
 =\mathbf P\{a\in B\}
 =m(B).
 \end{aligned}
\]
Thus $\psi$ preserves $m$.

Let us show that $(M,m,\psi)$ is ergodic.
Let $B\subset M$ be a Borel set invariant under $\psi$
modulo $m$, that is,
\[
 m(B\mathbin{\triangle}\psi^{-1}B)=0,
\]
where $\triangle$ denotes the symmetric difference of sets.
Since $a\circ\tau=\psi\circ a$ almost surely,
\[
 \begin{aligned}
 \tau^{-1}(a^{-1}(B))
 &=(a\circ\tau)^{-1}(B)
 =(\psi\circ a)^{-1}(B)
 =a^{-1}(\psi^{-1}(B))
 \end{aligned}
\]
modulo a $\mathbf P$-null set.
Taking inverse images preserves symmetric
differences, so
\[
 \begin{aligned}
 &\mathbf P\bigl(
 a^{-1}(B)\mathbin{\triangle}
 \tau^{-1}(a^{-1}(B))\bigr)
 =\mathbf P\bigl(
 a^{-1}(B\mathbin{\triangle}\psi^{-1}(B))\bigr)
 =m(B\mathbin{\triangle}\psi^{-1}(B))
 =0.
 \end{aligned}
\]
Thus $a^{-1}(B)$ is invariant under $\tau$
modulo $\mathbf P$.
Ergodicity of $\tau$ gives
\[
 m(B)=\mathbf P(a^{-1}(B))\in\{0,1\}.
\]

Applying Lemma~\ref{lem:projective-mixing},
we conclude that $m$ is supported on at
most two points.
Since $m=a_*\mathbf P$, this contradicts
the infinite essential range of $a$.
\end{proof}

\section{Longer products and finite-state classification}
\label{sec:routes}

We now prove NDI for every number of steps $k\ge2$
in two cases: finite-valued potential functions
on a mixing probability space, and measurable
functions of a Markov state under the strict
norm condition \eqref{eq:one-step-gap}.
For finite-state chains, we also prove that
primitivity is equivalent to NDI for every
nonconstant locally constant function and
every $k\ge2$.

\subsection{Finite-valued functions on a mixing base}

We first prove the remaining conclusion of
Theorem~\ref{thm:main-ndi}. 
As in Theorem~\ref{thm:main-ndi},
we allow an arbitrary invertible mixing
probability-preserving transformation.
The additional assumption is that $f$ has
finite essential range.
In particular, this includes every nonconstant
locally constant function for the first potential
\eqref{eq:first-potential}, since such a function
depends on finitely many coordinates taking
values in a finite alphabet.

\begin{proof}[Proof of Theorem~\ref{thm:main-ndi}
for functions with finite essential range]
Fix $\lambda\in\R$ and write
\[
 a_j(\omega)=c_\lambda(\tau^j\omega),
 \qquad j\in\Z,
\]
where $c_\lambda$ and $\tau$ are defined in
\eqref{eq:coefficient} and \eqref{eq:tau}.
The essential range $\mathcal C$ of $c_\lambda$
is finite, because $c_\lambda$ is an affine
function of $f$.
It contains at least two values by
\eqref{eq:energy-oscillation}.
There are therefore at most $|\mathcal C|^k$
possible sequences of $k$ consecutive
coefficient values.

We prove the result by induction on $k$.
For $k=2$, we use the two-step conclusion
proved in Section~\ref{sec:ndi}.
Let $k\ge3$ and suppose that the
$(k-1)$-step transformation law satisfies NDI.
Assume, towards a contradiction, 
that there are $\eta,\eta'\in\Prob(M)$ such that
\begin{equation}\label{eq:k-common-image}
 (T_{u_{k-1}}\circ\cdots\circ T_{u_0})_*\eta
 =\eta'
\end{equation}
for every $(u_0,\ldots,u_{k-1})\in\mathcal C^k$
with
\[
 \mathbf P\{a_0=u_0,\ldots,a_{k-1}=u_{k-1}\}>0.
\]
We distinguish the cases $\eta'=\delta_q$
and $\eta'\ne\delta_q$, where
$p,q$ are defined in
\eqref{eq:one-step-deterministic}.

\medskip
\noindent
\emph{Case 1: $\eta'=\delta_q$.}
Fix $(u_0,\ldots,u_{k-2})$ such that
\[
 \mathbf P\{a_0=u_0,\ldots,a_{k-2}=u_{k-2}\}>0.
\]
Since $\mathcal C$ is finite, there is
$b\in\mathcal C$ for which
\[
 \mathbf P\{a_0=u_0,\ldots,a_{k-2}=u_{k-2},
                         a_{k-1}=b\}>0.
\]
By \eqref{eq:one-step-deterministic},
$T_b^{-1}q=p$.
Applying $(T_b^{-1})_*$ to
\eqref{eq:k-common-image}, we obtain
\[
 (T_{u_{k-2}}\circ\cdots\circ T_{u_0})_*\eta
 =\delta_p.
\]
This holds for every sequence of $k-1$
coefficient values having positive probability.
Thus the $(k-1)$-step law has a deterministic
image, contradicting the induction hypothesis.

\medskip
\noindent
\emph{Case 2: $\eta'\ne\delta_q$.}
We first show that every sequence of $k-1$
coefficient values having positive probability
has exactly one possible next coefficient value.

Fix such a sequence
$w=(u_0,\ldots,u_{k-2})$ and set
\[
 \xi=(T_{u_{k-2}}\circ\cdots\circ T_{u_0})_*\eta.
\]
Suppose that two distinct values
$b,b'\in\mathcal C$ can follow this sequence
with positive probability.
Then \eqref{eq:k-common-image} gives
\[
 (T_b)_*\xi=\eta',
 \qquad
 (T_{b'})_*\xi=\eta'.
\]
Thus,
\[
 \begin{aligned}
 (T_{b'}T_b^{-1})_*\eta'
 &=(T_{b'})_*((T_b^{-1})_*\eta')
 =(T_{b'})_*\xi
 =\eta'.
 \end{aligned}
\]
By \eqref{eq:matrix-factorisation},
$T_{b'}T_b^{-1}=L_{b'-b}$.
Hence
\[
 (L_{b'-b})_*\eta'=\eta'.
\]
Since $b'\ne b$, this is invariance under
a nonzero translation in the coordinate
$z=\tan\phi$.
Applying Lemma~\ref{lem:translation}, we obtain
the Dirac measure at infinity in $z$, which
corresponds to $\phi=q=\pi/2\pmod\pi$,
giving $\eta'=\delta_q$, a contradiction.

Thus each sequence of $k-1$ coefficient values
having positive probability has at most
one possible next value.
It has at least one, because the probabilities
of its finitely many possible extensions sum
to its positive probability.
Therefore the next coefficient value is unique.

We now record the consecutive coefficients
in a finite-valued process
\begin{equation}\label{eq:finite-coefficient-block-process}
 W_n=(a_n,\ldots,a_{n+k-2}),
 \qquad n\in\Z.
\end{equation}
Each $W_n$ is a tuple of length $k-1$, where
the subscript $n$ denotes its starting index.
Let
\[
 \mathcal W
 =\{w\in\mathcal C^{k-1}:
                    \mathbf P\{W_0=w\}>0\}.
\]
For $w=(u_0,\ldots,u_{k-2})\in\mathcal W$,
denote its unique next coefficient by $s(w)$.
Define
\begin{equation}\label{eq:unique-block-continuation}
 \mathscr D(w)
 =(u_1,\ldots,u_{k-2},s(w)).
\end{equation}
Thus we delete the first coordinate and
append the uniquely determined next value.

Stationarity gives the same continuation rule
at every index. Hence
\begin{equation}\label{eq:deterministic-block-evolution}
 W_{n+1}=\mathscr D(W_n)
 \quad\text{almost surely},
 \quad
 W_n=\mathscr D^n(W_0)
 \quad\text{almost surely,}\, \,  n\ge0.
\end{equation}
Moreover, $\mathscr D$ maps $\mathcal W$
into itself: the positive-probability event
$\{W_0=w\}$ gives $W_1=\mathscr D(w)$,
and $W_1$ has the same distribution as $W_0$.
The initial tuple is random, but once it
is specified, every subsequent tuple is
uniquely determined.

Fix $w=(u_0,\ldots,u_{k-2})\in\mathcal W$
and put
\[
 \pi_w=\mathbf P\{W_0=w\}>0.
\]
By \eqref{eq:deterministic-block-evolution},
$W_n=\mathscr D^n(W_0)$ almost surely.
Hence mixing \eqref{eq:mix-prob} gives
\begin{equation}\label{eq:finite-block-mixing}
 \mathbf P\{W_0=w,W_n=w\}
 =
 \pi_w\mathbf1_{\{\mathscr D^n(w)=w\}}
 \longrightarrow\pi_w^2.
\end{equation}
The left-hand side can only equal
$0$ or $\pi_w$.
The positivity of the limit implies $\pi_w^2=\pi_w$, 
hence $\pi_w=1$.
In particular, $a_0=u_0$ almost surely,
contradicting \eqref{eq:energy-oscillation}.

Both cases are impossible, completing
the induction.
\end{proof}

\subsection{The universal locally constant classification}

We now prove Theorem~\ref{thm:main-classification}.
For an irreducible finite-state chain, we characterise
primitivity by NDI for all nonconstant locally constant
functions and all lengths $k\ge2$.
If the chain is periodic, we construct a potential
depending only on the current state for which
one fixed-length projective transformation is
the identity on every trajectory.

\begin{proof}[Proof of Theorem~\ref{thm:main-classification}] 



The equivalence of (a) and (b) is a standard
consequence of Perron--Frobenius theory
\cite{Seneta}.

Under (a), the stationary shift is mixing
in the sense of
\eqref{eq:mix-prob}--\eqref{eq:mix-test}.
Every locally constant function on $\Omega_P$
has finite range.
Moreover, full support of $\mathbf P$
makes a continuous nonconstant function
nonconstant almost everywhere.
We can therefore apply Theorem~\ref{thm:main-ndi}
for functions with finite essential range to obtain~(c).

It remains to prove that (c) fails when
the chain has period $d>1$.
Let
\[
 \mathcal A=C_0\sqcup\cdots\sqcup C_{d-1}
\]
be its cyclic decomposition, with transitions
from $C_j$ to $C_{j+1}$ modulo $d$.
Each class has stationary mass $1/d$.
We construct a potential constant on each class.
Its values will be chosen so that, at
$\lambda=0$, the scalar transfer matrix $M$
over one complete cycle satisfies
\[
 \tr M=0,\qquad \det M=1.
\]
These two identities give $M^2=-I$.
Indeed, writing
\[
 M=\begin{pmatrix}x&y\\z&-x\end{pmatrix},
 \qquad -x^2-yz=1,
\]
we obtain directly
\begin{equation}\label{eq:msquare}
 M^2=(x^2+yz)I=-I.
\end{equation}
Thus two complete cycles will induce the
identity transformation on the projective line.

We use the matrices $B(t)$ defined in
\eqref{eq:scalar-matrix-notation}.
Choose the scalar transfer coefficient $3$
on $d-1$ classes and a coefficient $b$,
to be determined, on the remaining class.
Put
\[
 D=B(3)^{d-1}.
\]
The trace of the product over a cycle is
\[
 \tr(DB(b))=bD_{11}+D_{12}-D_{21}.
\]
We choose $b$ to make this expression zero.

To compute the required entries of $D$,
define
\[
 s_0=0,\qquad s_1=1,\qquad
 s_{n+1}=3s_n-s_{n-1},\quad \text{for any}\quad n\ge1.
\]
Multiplication by $B(3)$ gives, by induction,
\[
 B(3)^n
 =\begin{pmatrix}
   s_{n+1}&-s_n\\
   s_n&-s_{n-1}
  \end{pmatrix},
 \qquad n\ge1.
\]
Also $s_1>s_0\ge0$, and
\[
 s_n>s_{n-1}\ge0
 \quad\Longrightarrow\quad
 s_{n+1}=3s_n-s_{n-1}>2s_n>s_n.
\]
Hence $s_{n+1}>s_n>0$ for every $n\ge1$.

In particular,
\[
 D_{11}=s_d>0,\qquad
 D_{21}=s_{d-1},\qquad
 D_{12}=-s_{d-1}.
\]
We therefore take
\[
 b=\frac{2s_{d-1}}{s_d}.
\]
Then $0<b<2$, so $b\ne3$, and
\begin{equation}\label{eq:periodic-zero-trace-choice}
 \tr(DB(b))
 =bs_d-2s_{d-1}
 =0.
\end{equation}

Define the potential function by
\[
 f(\omega)=
 \begin{cases}
 -b,&X_0(\omega)\in C_0,\\
 -3,&X_0(\omega)\in C_1\cup\cdots\cup C_{d-1}.
 \end{cases}
\]
This function depends only on the current state
and is therefore locally constant.
It is nonconstant almost everywhere because
$b\ne3$ and every cyclic class has mass $1/d$.
At $\lambda=0$, the scalar transfer coefficient
$\lambda-f$ equals $b$ on $C_0$ and $3$ on
the other classes.
Write
\[
 t_0=b,\qquad t_j=3\quad(1\le j\le d-1),
\]
with indices understood modulo $d$.

The projective transformations
\eqref{eq:route-laws} use $\tau=\sigma^{-1}$.
We use their phase-coordinate representation
\eqref{eq:route-laws-sec2}, conjugate to
scalar products of the matrices $B(t)$ by
\eqref{eq:phase-scalar-conjugacy}.
A transition of the reversed chain goes
from $C_j$ to $C_{j-1}$.
Starting in $C_j$, we therefore use the
successive coefficients
\[
 t_j,\ t_{j-1},\ldots,t_{j-d+1}.
\]
Although the states within a class remain
random, these coefficients are determined
by the initial class.

The scalar product over these $d$ steps is
\[
 M_j
 =B(t_{j-d+1})\cdots B(t_{j-1})B(t_j),
\]
where the rightmost factor is applied first.
For $j=0$, this gives
\[
 M_0=B(3)^{d-1}B(b)=DB(b).
\]
Hence $\tr M_0=0$ by
\eqref{eq:periodic-zero-trace-choice}.

We next verify the same identity for every
initial class.
Starting in $C_{j-1}$ rather than $C_j$
moves $B(t_j)$ from the right end of the
product to the left end.
Since $t_{j-d}=t_j$,
\[
 \begin{aligned}
 M_{j-1}
 &=B(t_j)B(t_{j-d+1})\cdots B(t_{j-1})
 =B(t_j)M_jB(t_j)^{-1}.
 \end{aligned}
\]
Thus the matrices $M_j$ are conjugate,
and $\tr M_j=0$ for every $j$.
Each factor has determinant one, so
$\det M_j=1$.
By \eqref{eq:msquare}, 
\[
 M_j^2=-I
 \quad \text{for every} \quad j=0,\ldots,d-1.
\]

After $d$ steps, the chain returns to its
initial cyclic class.
The next $d$ coefficients are therefore
the same, in the same order.
Thus the $2d$-step scalar product
is $M_j^2=-I$ on every trajectory starting
in $C_j$.
This holds for every initial class.

The matrix $-I$ induces the identity on
the projective line.
Using the conjugacy
\eqref{eq:phase-scalar-conjugacy},
we conclude that every $2d$-step projective
transformation at $\lambda=0$ is the identity.
The same is true in the original coordinates
of \eqref{eq:route-laws}.
Hence
\[
 \mu_0^{(2d)}=\delta_{\mathrm{Id}_M}.
\]
For any $\eta\in\Prob(M)$,
\[
 (\mathrm{Id}_M)_*\eta=\eta.
\]
Taking $\eta'=\eta$, we obtain a deterministic
image for $\mu_0^{(2d)}$, contrary to (c).
Thus (c) implies (a), completing the proof.
\end{proof}

\subsection{Functions of a general-state Markov chain}

We now prove Theorem~\ref{thm:main-markov-ndi} 
for the operator \eqref{eq:operator-introduction}
with the second potential
\eqref{eq:second-potential}.
The function $V$ is measurable 
and not constant $F$-almost everywhere.
We assume the strict one-step norm condition
\eqref{eq:one-step-gap}, but impose no
boundedness or moment assumption on $V$.

We use the chain $Y_n=X_{-n}$ from
\eqref{eq:evolution-shift}.
Its transition kernel is $K^\leftarrow$
in \eqref{eq:reversed-kernel}, and its
Markov operator is $P^\leftarrow$.
By \eqref{eq:reversed-operator},
\[
 \left\|P^\leftarrow\big|_{L^2_0(F)}\right\|
 =\rho<1.
\]
The potential process need not itself
be Markov.

We first use this norm bound to show that
the next coefficient cannot be determined
by the current state for almost every state.
We then apply this fact to the last
transformation in a product.

\begin{lemma}
\label{lem:successor}
Assume \eqref{eq:one-step-gap}, and let
$V$ be measurable 
and not constant $F$-almost everywhere.
For every $\lambda\in\R$, it is impossible
for the conditional distribution of
$c_\lambda(Y_1)$ given $Y_0=v$ to be
a point mass for $F$-almost every $v$.
\end{lemma}

\begin{proof}
Suppose, to the contrary, that this conditional
distribution is a point mass for almost every
state.
Thus, once $Y_0$ is specified, the value
of $c_\lambda(Y_1)$ is determined
almost surely.

The coefficient $c_\lambda$ need not be bounded
or square integrable.
Thus we define
\[
 h(v)=\arctan c_\lambda(v)
      -\int_S\arctan c_\lambda(w)\,dF(w).
\]
Nonconstancy of $c_\lambda$ and injectivity
of $\arctan$ imply that
$h\in L^2_0(F)\setminus\{0\}$.

Since $h(Y_1)$ is determined by $Y_0$,
we obtain
\[
 (P^\leftarrow h)(Y_0)
 =\mathbf E\{h(Y_1)\mid Y_0\}
 =h(Y_1)
 \qquad\text{almost surely}.
\]
Both $Y_0$ and $Y_1$ have distribution $F$.
Hence
\[
 \begin{aligned}
 \|P^\leftarrow h\|_{L^2(F)}^2
 &=\mathbf E\{|(P^\leftarrow h)(Y_0)|^2\}
 =\mathbf E\{|h(Y_1)|^2\}
 =\|h\|_{L^2(F)}^2.
 \end{aligned}
\]
On the other hand,
\[
 \|P^\leftarrow h\|_{L^2(F)}
 \le\rho\|h\|_{L^2(F)},
 \qquad \rho<1,\quad \|h\|_{L^2(F)}>0.
\]
This is a contradiction.
\end{proof}

The proof of Theorem~\ref{thm:main-markov-ndi}
uses Lemma~\ref{lem:successor}
and the fact that two distinct final coefficients
force a common image to be invariant under
a nonzero translation.
We treat the exceptional common image
$\delta_q$ separately.

\begin{proof}[Proof of Theorem~\ref{thm:main-markov-ndi}]
Fix $\lambda\in\R$ and write $c=c_\lambda$,
with $c_\lambda$ as in
\eqref{eq:coefficient} for the second
potential.
Recall $p,q$ from
\eqref{eq:one-step-deterministic}:
\[
 p=\pi/2-\beta,\qquad q=\pi/2\pmod\pi,
 \qquad T_b(p)=q,\qquad T_b^{-1}(q)=p
 \quad(b\in\R).
\]

Given $Y_0=v$, the next state has
distribution $K^\leftarrow(v,dw)$.
The conditional distribution of the next
coefficient is therefore its image under $c$
\begin{equation}\label{eq:conditional-coefficient-law}
 \mathbf P\{c(Y_1)\in B\mid Y_0=v\}
 =K^\leftarrow(v,c^{-1}(B)),
 \qquad B\subseteq\R\ \text{Borel}.
\end{equation}

\noindent \emph{Two steps.} By \eqref{eq:one-step-gap}, the stationary shift
on the two-sided trajectory space is mixing.
Theorem~\ref{thm:main-ndi}, applied to
$f=V\circ X_0$, therefore gives NDI for $k=2$.

\medskip

\noindent
\emph{The induction step.}
Let $k\ge3$, and suppose that we have proved NDI 
for $k-1$ steps.
Assume that there are
$\eta,\eta'\in\Prob(M)$ such that
\begin{equation}\label{eq:markov-k-step-common-image}
 (T_{c(Y_{k-1})}\circ\cdots
       \circ T_{c(Y_0)})_*\eta=\eta'
 \qquad\text{almost surely}.
\end{equation}

If $\eta'=\delta_q$, we apply the inverse
of the last transformation.
Since $T_b^{-1}q=p$ for every real $b$,
\[
 \begin{aligned}
 &(T_{c(Y_{k-2})}\circ\cdots
       \circ T_{c(Y_0)})_*\eta
 =(T_{c(Y_{k-1})}^{-1})_*\delta_q
 =\delta_p
 \qquad\text{almost surely}.
 \end{aligned}
\]
This is a deterministic image for the
$(k-1)$-step law, contradicting the
induction hypothesis.

It remains to consider $\eta'\ne\delta_q$.
By the Markov property of $Y_n$ and
the reversed kernel
\eqref{eq:reversed-kernel},
\begin{equation}\label{eq:last-coefficient-kernel}
 \mathbf P\{c(Y_{k-1})\in B
             \mid Y_0,\ldots,Y_{k-2}\}
 =K^\leftarrow(Y_{k-2},c^{-1}(B))
\end{equation}
for every Borel set $B\subseteq\R$.
Thus, after the preceding states have
been specified, the distribution of the
last coefficient depends only on
$Y_{k-2}$.

For almost every choice
$(v_0,\ldots,v_{k-2})$ under the law
of $(Y_0,\ldots,Y_{k-2})$, put
\[
 \xi
 =(T_{c(v_{k-2})}\circ\cdots
       \circ T_{c(v_0)})_*\eta.
\]
Once these states are fixed, $\xi$ is a
fixed probability measure.
Conditioning
\eqref{eq:markov-k-step-common-image}
on these states gives
\[
 (T_{c(w)})_*\xi=\eta'
 \qquad\text{for }
 K^\leftarrow(v_{k-2},dw)
 \text{-almost every }w.
\]
If the conditional distribution of the last
coefficient is not a point mass, we can
choose two distinct coefficient values
$b_1,b_2$ for which
\[
 (T_{b_1})_*\xi=(T_{b_2})_*\xi=\eta'.
\]
By Corollary~\ref{cor:equal-one-step}, we have
$\xi=\delta_p$, and hence $\eta'=\delta_q$
by \eqref{eq:one-step-deterministic}, a contradiction.

Thus the conditional distribution of the last coefficient
is a point mass almost surely.
Since $Y_{k-2}$ has distribution $F$,
\eqref{eq:last-coefficient-kernel} implies that, for
$F$-almost every $v$, there is $b(v)\in\R$ such that
\[
 K^\leftarrow(v,c^{-1}(B))=\delta_{b(v)}(B)
 \qquad\text{for every Borel set }B\subseteq\R.
\]
This contradicts Lemma~\ref{lem:successor}
and completes the induction.
\end{proof}

\section{Uniform Wasserstein separation}\label{sec:wasserstein}

The Wasserstein distance was defined in
\eqref{eq:wasserstein-definition}.  On compact $M$, it metrises weak
convergence.
Writing $D=\operatorname{diam}(M)=\pi/2$, we shall use
$W\le D$ and
\begin{equation}\label{eq:pushforward-lipschitz}
 W(F_*\eta,F_*\eta')\le\Lip(F)\,W(\eta,\eta').
\end{equation}
The latter follows by applying $F\times F$ to any coupling.

\begin{proposition}\label{thm:wasserstein}
Let $\Lambda\subset\R$ be compact and $(Z,\mathfrak m)$ a probability
space.  Let $F_{\lambda,z}$ be measurable families of continuous
projective diffeomorphisms such that, for $\mathfrak m$-almost every
$z$, the map $\lambda\mapsto F_{\lambda,z}$ is continuous in the
uniform topology on $M$, that is,
\begin{equation}\label{eq:parameter-continuity}
 \sup_{x\in M}d(F_{\lambda_n,z}x,F_{\lambda,z}x)\longrightarrow0
 \quad\text{if }\lambda_n\to\lambda,
 \quad\text{for almost every }z.
\end{equation}
Write
$\mu_\lambda=(z\mapsto F_{\lambda,z})_*\mathfrak m$.
Then the following are equivalent.
\begin{enumerate}
\item $\mu_\lambda$ satisfies NDI for every $\lambda\in\Lambda$.
\item There exists $d_*>0$ such that
\begin{equation}\label{eq:wasserstein-separation}
 \inf_{\lambda\in\Lambda}\inf_{\eta,\eta'\in\Prob(M)}
 \int_Z W((F_{\lambda,z})_*\eta,\eta')\,d\mathfrak m(z)\ge d_*.
\end{equation}
\end{enumerate}
\end{proposition}

\begin{proof}
Condition (2) immediately implies (1).
Assume (1), and suppose that (2) fails.
Then there are
$\lambda_n\in\Lambda$ and
$\eta_n,\eta_n'\in\Prob(M)$ such that
\begin{equation}\label{eq:ndi-uniform-zero-path-form}
 \int_Z
 W\bigl((F_{\lambda_n,z})_*\eta_n,\eta_n'\bigr)
 \,d\mathfrak m(z)\longrightarrow0.
\end{equation}
By compactness of $\Lambda$ and $\Prob(M)$,
after passing to a subsequence we may assume
\[
 \lambda_n\longrightarrow\lambda\in\Lambda,
 \qquad
 \eta_n\xrightarrow{w^*}\eta,
 \quad
 \eta_n'\xrightarrow{w^*}\eta'.
\]

For almost every $z$,
\eqref{eq:parameter-continuity} gives uniform
convergence of $F_{\lambda_n,z}$ to
$F_{\lambda,z}$ on $M$.
Together with weak convergence of $\eta_n$,
this implies
\[
 (F_{\lambda_n,z})_*\eta_n
 \xrightarrow{w^*}
 (F_{\lambda,z})_*\eta.
\]
Indeed, for every $\varphi\in C(M)$,
\begin{align*}
 &\left|
 \int_M\varphi(F_{\lambda_n,z}(x))\,d\eta_n(x)
 -\int_M\varphi(F_{\lambda,z}(x))\,d\eta(x)
 \right| \\
 &\quad
 \le
 \sup_{x\in M}
 |\varphi(F_{\lambda_n,z}(x))
       -\varphi(F_{\lambda,z}(x))|
 +
 \left|
 \int_M\varphi(F_{\lambda,z}(x))
                     \,d(\eta_n-\eta)(x)
 \right|.
\end{align*}
The first term tends to zero by uniform
continuity of $\varphi$, and the second
by weak convergence of $\eta_n$.

Since $M$ is compact, the Wasserstein
distance metrises weak convergence.
Thus, for almost every $z$,
\[
 W\bigl((F_{\lambda_n,z})_*\eta_n,\eta_n'\bigr)
 \longrightarrow
 W\bigl((F_{\lambda,z})_*\eta,\eta'\bigr).
\]
These distances are bounded by
$\operatorname{diam}(M)$.
Applying dominated convergence to
\eqref{eq:ndi-uniform-zero-path-form}, we obtain
\[
 \int_Z
 W\bigl((F_{\lambda,z})_*\eta,\eta'\bigr)
 \,d\mathfrak m(z)=0.
\]
Hence $(F_{\lambda,z})_*\eta=\eta'$
for $\mathfrak m$-almost every $z$,
contradicting NDI at $\lambda$.
\end{proof}

\begin{corollary}\label{cor:route-separation}
Fix a nonzero coupling and a compact interval of spectral
parameters.
The two-step laws of Theorem~\ref{thm:main-ndi}
and Corollary~\ref{cor:ergodic-infinite-range}
satisfy NDI throughout this interval and, equivalently,
the uniform Wasserstein separation estimate
\eqref{eq:wasserstein-separation}.
The same holds for every fixed $k\ge2$ under the hypotheses
of Theorem~\ref{thm:main-ndi} when $f$ has finite essential
range, and under the hypotheses of
Theorem~\ref{thm:main-markov-ndi}.
\end{corollary}

\begin{proof}
Use the stationary path law for $\mathfrak m$ and the fixed-length
transfer product for $F_{\lambda,z}$.  For each fixed finite trajectory segment, every
coefficient is finite and depends affinely on $\lambda$.
The matrices and their projective actions therefore depend continuously
on $\lambda$, uniformly over $M$ for that fixed trajectory segment.
Proposition~\ref{thm:wasserstein} therefore applies
with the corresponding NDI result.
\end{proof}

\begin{corollary}\label{cor:wasserstein-classification}
For a stationary irreducible finite-state Markov chain, primitivity is
equivalent to the following universal statement: for every nonconstant
locally constant function, every nonzero coupling, every fixed $k\ge2$
and every compact interval of spectral parameters, the corresponding transformation laws
have a strictly positive uniform Wasserstein separation.
\end{corollary}

\begin{proof}
Primitivity gives NDI by  Theorem~\ref{thm:main-classification} 
and Corollary~\ref{cor:route-separation} gives the uniform separation.
If the chain has period $d>1$, the potential constructed in the proof of 
Theorem~\ref{thm:main-classification}
makes the $2d$-step
transformation the identity at $\lambda=0$.  Taking $\eta'=\eta$ makes the
separation zero on any compact interval of spectral parameters containing zero.
\end{proof}

\section{Regularised projective energies}\label{sec:energies}

We use the one-dimensional energy construction of
Gorodetski--Kleptsyn--Monakov~\cite{GKM}, with the angular metric
$d$ of \eqref{eq:angular-distance} and arclength measure fixed in Section~\ref{sec:models}.
For $0<\gamma<1/2$ and $\varepsilon>0$ set
\begin{equation}\label{eq:regularising-kernel}
 \varphi_{\gamma,\varepsilon}(r)=
 \begin{cases}
 r^{-(1+\gamma)/2},&r\ge\varepsilon,\\
 \varepsilon^{-\gamma}r^{-(1-\gamma)/2},&0<r<\varepsilon.
 \end{cases}
\end{equation}
For $\nu\in\Prob(M)$ define
\begin{align}
 \mathcal R_{\gamma,\varepsilon}[\nu](y)
   &=\int_M\varphi_{\gamma,\varepsilon}(d(x,y))\,d\nu(x),
       \label{eq:regularised-profile}\\
 \Ereg_{\gamma,\varepsilon}(\nu)
   &=\|\mathcal R_{\gamma,\varepsilon}[\nu]\|_{L^2(M,\Leb)}^2,
       \label{eq:profile-energy}\\
 d\theta_{\gamma,\varepsilon}(\nu)
   &=\frac{\mathcal R_{\gamma,\varepsilon}[\nu]^2}
                  {\Ereg_{\gamma,\varepsilon}(\nu)}\,d\Leb .
       \label{eq:normalised-energy-measure}
\end{align}
The map $\nu\mapsto\mathcal R_{\gamma,\varepsilon}[\nu]$ is linear. Tonelli's theorem gives
\[
 \Ereg_{\gamma,\varepsilon}(\nu)
     =\iint K_{\gamma,\varepsilon}(x,z)\,d\nu(x)d\nu(z),
\]
where
\[
 K_{\gamma,\varepsilon}(x,z)
    =\int_M\varphi_{\gamma,\varepsilon}(d(x,y))
              \varphi_{\gamma,\varepsilon}(d(z,y))\,d\Leb(y).
\]
The Euclidean comparison kernel and its interaction energy are
\begin{align}
 U_{\gamma,\varepsilon}(r)
   &=\int_\R\varphi_{\gamma,\varepsilon}(|y|)
                     \varphi_{\gamma,\varepsilon}(|r-y|)\,dy,
       \label{eq:interaction-kernel}\\
 \mathcal E_{\gamma,\varepsilon}(\nu)
   &=\iint U_{\gamma,\varepsilon}(d(x,z))\,d\nu(x)d\nu(z).
       \label{eq:interaction-energy}
\end{align}

We set $\varphi_{\gamma,\varepsilon}(0)=+\infty$.
The regularised functions are interpreted up to Lebesgue null sets.
Lemma~\ref{lem:initial-energy} gives square
integrability, while positivity of the kernel
gives a nonzero norm.
Thus \eqref{eq:normalised-energy-measure}
defines a probability measure.

\begin{lemma}\label{lem:initial-energy}
For fixed $0<\gamma<1/2$ there is $C_\gamma<\infty$ such that
\begin{equation}\label{eq:universal-initial-energy}
 \Ereg_{\gamma,\varepsilon}(\nu)
       \le C_\gamma\varepsilon^{-\gamma},
       \qquad \text{for every}\quad 0<\varepsilon<1,\ \nu\in\Prob(M).
\end{equation}
In particular all regularised functions belong to $L^2(M,\Leb)$.
\end{lemma}

\begin{proof}
For a point mass the energy is the same at every location and equals
$2\int_0^{\pi/2}\varphi_{\gamma,\varepsilon}(r)^2\,dr$.
Its two pieces are bounded by
\[
 2\varepsilon^{-2\gamma}\int_0^\varepsilon r^{-1+\gamma}\,dr
       +2\int_\varepsilon^{\pi/2}r^{-1-\gamma}\,dr
 \le\frac4\gamma\varepsilon^{-\gamma}.
\]
For an arbitrary probability, apply Jensen's inequality to the
integral defining $\mathcal R_{\gamma,\varepsilon}[\nu]$ and then integrate in $y$.
For larger $\varepsilon$ the regularising kernel only decreases, so
square integrability holds for every $\varepsilon>0$.
\end{proof}

\begin{lemma}\label{lem:energy-toolkit}
The following estimates hold. Parts~\textup{(1)}--\textup{(3)} hold for every $0<\gamma<1/2$.
\begin{enumerate}
\item The kernel $U_{\gamma,\varepsilon}$ is finite and non-increasing.
      For some $a_\gamma,b_\gamma>0$,
\begin{equation}\label{eq:interaction-power-bounds}
 a_\gamma r^{-\gamma}\le U_{\gamma,\varepsilon}(r)
                 \le b_\gamma r^{-\gamma}\qquad(r\ge\varepsilon).
\end{equation}
\item For every $\delta>0$ there is $C_{\gamma,\delta}<\infty$ such that,
      for every probability $\nu$ and every $\varepsilon>0$,
\begin{align}
 \mathcal E_{\gamma,\varepsilon}(\nu)
     &\le(1+\delta)\Ereg_{\gamma,\varepsilon}(\nu)+C_{\gamma,\delta},
       \label{eq:energy-comparison-a}\\
 \Ereg_{\gamma,\varepsilon}(\nu)
     &\le(1+\delta)\mathcal E_{\gamma,\varepsilon}(\nu)+C_{\gamma,\delta}.
       \label{eq:energy-comparison-b}
\end{align}
\item For every diffeomorphism $F$,
\begin{equation}\label{eq:interaction-distortion}
 \mathfrak L(F)^{-\gamma}\mathcal E_{\gamma,\varepsilon}(\nu)
 \le\mathcal E_{\gamma,\varepsilon}(F_*\nu)
 \le\mathfrak L(F)^\gamma\mathcal E_{\gamma,\varepsilon}(\nu),
\end{equation}
where $\mathfrak L(F)$ is defined in \eqref{eq:lipschitz-size}.
\item For every $\delta>0$ and $L<\infty$ there exists
      $0 < \gamma_0(\delta,L) \leq 1/2$ such that, for each
      $0<\gamma<\gamma_0(\delta,L)$, there is a finite threshold
      $H_{\gamma,\delta,L}$ with
\begin{equation}\label{eq:approximate-equivariance}
 W(F_*\theta_{\gamma,\varepsilon}(\nu),
                 \theta_{\gamma,\varepsilon}(F_*\nu))<\delta
\end{equation}
      whenever $\mathfrak L(F)\le L$ and
      $\Ereg_{\gamma,\varepsilon}(\nu)>H_{\gamma,\delta,L}$.
      The threshold is uniform in $\varepsilon>0$ and $F$.
\end{enumerate}
\end{lemma}

\begin{proof}
Part~(1) is \cite[Proposition~4.6]{GKM}.
Part~(2) is the all-energy form of
\cite[Proposition~4.10 and Remark~4.12]{GKM}.  The arclength
normalisation ensures that the local Euclidean comparison kernel is
exactly \eqref{eq:interaction-kernel}.
For part~(3), $$\mathfrak L(F)^{-1}d(x,z)\le
d(Fx,Fz)\le\mathfrak L(F)d(x,z).$$
The scaling estimate in \cite[Proposition~4.6(3)]{GKM} and monotonicity
give, pointwise,
\[
 \mathfrak L(F)^{-\gamma}U_{\gamma,\varepsilon}(d(x,z))
 \le U_{\gamma,\varepsilon}(d(Fx,Fz))
 \le\mathfrak L(F)^\gamma U_{\gamma,\varepsilon}(d(x,z)).
\]
Integration in $\nu(x)\nu(z)$ proves
\eqref{eq:interaction-distortion}, with exactly the same $\varepsilon$.
This is also \cite[Proposition~4.13]{GKM}.

Part~(4) is \cite[Proposition~4.21]{GKM} on the angular
Riemannian circle.  Apply it with the strict size bound $L+1$ to
cover all maps with $\mathfrak L(F)\le L$.
Its exponent is chosen after $\delta,L$, and its high-energy threshold
is then uniform in the map, the probability and $\varepsilon$.
These are pointwise analytic estimates.  We do not use an averaged
energy-contraction theorem from \cite{GKM} for the Markov evolution.
\end{proof}

\begin{lemma}\label{lem:squares}
Let $r$ and $s$ be nonzero nonnegative functions in $L^2(M,\Leb)$.
Then
\begin{equation}\label{eq:square-stability}
 \left\|\frac{r^2}{\|r\|_2^2}
              -\frac{s^2}{\|s\|_2^2}\right\|_{L^1}
 \le\frac{4\|r-s\|_2}{\min\{\|r\|_2,\|s\|_2\}}.
\end{equation}
The Wasserstein distance between the corresponding probabilities is
at most $D$ times this $L^1$ distance.
\end{lemma}

\begin{proof}
The reverse triangle inequality gives
$|\|r\|_2-\|s\|_2|\le\|r-s\|_2$ and hence
\[
 \left\|\frac r{\|r\|_2}-\frac s{\|s\|_2}\right\|_2
      \le\frac{2\|r-s\|_2}{\|r\|_2}.
\]
The $L^1$ norm of the difference of the squares of two unit
$L^2$ functions is at most twice their $L^2$ distance, by
Cauchy--Schwarz.  This proves \eqref{eq:square-stability}.
For the Wasserstein estimate, couple the common part of the two
density measures identically and couple their remaining mass arbitrarily.
The cost is at most $D$ times that remaining mass, which is bounded
by the displayed $L^1$ distance.
\end{proof}

\begin{lemma}\label{lem:ball-energy}
For $r>\varepsilon$ and $x\in M$,
\begin{equation}\label{eq:ball-energy}
 \nu(B_r(x))
   \le\left(\frac{2^\gamma}{a_\gamma}
                 \mathcal E_{\gamma,\varepsilon}(\nu)\right)^{1/2}
                  r^{\gamma/2}.
\end{equation}
\end{lemma}

\begin{proof}
For $\phi,\psi\in B_r(x)$, the triangle inequality
gives $d(\phi,\psi)<2r$.
By \eqref{eq:interaction-energy}, monotonicity
and \eqref{eq:interaction-power-bounds}, since
$2r>\varepsilon$, we obtain
\[
 \begin{aligned}
 \mathcal E_{\gamma,\varepsilon}(\nu)
 &\ge
 \iint_{B_r(x)\times B_r(x)}
 U_{\gamma,\varepsilon}(d(\phi,\psi))\,
 d\nu(\phi)d\nu(\psi)\\
 &\ge U_{\gamma,\varepsilon}(2r)\nu(B_r(x))^2
 \ge a_\gamma(2r)^{-\gamma}\nu(B_r(x))^2.
 \end{aligned}
\]
Rearrange this inequality.
\end{proof}

\section{Regularisation for bounded Markov potentials}
\label{sec:markov-evolution}
\label{sec:regularisation}

We now prove Theorem~\ref{thm:main-regularity}.
We first express the distributions of directions
through the two-step transformations along
a Markov trajectory and identify the distributions
corresponding to the Cauchy conditions
\eqref{eq:cauchy-data}.
We then obtain a recursion for the conditional
distributions given the state of the chain.
This recursion allows us to separate the effect
of a projective transformation from the subsequent
averaging over the Markov transition.

We use the regularised energies and the analytic
estimates of Gorodetski--Kleptsyn--Monakov
\cite{GKM}, recorded in Section~\ref{sec:energies}.
The regularisation theorem of \cite{GKM} does not
apply directly.
In the averaging considered there, each random
transformation acts on the same probability measure.
Here the direction of the solution and the next
transformation depend on the same Markov trajectory.
The conditional distribution of the direction
therefore depends on the current state.
We apply the analytic estimates to these conditional
distributions and use the strict norm condition
to estimate the decrease in energy under
Markov averaging.

Our main step is to prove that sufficiently
large regularised energy decreases over
three steps.
Iteration gives the finite-step bound
and the conclusions for stationary
distributions.

Throughout this section assume
\begin{equation}\label{eq:regularity-assumptions}
 V\in L^\infty(F),\qquad
 \essosc_F V>0,\qquad
 \rho=\left\|P\big|_{L^2_0(F)}\right\|<1.
\end{equation}
Fix $\beta\in(0,\pi)$ and a compact interval
$\Lambda$ of spectral parameters.
Let $L$ be the uniform bound
\eqref{eq:uniform-lipschitz}, and let $d_*>0$
be the separation constant for the two-step
laws from Corollary~\ref{cor:route-separation}.
All constants may depend on these fixed data,
including $F,K,V$, but are uniform in
$\lambda\in\Lambda$, steps, the regularisation
scale and the initial joint distribution
with state distribution $F$.

We use the reversed chain $Y_n=X_{-n}$
from \eqref{eq:evolution-shift} and write
$T_{\lambda,v}=T_{c_\lambda(v)}$.

\subsection{Finite-step and conditional distributions}

We construct the distributions used below from consecutive
two-step transformations along the same Markov trajectory,
each with law $\mu_\lambda^{(2)}$ satisfying NDI.
We identify the resulting Cauchy distributions with those in
\eqref{eq:cauchy-phase-laws} and \eqref{eq:intro-joint-laws},
without assuming uniqueness of a stationary joint distribution.
For $i\ge0$, put
\begin{equation}\label{eq:translated-two-step}
 F_{\lambda,i}^{(2)}(\omega)
 =F_{\lambda,\tau^{2i}\omega}^{(2)}
 =T_{c_\lambda(\tau^{2i+1}\omega)}
       \circ T_{c_\lambda(\tau^{2i}\omega)},
 \qquad
 \mu_{\lambda,i}^{(2)}
 =\operatorname{Law}(F_{\lambda,i}^{(2)}),
\end{equation}
where $\tau=\sigma^{-1}$ as in \eqref{eq:tau}.
Since $\mathbf P$ is invariant under $\tau$,
for every Borel set
$\mathcal M\subset\operatorname{Diff}^{\infty}(M)$,
\begin{equation}\label{eq:two-step-shift-invariance}
 \begin{aligned}
 \mathbf P\{F_{\lambda,i}^{(2)}\in\mathcal M\}
 &=\mathbf P\bigl((\tau^{2i})^{-1}
       \{\omega:F_{\lambda,\omega}^{(2)}
                         \in\mathcal M\}\bigr)
 =\mathbf P\{F_{\lambda,\omega}^{(2)}
                         \in\mathcal M\}.
 \end{aligned}
\end{equation}
Thus $\mu_{\lambda,i}^{(2)}=\mu_\lambda^{(2)}$.
The transformations in \eqref{eq:translated-two-step}
have the same distribution, but remain dependent
through the Markov trajectory.

Let the initial direction have distribution
$\nu_0$, independently of the chain.
At even steps its distribution is
\begin{equation}\label{eq:even-phase-laws}
 \nu_{2m,\lambda}
 =\mathbf E_{\mathbf P}\left[
   \bigl(F_{\lambda,m-1}^{(2)}\circ\cdots
             \circ F_{\lambda,0}^{(2)}\bigr)_*\nu_0
             \right],
 \qquad m\ge0,
\end{equation}
where the empty product is the identity.
At odd steps we apply one further Riccati map:
\begin{equation}\label{eq:odd-phase-laws}
 \nu_{2m+1,\lambda}
 =\mathbf E_{\mathbf P}\left[
  \bigl(T_{c_\lambda(\tau^{2m}\omega)}
       \circ F_{\lambda,m-1}^{(2)}\circ\cdots
       \circ F_{\lambda,0}^{(2)}\bigr)_*\nu_0
             \right].
\end{equation}
All transformations in these formulas are evaluated
along the same trajectory.
Choosing $\nu_0=\delta_{\Phi_0^{\rm C}}$, the
Dirac probability at the initial Cauchy phase,
gives exactly the distributions
\eqref{eq:cauchy-phase-laws} of the recursion
\eqref{eq:actual-phase-recursion}.

We can allow an arbitrary initial joint distribution
with state distribution $F$, evolved by
\eqref{eq:skew-transition}.
The direction satisfies
\begin{equation}\label{eq:markov-phase-recursion}
 \Phi_{n+1}=T_{\lambda,Y_n}\Phi_n.
\end{equation}
We disintegrate the joint distribution as
\begin{equation}\label{eq:phase-disintegration}
 \mathcal P_{n,\lambda}(d\phi,dv)
 =\mu_{n,v,\lambda}(d\phi)\,dF(v),
 \qquad
 \nu_{n,\lambda}
 =\int_S\mu_{n,v,\lambda}\,dF(v).
\end{equation}
For the Cauchy initial condition,
$\mu_{0,v,\lambda}=\delta_{\Phi_0^{\rm C}}$
for $F$-almost every $v$.

The chain $Y_n$ has transition kernel
$K^\leftarrow$.
When we condition $Y_n=v$ on $Y_{n+1}=w$,
the identity \eqref{eq:reversed-kernel} gives
the conditional kernel $K(w,dv)$.
This is why the original forward operator
appears in the following recursion.

\begin{proposition}
\label{prop:fibre-recursion}
For $F$-almost every $w$,
\begin{equation}\label{eq:fibre-recursion}
 \mu_{n+1,w,\lambda}
 =\int_S(T_{\lambda,v})_*\mu_{n,v,\lambda}
                                      \,K(w,dv).
\end{equation}
\end{proposition}

\begin{proof}
For a bounded measurable test function $H$
on $M\times S$, the transition
\eqref{eq:skew-transition} gives
\[
 \begin{aligned}
 \mathbf E H(\Phi_{n+1},Y_{n+1})
 &=\int_S dF(v)\int_S K^\leftarrow(v,dw)
       \int_M H(T_{\lambda,v}\phi,w)
                         \,d\mu_{n,v,\lambda}(\phi)\\
 &=\int_S dF(w)\int_S K(w,dv)
       \int_M H(T_{\lambda,v}\phi,w)
                         \,d\mu_{n,v,\lambda}(\phi),
 \end{aligned}
\]
where the second equality follows from
\eqref{eq:reversed-kernel}.
By the uniqueness part of Rokhlin's disintegration theorem, we obtain
\eqref{eq:fibre-recursion}.
\end{proof}

We shall also apply the regularity estimates
to stationary joint distributions.

\begin{lemma}
\label{lem:stationary-existence}
For every fixed $\lambda$, the transition
\eqref{eq:skew-transition} has a stationary
joint distribution with state distribution $F$.

This conclusion requires neither boundedness
of $V$ nor the norm condition
\eqref{eq:one-step-gap}.
\end{lemma}

We give a self-contained proof of
Lemma~\ref{lem:stationary-existence}
in Appendix~\ref{app:proofs}.

\subsection{The decrease under Markov averaging}

Fix $0<\gamma<1/2$ and $0<\varepsilon<1$.
We apply the regularisation
\eqref{eq:regularised-profile} to the conditional
distribution $\mu_{n,v,\lambda}$ of
\eqref{eq:phase-disintegration} before and after the projective
transformation
\begin{equation}\label{eq:markov-profiles}
 \begin{aligned}
 R_n(v,z)
 &=\int_M\varphi_{\gamma,\varepsilon}
                   (d(\phi,z))\,d\mu_{n,v,\lambda}(\phi),\\
 S_n(v,z)
 &=\int_M\varphi_{\gamma,\varepsilon}
                   (d(T_{\lambda,v}\phi,z))
                              \,d\mu_{n,v,\lambda}(\phi).
 \end{aligned}
\end{equation}
Thus $S_n(v,\cdot)$ is the regularised function
associated with $(T_{\lambda,v})_*\mu_{n,v,\lambda}$.
We suppress $\lambda,\gamma,\varepsilon$
in the notation and set
\begin{equation}\label{eq:markov-energies}
 \begin{aligned}
 A_n^2
 &=\int_S\int_M R_n(v,z)^2\,d\Leb(z)\,dF(v),\\
 B_n^2
 &=\int_S\int_M S_n(v,z)^2\,d\Leb(z)\,dF(v).
 \end{aligned}
\end{equation}
Both functions in \eqref{eq:markov-profiles}
are measurable, since they are integrals
of nonnegative measurable functions against
measurable probability kernels.
By \eqref{eq:profile-energy} and
\eqref{eq:markov-profiles}, we have
\begin{equation}\label{eq:conditional-profile-energies}
 A_n^2
 =\int_S\Ereg_{\gamma,\varepsilon}
       (\mu_{n,v,\lambda})\,dF(v),
 \qquad
 B_n^2
 =\int_S\Ereg_{\gamma,\varepsilon}
       \bigl((T_{\lambda,v})_*\mu_{n,v,\lambda}\bigr)\,dF(v).
\end{equation}
By Lemma~\ref{lem:initial-energy}, the integrals in
\eqref{eq:conditional-profile-energies} are finite and satisfy
\[
 A_n^2,\ B_n^2\le C_\gamma\varepsilon^{-\gamma}.
\]
In particular, $R_n,S_n\in L^2(F\otimes\Leb)$.

Applying Tonelli's theorem to
\eqref{eq:fibre-recursion}, we obtain
\begin{equation}\label{eq:profile-recursion}
 R_{n+1}(v,z)
 =\int_S S_n(w,z)\,K(v,dw)
 =(PS_n)(v,z)
\end{equation}
for $(F\otimes\Leb)$-almost every $(v,z)$.
Here $P$ acts only on the state variable.

We use $\|\cdot\|_2$ for the norm in
$L^2(F\otimes\Leb)$.
For functions of $z$ alone, we retain the notation
$\mathscr H=L^2(M,\Leb)$.
For a fixed state $v$, we abbreviate
$R_n(v,\cdot)$ and $S_n(v,\cdot)$
to $R_n(v)$ and $S_n(v)$, respectively.
Their $\mathscr H$-norms are taken
in the variable $z$.
Define
\[
 \overline S_n(z)=\int_S S_n(v,z)\,dF(v),
 \qquad
 S_n^\perp(v,z)=S_n(v,z)-\overline S_n(z).
\]
By Jensen's inequality,
$\|\overline S_n\|_{\mathscr H}^2\le B_n^2$,
so these functions are square integrable.

The difference $B_n^2-A_{n+1}^2$ is the decrease
in energy under Markov averaging.
We show that a small decrease forces
$S_n(v,\cdot)$ to be close, on average over
the states, to the single function $\overline S_n$.
The same function then approximates
$R_{n+1}(v,\cdot)$.

\begin{proposition}
\label{prop:markov-defect}
Under \eqref{eq:one-step-gap},
\begin{equation}\label{eq:markov-defect}
 D_n^{\rm M}:=B_n^2-A_{n+1}^2
 =\|S_n^\perp\|_2^2-\|PS_n^\perp\|_2^2
 \ge(1-\rho^2)\|S_n^\perp\|_2^2.
\end{equation}
Moreover,
\begin{equation}\label{eq:input-collapse}
 \|R_{n+1}-\overline S_n\|_2
 \le\rho\|S_n^\perp\|_2.
\end{equation}
\end{proposition}

\begin{proof}
For Lebesgue-almost every $z$,
the function $S_n^\perp(\cdot,z)$ belongs to
$L^2_0(F)$.
The norm condition \eqref{eq:one-step-gap}
therefore gives
\[
 \int_S|PS_n^\perp(v,z)|^2\,dF(v)
 \le\rho^2\int_S|S_n^\perp(v,z)|^2\,dF(v).
\]
Integrating in $z$, we obtain
\[
 \|PS_n^\perp\|_2^2
 \le\rho^2\|S_n^\perp\|_2^2.
\]

Since $P$ preserves constants,
\eqref{eq:profile-recursion} gives
\[
 R_{n+1}(v,z)
 =\overline S_n(z)+(PS_n^\perp)(v,z).
\]
Both $S_n^\perp(\cdot,z)$ and
$PS_n^\perp(\cdot,z)$ have zero $F$-mean.
The mixed terms therefore vanish when we
expand the squares and integrate:
\[
 \begin{aligned}
 B_n^2
 &=\|\overline S_n\|_{\mathscr H}^2
                         +\|S_n^\perp\|_2^2, \quad 
 A_{n+1}^2
 =\|\overline S_n\|_{\mathscr H}^2
                         +\|PS_n^\perp\|_2^2.
 \end{aligned}
\]
Subtracting proves the identity and inequality
in \eqref{eq:markov-defect}.
Finally,
$R_{n+1}-\overline S_n=PS_n^\perp$,
which proves \eqref{eq:input-collapse}.
\end{proof}

\subsection{Three-step contraction}

We next estimate the change in energy caused
by the projective transformation.
Using the comparisons
\eqref{eq:energy-comparison-a}--\eqref{eq:energy-comparison-b}
and the distortion estimate \eqref{eq:interaction-distortion}, we show
that this change has arbitrarily small
relative size when $\gamma$ is sufficiently
small, up to an additive constant independent
of $\varepsilon$.

\Needspace{8\baselineskip}
\begin{lemma}
\label{lem:fibre-energy-stability}
For every $0<\zeta<1$, after choosing
$\gamma>0$ sufficiently small,
\begin{equation}\label{eq:averaged-energy-stability}
 |B_n^2-A_n^2|\le\zeta A_n^2+C_\zeta
\end{equation}
with $C_\zeta<\infty$ uniform in
$n,\lambda,\varepsilon$ and the initial fibres.
\end{lemma}

\begin{proof}
Fix $\zeta\in(0,1)$.
Choose $\delta>0$ and then $\gamma>0$ so that
\begin{equation}\label{eq:fibre-stability-choices}
 q:=(1+\delta)^2L^\gamma\le1+\zeta,
\end{equation}
where $L$ is the uniform Lipschitz bound
in \eqref{eq:uniform-lipschitz}.
Such a choice is possible by first taking
$(1+\delta)^2<1+\zeta$ and then decreasing $\gamma$.
Write $C_0=C_{\gamma,\delta}$ for the common
constant in the two comparisons
\eqref{eq:energy-comparison-a} and
\eqref{eq:energy-comparison-b}.
Fix a state $v$ for which the uniform Lipschitz
bound holds, and abbreviate
$\mu=\mu_{n,v,\lambda}$ and $T=T_{\lambda,v}$.
Both $T$ and $T^{-1}$ have size at most $L$.

Applying the energy comparisons
\eqref{eq:energy-comparison-a}--\eqref{eq:energy-comparison-b}
and the distortion estimate \eqref{eq:interaction-distortion}
consecutively, we obtain
\begin{align*}
 \Ereg_{\gamma,\varepsilon}(T_*\mu)
 &\le(1+\delta)\mathcal E_{\gamma,\varepsilon}(T_*\mu)+C_0\\
 &\le(1+\delta)L^\gamma
                  \mathcal E_{\gamma,\varepsilon}(\mu)+C_0\\
 &\le q\Ereg_{\gamma,\varepsilon}(\mu)
             +\bigl((1+\delta)L^\gamma+1\bigr)C_0.
\end{align*}
Put $C_1=((1+\delta)L^\gamma+1)C_0$.
Applying precisely the same calculation
to $T^{-1}$ and $T_*\mu$ gives
\[
 \Ereg_{\gamma,\varepsilon}(\mu)
 \le q\Ereg_{\gamma,\varepsilon}(T_*\mu)+C_1.
\]
With $E_v=\Ereg_{\gamma,\varepsilon}(\mu)$
and $E'_v=\Ereg_{\gamma,\varepsilon}(T_*\mu)$,
using $q^{-1}\ge1-\zeta$ and $C_1/q\le C_1$,
the two energy comparisons give
\[
 |E'_v-E_v|\le\zeta E_v+C_1,
\]
uniformly in $v,n,\lambda,\varepsilon$.

Integrating the pointwise inequality and using
\eqref{eq:conditional-profile-energies} and $F(S)=1$,
we obtain
\[
 |B_n^2-A_n^2|
 \le\int_S|E'_v-E_v|\,dF(v)
 \le\zeta A_n^2+C_1.
\]
This is \eqref{eq:averaged-energy-stability}
with $C_\zeta=C_1$.
\end{proof}

We now combine the two energy estimates.
By \eqref{eq:averaged-energy-stability},
applying a projective transformation changes
a sufficiently large energy by only a small
relative amount.
Markov averaging decreases the energy.
If the total decrease over three steps is
substantial, it outweighs the possible increases
caused by the projective transformations.
We therefore need to exclude the alternative
in which the energy remains large and all three
decreases under Markov averaging are small.

We explain why we use three steps.
By \eqref{eq:markov-defect} and
\eqref{eq:input-collapse}, a small decrease
at step $j$ makes both $S_j$ and $R_{j+1}$
close to the same function $\overline S_j$,
which does not depend on the state.
For $j=n,n+1,n+2$, these estimates give
\[
 \begin{aligned}
 R_{n+1}&\approx\overline S_n,\\
 S_{n+1}&\approx\overline S_{n+1}
                         \approx R_{n+2},\\
 S_{n+2}&\approx\overline S_{n+2}.
 \end{aligned}
\]
Here the approximations are in
$L^2(F\otimes\Leb)$, with errors small
relative to $A_n$.
The first averaging provides a function
independent of the state for the beginning
of the next projective step.
The second provides such a function for
the end of that step and the beginning
of the following one.
The third provides it for the final distribution.
Thus three averaging estimates allow us
to compare two consecutive projective steps.

We turn the three functions
$\overline S_n$, $\overline S_{n+1}$ and
$\overline S_{n+2}$ into probability measures
by squaring and normalising.
Denote the resulting probabilities by
$\eta^{(1)},\eta^{(2)},\eta^{(3)}$, respectively.
By Lemma~\ref{lem:squares}, we convert the
$L^2$ estimates into Wasserstein estimates.
Using \eqref{eq:approximate-equivariance},
we then show that a one-step projective
transformation sends the first probability
close to the second, and the second close
to the third, on average over the states.
Composing the two transformations along
the Markov chain gives
\[
 \int W(F_*\eta^{(1)},\eta^{(3)})
                         \,d\mu_\lambda^{(2)}(F)<d_*,
\]
contrary to \eqref{eq:wasserstein-separation}.

\begin{proposition}
\label{prop:three-step}
There are $0<\gamma<1/2$, $0<\kappa<1$
and $C_3<\infty$ such that
\begin{equation}\label{eq:three-step-contraction}
 A_{n+3}^2\le\max\{\kappa A_n^2,C_3\}
\end{equation}
for every $n\ge0$, every $\lambda\in\Lambda$,
every $0<\varepsilon<1$ and every initial
law allowed in \eqref{eq:skew-transition}.
\end{proposition}

\begin{proof}
Put 
\begin{equation}\label{eq:s}
s=1-\rho^2>0, \quad D=\operatorname{diam}(M)=\pi/2.
\end{equation}
First choose $\zeta>0$ so small that
\begin{equation}\label{eq:zeta-choice}
 \zeta\le\min\left\{\frac{s}{40},\frac1{256}\right\},
 \qquad
 (L+1)\left(16D(L+1)\zeta^{1/4}
                      +\frac{40D\zeta^{1/2}}s\right)<\frac{d_*}3.
\end{equation}
Next choose $\delta>0$ with $(L+1)\delta<d_*/3$.
Choose $\gamma>0$ below both thresholds required by
Lemma~\ref{lem:fibre-energy-stability},
with this $\zeta$, and
Lemma~\ref{lem:energy-toolkit}(4),
with this $\delta$ and $L$.
This order fixes the exponent before
the energy thresholds are chosen.
Let $C_\zeta$ be the constant in
\eqref{eq:averaged-energy-stability}, and let
$H_{\gamma,\delta,L}$ be the threshold in
\eqref{eq:approximate-equivariance}.
Choose
\begin{equation}\label{eq:large-energy-threshold}
 H\ge\max\left\{\frac{C_\zeta}{\zeta},
                         16(H_{\gamma,\delta,L}+1)\right\}.
\end{equation}
Fix $n,\lambda,\varepsilon$ and put $a=A_n^2$.
To simplify the indices within this proof, write
\[
 \begin{gathered}
 R_t=R_{n+t},\qquad S_t=S_{n+t},\qquad
 \overline S_t=\overline S_{n+t},\\
 A_t=A_{n+t},\qquad B_t=B_{n+t}.
 \end{gathered}
\]
In particular, $A_0^2=a$.
Assume $a\ge H$, where $H$ is chosen in
\eqref{eq:large-energy-threshold}.
We shall prove
$A_3^2<(1-\zeta)A_0^2=(1-\zeta)a$.
Write
\[
 D_t^{\rm M}=B_t^2-A_{t+1}^2,
 \qquad t=0,1,2.
\]
The exact three-step identity is
\begin{equation}\label{eq:three-step-energy-balance}
 A_3^2=a+\sum_{t=0}^2(B_t^2-A_t^2)
                         -\sum_{t=0}^2D_t^{\rm M}.
\end{equation}
The changes caused by the projective transformations
have small relative size by
Lemma~\ref{lem:fibre-energy-stability}.
We split according to the total decrease
under Markov averaging.

By \eqref{eq:markov-defect} and
\eqref{eq:averaged-energy-stability},
\begin{equation}\label{eq:tower-upper-step}
 A_{t+1}^2\le B_t^2\le(1+\zeta)A_t^2+C_\zeta.
\end{equation}
Since $C_\zeta\le\zeta a$, iteration
from $A_0^2=a$ gives
\[
 A_t^2\le\bigl(2(1+\zeta)^t-1\bigr)a\le2a,
 \qquad 0\le t\le3.
\]

\smallskip
\noindent\textit{Case I: the decrease under
Markov averaging is substantial.}
If $\sum_{t=0}^2D_t^{\rm M}>10\zeta a$, then
\eqref{eq:averaged-energy-stability}
and $A_t^2\le2a$ give
\[
 \sum_{t=0}^2(B_t^2-A_t^2)
 \le\zeta\sum_{t=0}^2 A_t^2+3C_\zeta
 \le9\zeta a.
\]
Subtracting $\sum_{t=0}^2D_t^{\rm M}$ in
\eqref{eq:three-step-energy-balance}
gives $A_3^2<(1-\zeta)a$.

\smallskip
\noindent\textit{Case II: the decrease under
Markov averaging is small.}

Assume 
\begin{equation}\label{eq:dm}
\sum_{t=0}^2D_t^{\rm M}\le10\zeta a.
\end{equation}
For $a\ge H$, the inequality $A_3^2<(1-\zeta)a$ gives
\eqref{eq:three-step-contraction} with $\kappa=1-\zeta$.
We therefore suppose, for a contradiction, that
\begin{equation}\label{eq:assumed-small-loss}
 A_3^2\ge(1-\zeta)a.
\end{equation}
Put $q=1+\zeta$. Applying \eqref{eq:tower-upper-step}
successively, we obtain, for $t=0,1,2$,
\[
 A_3^2\le q^{3-t}A_t^2+C_\zeta\sum_{j=0}^{2-t}q^j
       \le q^{3-t}A_t^2+3\zeta q^2a,
\]
where we used $C_\zeta\le\zeta a$ and
$\sum_{j=0}^{2-t}q^j\le3q^2$. Hence
\[
 A_t^2\ge
 \frac{1-\zeta-3\zeta q^2}{q^{3-t}}\,a
 \ge\frac{1-\zeta-3\zeta q^2}{q^3}\,a
 \ge\frac a2,\qquad t=0,1,2,
\]
since $\zeta\le1/100$. For $t=3$, the same lower bound follows
directly from \eqref{eq:assumed-small-loss}.
Also, for $0\le t\le2$,
\[
 B_t^2\ge A_{t+1}^2,\qquad
 B_t^2\le(1+\zeta)2a+\zeta a\le3a.
\]
We have proved
\begin{equation}\label{eq:tower-comparability}
 \frac a2\le A_t^2\le2a, \quad 0\le t\le3,\qquad
 \frac a2\le B_t^2\le3a, \quad 0\le t\le2.
\end{equation}

Each decrease $D_t^{\rm M}$ is nonnegative,
so the small-loss assumption \eqref{eq:dm} gives
$D_t^{\rm M}\le10\zeta a$ for $t=0,1,2$.
Applying \eqref{eq:markov-defect} and
\eqref{eq:input-collapse}, with $s$ of \eqref{eq:s}, we obtain
\begin{equation}\label{eq:profile-collapse}
 \|S_t-\overline S_t\|_2^2
 \le\frac{10\zeta}s\,a,\qquad
 \|R_{t+1}-\overline S_t\|_2^2
 \le\frac{10\zeta}s\,a,\qquad 0\le t\le2.
\end{equation}
Since $\int_S(S_t(v,z)-\overline S_t(z))\,dF(v)=0$,
expanding the square gives
\[
 B_t^2=\|\overline S_t\|_{\mathscr H}^2
              +\|S_t-\overline S_t\|_2^2.
\]
Using \eqref{eq:tower-comparability},
\eqref{eq:profile-collapse} and $\zeta\le s/40$, we obtain
\[
 \frac a4\le\frac a2-\frac{10\zeta}s\,a
 \le\|\overline S_t\|_{\mathscr H}^2\le B_t^2\le3a.
\]
Since $a=A_0^2$, this gives
\begin{equation}\label{eq:common-profile-size}
 \frac{A_0}2\le\|\overline S_t\|_{\mathscr H}
       \le\sqrt3 A_0,\qquad 0\le t\le2.
\end{equation}

For $t=1,2,3$, define the probability measures
\begin{equation}\label{eq:common-theta}
 d\eta^{(t)}
 =\frac{\overline S_{t-1}^{\,2}}
        {\|\overline S_{t-1}\|_{\mathscr H}^2}\,d\Leb.
\end{equation}
For $t=1,2$, let $G_t$ consist of states $v$
for which both
\[
 \|R_t(v)-\overline S_{t-1}\|_{\mathscr H}
       \le\zeta^{1/4}A_0,\qquad
 \|S_t(v)-\overline S_t\|_{\mathscr H}
       \le\zeta^{1/4}A_0.
\]
Put $G=G_1\cap G_2$.
For example, Chebyshev's inequality gives
\begin{align*}
 F\{v:\|R_1(v)-\overline S_0\|_{\mathscr H}
                         >\zeta^{1/4}A_0\}
 &\le\frac{1}{\zeta^{1/2}A_0^2}
       \int_S\|R_1(v)-\overline S_0\|_{\mathscr H}^2\,dF(v)\\
 &\le\frac{10\zeta^{1/2}}s,
\end{align*}
where the last inequality uses
\eqref{eq:profile-collapse} and $A_0^2=a$.
Each of the other three errors has the same bound.
The complement of $G$ is their union, so
\begin{equation}\label{eq:bad-state-mass}
 F(G^c)\le\frac{40\zeta^{1/2}}s.
\end{equation}
For $v\in G$ and $t=1,2$, the definition of $G$ and
\eqref{eq:common-profile-size} give
\[
 \begin{aligned}
 \|R_t(v)\|_{\mathscr H}
 &\ge\|\overline S_{t-1}\|_{\mathscr H}
       -\|R_t(v)-\overline S_{t-1}\|_{\mathscr H}
 \ge A_0/2-\zeta^{1/4}A_0\ge A_0/4,\\
 \|S_t(v)\|_{\mathscr H}
 &\ge\|\overline S_t\|_{\mathscr H}
       -\|S_t(v)-\overline S_t\|_{\mathscr H}
 \ge A_0/2-\zeta^{1/4}A_0\ge A_0/4,
 \end{aligned}
\]
where $\zeta^{1/4}\le1/4$ by our choice \eqref{eq:zeta-choice}.

By \eqref{eq:normalised-energy-measure} and
\eqref{eq:markov-profiles},
\[
 \begin{aligned}
 d\theta_{\gamma,\varepsilon}(\mu_{n+t,v,\lambda})(z)
 &=\frac{R_t(v,z)^2}{\|R_t(v)\|_{\mathscr H}^2}\,d\Leb(z),\\
 d\theta_{\gamma,\varepsilon}
       ((T_{\lambda,v})_*\mu_{n+t,v,\lambda})(z)
 &=\frac{S_t(v,z)^2}{\|S_t(v)\|_{\mathscr H}^2}\,d\Leb(z).
 \end{aligned}
\]
We apply Lemma~\ref{lem:squares} first with
$(r_1,r_2)=(R_t(v),\overline S_{t-1})$, and then with
$(r_1,r_2)=(S_t(v),\overline S_t)$.
In both cases, the two norms are at least $A_0/4$ and
$\|r_1-r_2\|_{\mathscr H}\le\zeta^{1/4}A_0$.
Thus the normalised densities satisfy
\[
 \left\|\frac{r_1^2}{\|r_1\|_{\mathscr H}^2}
              -\frac{r_2^2}{\|r_2\|_{\mathscr H}^2}\right\|_{L^1}
 \le\frac{4\zeta^{1/4}A_0}{A_0/4}=16\zeta^{1/4}.
\]
The Wasserstein distance is at most $D$ times this $L^1$ norm.
By \eqref{eq:common-theta}, we obtain
\begin{align}
 W\bigl(\theta_{\gamma,\varepsilon}
             (\mu_{n+t,v,\lambda}),\eta^{(t)}\bigr)
 &\le16D\zeta^{1/4},
                  \label{eq:input-theta-close}\\
 W\bigl(\theta_{\gamma,\varepsilon}
             ((T_{\lambda,v})_*\mu_{n+t,v,\lambda}),
                               \eta^{(t+1)}\bigr)
 &\le16D\zeta^{1/4},
                  \label{eq:output-theta-close}
\end{align}
for $t=1,2$ and $v\in G$.

Also, since $A_0^2 = a \geq H$, \eqref{eq:large-energy-threshold} gives
\[
 \Ereg_{\gamma,\varepsilon}(\mu_{n+t,v,\lambda})
 =\|R_t(v)\|_{\mathscr H}^2
 \ge A_0^2/16\ge H/16>H_{\gamma,\delta,L}.
\]
We can therefore apply \eqref{eq:approximate-equivariance}.
For this calculation, write $T=T_{\lambda,v}$,
$\mu=\mu_{n+t,v,\lambda}$ and $\theta=\theta_{\gamma,\varepsilon}$.
By the triangle inequality and \eqref{eq:pushforward-lipschitz},
\[
 \begin{aligned}
 W(T_*\eta^{(t)},\eta^{(t+1)})
 &\le W(T_*\eta^{(t)},T_*\theta(\mu))
       +W(T_*\theta(\mu),\theta(T_*\mu))
 +W(\theta(T_*\mu),\eta^{(t+1)})\\
 &\le L W(\eta^{(t)},\theta(\mu))
       +\delta+W(\theta(T_*\mu),\eta^{(t+1)})\\
 &\le16D(L+1)\zeta^{1/4}+\delta,
 \end{aligned}
\]
where the factor $L$ comes from
\eqref{eq:uniform-lipschitz}.
On $G^c$, the Wasserstein distance is at most $D$.
Thus, using \eqref{eq:bad-state-mass}, we obtain for $t=1,2$,
\begin{equation}\label{eq:one-step-common-transport}
 \int_S W((T_{\lambda,v})_*\eta^{(t)},
                              \eta^{(t+1)})\,dF(v)
 \le16D(L+1)\zeta^{1/4}
       +\frac{40D\zeta^{1/2}}s+\delta.
\end{equation}
For a stationary consecutive pair $(v,w)$,
the triangle inequality gives
\[
 \begin{aligned}
 W((T_{\lambda,w}T_{\lambda,v})_*\eta^{(1)},
                                      \eta^{(3)})
 &\le L W((T_{\lambda,v})_*\eta^{(1)},\eta^{(2)})
 +W((T_{\lambda,w})_*\eta^{(2)},\eta^{(3)}).
 \end{aligned}
\]
Integrate against the consecutive-pair
distribution $dF(v)K^\leftarrow(v,dw)$.
Both states have distribution $F$.
By \eqref{eq:one-step-common-transport},
\eqref{eq:zeta-choice} and
$(L+1)\delta<d_*/3$,
\[
 \int W(F_*\eta^{(1)},\eta^{(3)})
                      \,d\mu_\lambda^{(2)}(F)
 <\frac{2d_*}3<d_*.
\]
This contradicts \eqref{eq:wasserstein-separation},
proving $A_3^2<(1-\zeta)A_0^2$ for
$A_0^2\ge H$ in Case II as well.

Set $\kappa=1-\zeta$.
For $a<H$, iterating
\eqref{eq:tower-upper-step} bounds $A_3^2$ by
\[
 C_3=(1+\zeta)^3H
       +C_\zeta\bigl(1+(1+\zeta)+(1+\zeta)^2\bigr).
\]
For $a\ge H$, the argument gives $A_3^2<\kappa a$.
Returning to the original indices,
we obtain \eqref{eq:three-step-contraction}.
\end{proof}

\subsection{Finite-step cutoff and limits}

\begin{proof}[Proof of Theorem~\ref{thm:main-regularity}(1) and the
conclusion for Ces\`aro averages in (2)]
Use $\gamma$, $\kappa$, $C_3$ from Proposition~\ref{prop:three-step}.
For any initial probability kernel, Lemma~\ref{lem:initial-energy}
gives $A_0^2\le C_\gamma\varepsilon^{-\gamma}$.
Iterating \eqref{eq:three-step-contraction} gives
\[
 A_{3m}^2\le
        \max\{\kappa^m C_\gamma\varepsilon^{-\gamma},C_3\}.
\]
For the remaining zero, one or two steps use
\eqref{eq:tower-upper-step}.  Enlarging a fixed constant yields
\begin{equation}\label{eq:energy-cutoff}
 A_N^2\le C_{\mathrm{en}}\bigl(1+\kappa^{\lfloor N/3\rfloor}
                                         \varepsilon^{-\gamma}\bigr),
                    \qquad 0<\varepsilon<1.
\end{equation}
By \eqref{eq:regularised-profile}, \eqref{eq:phase-disintegration}, \eqref{eq:markov-profiles}, and
Tonelli's theorem,
\[
 \mathcal R_{\gamma,\varepsilon}
       [\nu_{N,\lambda}](z)
 =\int_S R_N(v,z)\,dF(v).
\]
Applying Jensen's inequality for each $z$
and then integrating, we obtain
\[
 \begin{aligned}
 \Ereg_{\gamma,\varepsilon}(\nu_{N,\lambda})
 &=\int_M\left|\int_S R_N(v,z)\,dF(v)\right|^2
                                      d\Leb(z)\\
 &\le\int_M\int_S|R_N(v,z)|^2\,dF(v)\,d\Leb(z)
 =A_N^2.
 \end{aligned}
\]      
Use \eqref{eq:energy-comparison-a}, with for example $\delta=1$,
to obtain the same bound for $\mathcal E_{\gamma,\varepsilon}$.

Choose $\kappa_0\in(\kappa^{1/\gamma},1)$.
For $m=\lfloor N/3\rfloor\ge1$ set
$\varepsilon_N=\kappa^{m/\gamma}$.
Then $\kappa^m\varepsilon_N^{-\gamma}=1$ and
$\varepsilon_N<\kappa_0^m$.
The interaction energy is uniformly bounded at this scale.

Lemma~\ref{lem:ball-energy} gives \eqref{eq:main-finite-holder}
with $\alpha=\gamma/2$,
\[
 \nu_{N,\lambda}(B_r(x))\le Cr^\alpha, \quad \text{for every} \quad r>\kappa_0^m.
\]
For $m=0$ this condition is $r>1$, where the estimate is immediate
after increasing $C$ to at least one, since every probability
has total mass one.

For fixed $r>0$, only finitely many indices fail the cutoff condition.
Consequently the Ces\`aro averages assign mass at most
$Cr^\alpha+o(1)$ to $B_r(x)$.
For a weak limit $\nu_\lambda$ along a subsequence of Ces\`aro
averages, weak convergence and openness of the ball give
\begin{equation}\label{eq:main-limit-holder}
 \nu_\lambda(B_r(x))
 \le\liminf_N\frac1N\sum_{n=0}^{N-1}\nu_{n,\lambda}(B_r(x))
 \le Cr^\alpha.
\end{equation}
The liminf is taken along the convergent subsequence.
The constants are uniform in $\lambda\in\Lambda$ and in the initial joint distribution.
\end{proof}

\begin{corollary}\label{cor:stationary-energy}
Every stationary joint law of \eqref{eq:skew-transition} with state distribution $F$ has a phase distribution satisfying
\eqref{eq:main-limit-holder}.  Moreover, for the exponent $\gamma$ in
Proposition~\ref{prop:three-step}, its stationary fibres satisfy
\begin{equation}\label{eq:stationary-fibre-energy}
 \int_S\iint_{M\times M}d(\phi,\psi)^{-\gamma}\,
       d\mu_{v,\lambda}(\phi)d\mu_{v,\lambda}(\psi)\,dF(v)<\infty,
\end{equation}
with a bound uniform for $\lambda\in\Lambda$ and all stationary joint
laws.  In particular almost every stationary fibre is H\"older with
exponent $\gamma/2$, with a fibre-dependent constant.
\end{corollary}

\begin{proof}
Start the evolution with a stationary joint law.
Its phase distribution $\nu_{N,\lambda}$ is the same for every $N$.
Letting $N\to\infty$ in \eqref{eq:main-finite-holder} proves the
claim about the phase distribution at every positive radius.

For each fixed $0<\varepsilon<1$, stationarity gives
$A_3^2=A_0^2$. Hence \eqref{eq:three-step-contraction} yields
\[
 A_0^2\le\max\{\kappa A_0^2,C_3\}.
\]
Since $\kappa<1$, we obtain $A_0^2\le C_3$.
The energy $A_0^2$ may depend on $\varepsilon$, but $C_3$ does not.
Applying \eqref{eq:energy-comparison-a} with $\delta=1$
and integrating in $F$, we obtain
\[
 \begin{aligned}
 \int_S\mathcal E_{\gamma,\varepsilon}(\mu_{v,\lambda})\,dF(v)
 &\le2\int_S\Ereg_{\gamma,\varepsilon}(\mu_{v,\lambda})\,dF(v)
          +C_{\gamma,1}
 =2A_0^2+C_{\gamma,1}\le2C_3+C_{\gamma,1},
 \end{aligned}
\]
where the equality uses \eqref{eq:markov-energies}.
By \eqref{eq:interaction-energy}, this is
\[
 \int_S\iint_{M\times M}
 U_{\gamma,\varepsilon}(d(\phi,\psi))\,
 d\mu_{v,\lambda}(\phi)d\mu_{v,\lambda}(\psi)\,dF(v)
 \le2C_3+C_{\gamma,1},
\]
where 
$C_3, C_{\gamma,1}$ are independent of $\varepsilon$.

Put $p=(1+\gamma)/2$. By \eqref{eq:regularising-kernel},
\[
 \varphi_{\gamma,\varepsilon}(r)
 =r^{-p}\min\{1,(r/\varepsilon)^\gamma\}
 \uparrow r^{-p}\qquad(\varepsilon\downarrow0,\ r>0).
\]
It follows from \eqref{eq:interaction-kernel} and monotone
convergence that, for $r>0$,
\[
 \begin{aligned}
 U_{\gamma,\varepsilon}(r)
 &\uparrow\int_\R|y|^{-p}|r-y|^{-p}\,dy
 =r^{1-2p}\int_\R|x|^{-p}|1-x|^{-p}\,dx
   =u_\gamma r^{-\gamma},
 \end{aligned}
\]
where we used $y=rx$ and $1-2p=-\gamma$.
The constant $u_\gamma$ is positive and finite:
$p<1$ makes the singularities at $0$ and $1$ integrable,
and $2p=1+\gamma>1$ gives integrability at infinity.
For $r=0$, the limiting integral is
$\int_\R|y|^{-1-\gamma}\,dy=+\infty$.

The conditional measures are fixed as $\varepsilon$ decreases.
Applying monotone convergence to the state and direction integrals,
we therefore obtain
\[
 u_\gamma\int_S\iint_{M\times M}
 d(\phi,\psi)^{-\gamma}\,
 d\mu_{v,\lambda}(\phi)d\mu_{v,\lambda}(\psi)\,dF(v)
 \le2C_3+C_{\gamma,1}.
\]
Dividing by $u_\gamma$ proves \eqref{eq:stationary-fibre-energy}.

For each conditional phase measure $\mu_{v,\lambda}$, define
\[
 I_v:=\iint_{M\times M}d(\phi,\psi)^{-\gamma}\,
                  d\mu_{v,\lambda}(\phi)d\mu_{v,\lambda}(\psi).
\]
This is its angular $\gamma$-energy: pairs of nearby phases contribute
with the weight $d(\phi,\psi)^{-\gamma}$.
The bound \eqref{eq:stationary-fibre-energy} says that
$\int_S I_v\,dF(v)$ is uniformly bounded, so $I_v$ is finite for
$F$-almost every $v$.  If $\phi,\psi$ belong to the same angular ball
$B_r(x)$, then $d(\phi,\psi)\le2r$.  Restricting the energy
integral to this ball therefore gives
\[
 I_v\ge(2r)^{-\gamma}
                \mu_{v,\lambda}(B_r(x))^2,
 \qquad
 \mu_{v,\lambda}(B_r(x))
       \le2^{\gamma/2}I_v^{1/2}r^{\gamma/2}.
\]
The H\"older constants of the fibres,
$C(v)=2^{\gamma/2}I_v^{1/2}$, satisfy
\[
 \int_S C(v)^2\,dF(v)
 =2^\gamma\int_S I_v\,dF(v),
\]
with a uniform bound.
\end{proof}

\begin{remark}
The recursion \eqref{eq:profile-recursion} uses the original Markov
operator $P$ because the chain $\{X_n\}$ was first read backwards and
then conditioned on the next state of $\{Y_n\}$.  No identity $P=P^*$
is used.  If one starts with an abstract forward-driven phase process
instead, the recursion for the regularised functions uses $P^*$, with the same
norm bound.  The proof covers that process as well.
\end{remark}

\section{Sharpness of the regularity hypotheses}\label{sec:sharpness}

\subsection{Proof of Proposition~\ref{prop:unbounded-obstruction}}

The counterexample comes from a competition between
the probability of a long stay at a large potential value
and the contraction of directions during that stay.
The probability decreases geometrically with the number
of steps, at a rate independent of the state, while
the contraction becomes arbitrarily strong.
For any proposed positive H\"older exponent, we can therefore
choose a state for which the probability decreases more
slowly than the corresponding power of the radius
of the resulting ball.
The mass in that ball then violates the proposed
H\"older bound.
A small stationary weight does not prevent this:
we first fix the state and then let the number
of steps grow.

Fix $0<p<1$ and put
\[
 v_j=2^j+3,\qquad
 F_j=Z^{-1}2^{-j^2},\qquad
 Z=\sum_{j\ge1}2^{-j^2}.
\]
On $S=\{v_j:j\ge1\}$, put
$F(\{v_j\})=F_j$ and $V(v)=v$, and define
\begin{equation}\label{eq:sticky-kernel}
 K(v_i,\{v_j\})
   =p\mathbf1_{\{i=j\}}+(1-p)F_j.
\end{equation}
This transition retains the state with probability $p$
and draws a state with distribution $F$ with probability $1-p$.
Since
\[
 F_iK(v_i,\{v_j\})
   =pF_i\mathbf1_{\{i=j\}}+(1-p)F_iF_j
   =F_jK(v_j,\{v_i\}),
\]
the probability $F$ is stationary and the chain
is reversible.
Its Markov operator satisfies
\[
 (Ph)(v_i)
   =ph(v_i)+(1-p)\int_S h\,dF.
\]
On $L^2_0(F)$ this is multiplication by $p$, and hence
\[
 \left\|P\big|_{L^2_0(F)}\right\|=p<1.
\]
The stationary weights decrease sufficiently rapidly
to give every positive potential moment
\[
 \int_S |V|^\delta\,dF
   =\sum_{j\ge1}F_jv_j^\delta
   \le C_\delta\sum_{j\ge1}2^{-j^2+\delta j}
   <\infty, 
 \qquad \text{for any} \quad \delta>0.
\]

We next describe the phase evolution while
the chain is stuck at $v_j$.
At $\lambda=0$, each such step applies
$T_j:=T_{c_0(v_j)}$.
By \eqref{eq:phase-scalar-conjugacy}, this map is
conjugate to the projective action of
\[
 B(-v_j)
   =\begin{pmatrix}-v_j&-1\\1&0\end{pmatrix},
\]
where $B$ is defined in
\eqref{eq:scalar-matrix-notation}.
Put
\[
 \Lambda_j=\frac{v_j+\sqrt{v_j^2-4}}2>1.
\]
The eigenvalues of $B(-v_j)$ are
$-\Lambda_j$ and $-\Lambda_j^{-1}$, with respective
eigenvectors
\[
 e_j^+=\binom{1}{-\Lambda_j^{-1}},
 \qquad
 e_j^-=\binom{1}{-\Lambda_j}.
\]
Thus the attracting and repelling fixed points
of $T_j$ are
\[
 a_j=J_\beta^{-1}[1:-\Lambda_j^{-1}],
 \qquad
 r_j=J_\beta^{-1}[1:-\Lambda_j].
\]
Since $v_j$ is strictly increasing, so is $\Lambda_j$.
In particular,
\[
 \Lambda_j\longrightarrow\infty,
\]
and the repelling points $r_j$ are distinct.

The eigenvectors also give the contraction rate
without further matrix calculations.
Every phase $\phi\ne r_j$ has a unique representation
\[
 J_\beta\phi=[e_j^++t e_j^-],
 \qquad t\in\R.
\]
After $m$ iterations,
\[
 J_\beta(T_j^m\phi)
   =[e_j^++\Lambda_j^{-2m}t e_j^-].
\]
On a compact set avoiding $r_j$, the coordinate $t$
is bounded.
The map
$t\mapsto J_\beta^{-1}[e_j^++t e_j^-]$
is smooth near $t=0$ and takes $0$ to $a_j$.
Therefore, for every compact
$E_j\subset M\setminus\{r_j\}$, there is $C_j>0$
such that
\begin{equation}\label{eq:hyperbolic-concentration}
 T_j^m(E_j)
   \subset B_{C_j\Lambda_j^{-2m}}(a_j)
\end{equation}
for all sufficiently large $m$.
This is the concentration produced by a long
sequence of retention steps at $v_j$.

Now take any stationary joint distribution with
state distribution $F$.
Write its conditional phase distributions as $\mu_j$
and its phase distribution as
\[
 \nu=\sum_jF_j\mu_j.
\]
Lemma~\ref{lem:stationary-existence} gives such
a joint distribution without a boundedness
assumption on $V$.

After one projective step, the phase distribution is
$\sum_iF_i(T_i)_*\mu_i$.
Stationarity therefore gives
\[
 \nu=\sum_iF_i(T_i)_*\mu_i.
\]
Using \eqref{eq:fibre-recursion} and
\eqref{eq:sticky-kernel}, we obtain
\[
 \begin{aligned}
 \mu_j
   &=\sum_iK(v_j,\{v_i\})(T_i)_*\mu_i
   =p(T_j)_*\mu_j+(1-p)\nu.
 \end{aligned}
\]
We repeatedly substitute this identity into its
first term.
For every $m\ge0$,
\[
 \mu_j
   =(1-p)\sum_{\ell=0}^{m}p^\ell(T_j^\ell)_*\nu
     +p^{m+1}(T_j^{m+1})_*\mu_j,
\]
where $T_j^0$ is the identity.
All measures on the right are nonnegative, so
\begin{equation}\label{eq:sticky-lower-bound}
 \mu_j\ge(1-p)p^m(T_j^m)_*\nu.
\end{equation}
The factor $p^m$ accounts for $m$ consecutive
decisions to retain the state.
Thus \eqref{eq:sticky-lower-bound} gives the amount
of mass to which we can apply the contraction
in \eqref{eq:hyperbolic-concentration}.

Since the $r_j$ are distinct, $\nu=\delta_{r_j}$
can hold for at most one index.
For every other $j$, choose a compact set
$E_j\subset M\setminus\{r_j\}$ with $\nu(E_j)>0$.
Combining \eqref{eq:sticky-lower-bound} and
\eqref{eq:hyperbolic-concentration}, we obtain
\[
 \begin{aligned}
 \nu(B_{C_j\Lambda_j^{-2m}}(a_j))
   &\ge F_j\mu_j(B_{C_j\Lambda_j^{-2m}}(a_j))
   \ge F_j(1-p)p^m\nu(E_j).
 \end{aligned}
\]
Suppose that $\nu$ satisfied a H\"older estimate
\[
 \nu(B_r(x))\le Cr^\eta
 \qquad \text{for}\quad  x\in M,\ r>0
\]
for some $C<\infty$ and $\eta>0$.
Then
\[
 F_j(1-p)p^m\nu(E_j)
   \le CC_j^\eta\Lambda_j^{-2\eta m},
\]
or equivalently,
\[
 (p\Lambda_j^{2\eta})^m
   \le\frac{CC_j^\eta}{F_j(1-p)\nu(E_j)}.
\]
Since $\Lambda_j\to\infty$, choose one such $j$
with $p\Lambda_j^{2\eta}>1$.
For this fixed $j$, the left side tends to infinity
as $m\to\infty$, while the right side remains finite,
a contradiction.

This comparison identifies the role of unboundedness.
The probability $p$ of retaining the state is fixed,
whereas the contraction factors $\Lambda_j^{-2}$
tend to zero.
The small weights $F_j$ ensure finiteness of all
potential moments, but they do not change the
restriction on $\eta$:
we first fix $j$ and then let the number $m$ of retention
steps tend to infinity.

We also verify the stated moment bounds for the
projective transformations.
For the projective action $F_A$ of
$A\in\SL(2,\R)$, the angular derivative at a line
represented by a unit vector $e$ is
\[
 |F_A'([e])|=\|Ae\|^{-2}.
\]
Applying this identity to $A$ and $A^{-1}$ gives
\[
 \mathfrak L(F_A)
   \le\max\{\|A\|^2,\|A^{-1}\|^2\} = \|A\|^2.
\]
The entries of $A(c)$ and its inverse in
\eqref{eq:matrix-factorisation} grow at most
linearly in $|c|$.
Since $c_0(v_j)$ is affine in $v_j$,
\[
 \mathfrak L(T_j)\le C_\beta(1+v_j)^2,
\]
and therefore
\[
 \mathbf E_{\mathbf P}\,\{\mathfrak L(T_{c_0(X_0)})^\delta\}
   =\sum_jF_j\mathfrak L(T_j)^\delta
   \le C_{\beta,\delta}
          \sum_{j\ge1}2^{-j^2+2\delta j}
   <\infty,
 \qquad \delta>0.
\]
Nevertheless, at the repelling fixed point,
\[
 |T_j'(r_j)|=\Lambda_j^2,
 \qquad
 \mathfrak L(T_j)\ge\Lambda_j^2\longrightarrow\infty,
\]
so the uniform bound \eqref{eq:uniform-lipschitz} fails.

Finally, fix any initial direction.
Suppose that, for some $C<\infty$, $\alpha>0$
and positive sequence $r_N\to0$,
\[
 \nu_{N,0}(B_r(x))\le Cr^\alpha,
 \qquad
 N\ge1,\ x\in M,\ r>r_N.
\]
Start the Ces\`aro construction of
Lemma~\ref{lem:stationary-existence}
from this initial direction, with state
distribution $F$.
Along a subsequence, the direction Ces\`aro
averages converge weakly to the direction
distribution $\nu$ of a stationary joint distribution.

For each fixed $r>0$, choose $N_0(r)$
so that $r_n<r$ for all $n\ge N_0(r)$.
Then
\[
 \overline\nu_{N,0}(B_r(x))
 \le\frac{N_0(r)}N+Cr^\alpha.
\]
Since $B_r(x)$ is open, the weak convergence
gives
\[
 \nu(B_r(x))
 \le\liminf_{j\to\infty}
       \overline\nu_{N_j,0}(B_r(x))
 \le Cr^\alpha.
\]
This contradicts the failure of every positive
H\"older exponent proved for all stationary
joint laws.

\begin{remark}
The example in
Proposition~\ref{prop:unbounded-obstruction}
has the strict one-step norm bound,
NDI for every $k\ge2$ by
Theorem~\ref{thm:main-markov-ndi},
and the uniform Wasserstein separation of
Corollary~\ref{cor:route-separation}.
Moreover, every positive moment of the
Lipschitz size of its two-step transformation
is finite.
Indeed, submultiplicativity, the Cauchy--Schwarz inequality
and stationarity give
\[
 \mathbf E_{\mathbf P}\,\{
 \mathfrak L\bigl(T_{c_0(X_{-1})}\circ T_{c_0(X_0)}\bigr)^\delta\}
 \le
 \mathbf E_{\mathbf P}\,\{
 \mathfrak L(T_{c_0(X_0)})^{2\delta}\}
 <\infty.
\]
Nevertheless, the direction distribution under every
stationary joint law fails to be H\"older with any
positive exponent.
Thus boundedness in Theorem~\ref{thm:main-regularity}
cannot be replaced by finiteness of all positive
moments of the potential or of the Lipschitz size
of the two-step transformations, even when the
strict one-step norm condition, NDI and uniform
Wasserstein separation hold.
\end{remark}

\subsection{Proof of Proposition~\ref{prop:doubling-resonance}}

The potential \eqref{eq:doubling-sampling} can be written as
$V(X_n)$, where $X_n=(b_n,b_{n+1})$ and $V(i,j)=i+j-1$.
In the order $00,01,10,11$, the transition matrix of $X_n$
and its square are
\begin{equation}\label{eq:doubling-transition-matrix}
 Q=\frac12
 \begin{pmatrix}
 1&1&0&0\\
 0&0&1&1\\
 1&1&0&0\\
 0&0&1&1
 \end{pmatrix},
 \qquad
 Q^2=\frac14
 \begin{pmatrix}
 1&1&1&1\\
 1&1&1&1\\
 1&1&1&1\\
 1&1&1&1
 \end{pmatrix}.
\end{equation}
Its stationary distribution is $\widetilde\pi(i,j)=1/4$.
Every entry of $Q^2$ is positive, so $Q$ is primitive.
Let $\widehat P$ be the Markov operator
\eqref{eq:finite-markov-operator} corresponding to $Q$.
For $h(i,j)=i-1/2$,
\[
 (\widehat P h)(i,j)=j-1/2,
 \qquad
 \|\widehat P h\|_{L^2(\widetilde\pi)}
 =\|h\|_{L^2(\widetilde\pi)}>0.
\]
Since $\widehat P$ is a contraction and
$\widehat P^2 h=0$ for every $h\in L^2_0(\widetilde\pi)$,
\[
 \left\|\widehat P\big|_{L^2_0(\widetilde\pi)}\right\|=1,
 \qquad
 \left\|\widehat P^2\big|_{L^2_0(\widetilde\pi)}\right\|=0.
\]

At $\lambda=0$, we shall prove that all scalar transfer
products belong, up to sign, to one fixed finite family.
The projective images of any fixed initial direction
then belong to one fixed finite set, independently of
the number of steps.
This will directly exclude every H\"older estimate
with a cutoff tending to zero.
We shall subsequently obtain $L(0)=0$ from the same
finite family and prove positivity at every
$\lambda\ne0$ by considering products between
successive returns of the binary digits to zero.

For consecutive digits $i,j\in\{0,1\}$, the potential
\eqref{eq:doubling-sampling} gives, at $\lambda=0$,
the scalar transfer matrix
\[
 A_{ij}=B(1-i-j),
\]
where $B$ is defined in
\eqref{eq:scalar-matrix-notation}.
We choose coordinates depending on the current digit:
a matrix $J_i$ at the beginning of a transition
$i\to j$, and $J_j$ at its end.
In these coordinates the transition matrix is
$J_j^{-1}A_{ij}J_i$.
We shall arrange that all four such matrices are,
up to sign, powers of a single matrix of finite order.

Put
\[
 A=B(-1)=\begin{pmatrix}-1&-1\\1&0\end{pmatrix},
 \quad
 R=B(0)=\begin{pmatrix}0&-1\\1&0\end{pmatrix},
 \quad
 J_0=R,\quad J_1=I.
\]
Direct multiplication gives
\begin{equation}\label{eq:doubling-four-transitions}
 \begin{aligned}
 J_0^{-1}A_{00}J_0&=-A^{-1},
 &J_1^{-1}A_{01}J_0&=-I,\\
 J_0^{-1}A_{10}J_1&=I,
 &J_1^{-1}A_{11}J_1&=A.
 \end{aligned}
\end{equation}
Moreover,
\begin{equation}\label{eq:doubling-finite-order}
 A^2=\begin{pmatrix}0&1\\-1&-1\end{pmatrix},
 \qquad A^3=I.
\end{equation}
Thus, if $\epsilon_{ij}=-1$ for $i=0$ and
$\epsilon_{ij}=1$ for $i=1$, then
\[
 A_{ij}=\epsilon_{ij}J_jA^{i+j-1}J_i^{-1}.
\]

The reason for choosing different coordinates at
the two digits is that the intermediate coordinate
changes cancel along a path.
For example, up to an overall sign, the product
of two consecutive transition matrices is
\[
 \begin{aligned}
 &\bigl(J_{b_2}A^{b_1+b_2-1}J_{b_1}^{-1}\bigr)
  \bigl(J_{b_1}A^{b_0+b_1-1}J_{b_0}^{-1}\bigr)
   =J_{b_2}A^{b_0+2b_1+b_2-2}J_{b_0}^{-1}.
 \end{aligned}
\]
The signs do not affect the projective action.
For an arbitrary binary path $b_0,\ldots,b_N$,
the same cancellation gives
\begin{equation}\label{eq:doubling-telescope}
 \begin{aligned}
 \Pi_N
   &:=A_{b_{N-1}b_N}\cdots A_{b_0b_1}
     =\epsilon_NJ_{b_N}A^{k_N}J_{b_0}^{-1},\\
 k_N
   &
   \equiv\sum_{n=0}^{N-1}(b_n+b_{n+1}-1)\pmod3,
 \end{aligned}
\end{equation}
where $\epsilon_N\in\{-1,1\}$ and
$k_N\in\{0,1,2\}$.
Thus the product depends, apart from its sign,
only on the first digit, the last digit and
the sum of the exponents modulo three.
In particular, every $\Pi_N$ belongs to
\[
 \left\{\epsilon J_jA^kJ_i^{-1}:
   \epsilon\in\{-1,1\},\
   i,j\in\{0,1\},\
   k\in\{0,1,2\}\right\}.
\]

We now apply \eqref{eq:doubling-telescope}
to the backward Riccati evolution
\eqref{eq:actual-phase-recursion}
with the Cauchy boundary conditions \eqref{eq:cauchy-data}.
Set $d_n=b_{1-n}$.
Then, using \eqref{eq:tau},
\[
 f(\tau^n\omega)
   =b_{-n}+b_{1-n}-1
   =d_n+d_{n+1}-1.
\]
The coefficients read in reverse order therefore
have the same two-digit form as those in
\eqref{eq:doubling-telescope}.
Let $\Phi_0^{\rm C}$ be the fixed initial phase
and put $\ell_0=J_\beta\Phi_0^{\rm C}$ with $J_\beta$ given in \eqref{eq:phase-coordinate-matrix}.
By \eqref{eq:phase-scalar-conjugacy},
\[
 J_\beta\Phi_N
   =A_{d_{N-1}d_N}\cdots A_{d_0d_1}\ell_0,
\]
where the matrices act on projective lines.
Consequently, every phase distribution $\nu_{N,0}$
is supported on the same finite set
\begin{equation}\label{eq:doubling-finite-support}
 \mathscr O_{\Phi_0^{\rm C}}
   =\left\{
      J_\beta^{-1}J_jA^kJ_i^{-1}\ell_0:
      i,j\in\{0,1\},\ k\in\{0,1,2\}
     \right\},
 \qquad
 \#\mathscr O_{\Phi_0^{\rm C}}\le12.
\end{equation}
For each $N$, at least one point
$x_N\in\mathscr O_{\Phi_0^{\rm C}}$ therefore satisfies
\[
 \nu_{N,0}(\{x_N\})\ge\frac1{12}.
\]
Suppose that, for some $C<\infty$, $\alpha>0$
and positive sequence $r_N\to0$,
\[
 \nu_{N,0}(B_r(x))\le Cr^\alpha
 \quad \text{for}\quad  N\ge1,\ x\in M,\ r>r_N.
\]
Taking $x=x_N$ and $r=2r_N$, we obtain
\[
 \frac1{12}
   \le\nu_{N,0}(B_{2r_N}(x_N))
   \le C(2r_N)^\alpha\longrightarrow0,
\]
a contradiction.
Since the set in \eqref{eq:doubling-finite-support}
is finite and closed, every weak limit of the
phase distributions or their Ces\`aro averages
is supported on that set as well.

Note that we have not used the equality $L(0)=0$:
the failure of H\"older regularity follows directly
from the fixed finite support
\eqref{eq:doubling-finite-support}.

We next obtain $L(0)=0$ as a separate consequence
of \eqref{eq:doubling-telescope}.
Let $C_{\mathrm{bd}}$ be the maximum of the
Euclidean operator norms of the matrices in
the finite family in \eqref{eq:doubling-telescope}.
Every $\Pi_N$ has determinant one, so
\begin{equation}\label{eq:doubling-two-sided-norm}
 1\le\|\Pi_N\|\le C_{\mathrm{bd}},
 \qquad
 0\le\frac1N\log\|\Pi_N\|
   \le\frac{\log C_{\mathrm{bd}}}{N}
   \longrightarrow0.
\end{equation}
These bounds hold for every binary path.
In particular, the definition
\eqref{eq:doubling-lyapunov-definition}
gives $L(0)=0$.

It remains to prove that $L(\lambda)>0$
for every $\lambda\ne0$.
We group the scalar transfer matrices between
successive returns of the Bernoulli digits to zero.
The digits following each return are independent
of the preceding digits and have the same
Bernoulli distribution.
Thus the successive return products are
independent and identically distributed.

The paths $00$, $010$ and $0110$ between
successive returns have positive probabilities.
Their products are, respectively,
\[
 U=B(\lambda+1),\qquad
 C_0=B(\lambda)^2,\qquad
 C_1=B(\lambda)B(\lambda-1)B(\lambda).
\]
They therefore belong to the support of the
return-product distribution.

To apply Furstenberg's theorem, we verify that
the group generated by the return support
is noncompact and strongly irreducible.

This group contains $U$ and
\[
 V=C_0^{-1}C_1
   =B(\lambda)^{-1}B(\lambda-1)B(\lambda)
   =\begin{pmatrix}\lambda&-1\\\lambda+1&-1\end{pmatrix}.
\]
A further multiplication gives
\[
 UV^{-1}
   =\begin{pmatrix}0&1\\-1&1\end{pmatrix}
   =B(1)^{-1}.
\]
Hence the group also contains
\[
 N_\lambda:=UB(1)^{-1}
   =\begin{pmatrix}1&\lambda\\0&1\end{pmatrix}.
\]
For $\lambda\ne0$, this is a nontrivial parabolic
transformation, and
\[
 N_\lambda^m
   =\begin{pmatrix}1&m\lambda\\0&1\end{pmatrix}.
\]
Its powers are unbounded, so the group is noncompact.

Suppose that a nonempty finite set of projective
directions were invariant under the group.
It would be invariant under $N_\lambda$.
By Lemma~\ref{lem:parabolic-finite-set},
it would therefore consist only of the unique
fixed point $[1:0]$ of $N_\lambda$.
But
\[
 U[1:0]=[\lambda+1:1]\ne[1:0].
\]
Thus this singleton is not invariant under $U$.
We have excluded every nonempty finite invariant
set and proved strong irreducibility for all
$\lambda\ne0$.

The return time $\tau$ satisfies
$\mathbf P\{\tau=r\}=2^{-r}$, $r\ge1$,
since a return after $r$ steps consists of $r-1$
consecutive ones followed by a zero.
Thus $\mathbf E\tau=2$.
At a fixed spectral parameter, each one-step
matrix and its inverse have uniformly bounded
norms.
The logarithmic norm of a return product is
therefore bounded above by a constant times
the number of steps in its return block and is integrable.
Furstenberg's theorem~\cite{Furstenberg}
now gives a positive Lyapunov exponent per return.

Finally, we pass from returns to individual
scalar steps.
The number of completed returns among the first $N$ steps,
divided by $N$, tends almost surely to $1/2$.
The initial segment before the first zero
is finite almost surely, and the unfinished
final excursion has length $o(N)$ almost surely.
The bounds on one-step matrices and their
inverses imply that inserting or removing
these segments changes the logarithmic norm
by $o(N)$.
Hence the Lyapunov exponent per scalar step
is one half of the positive return exponent.
This proves $L(\lambda)>0$ for every
$\lambda\ne0$.

\section*{Acknowledgements}
I am grateful to I.~Krasovsky for useful discussions and for patiently
listening to all of this countless times.

\appendix
\section{Auxiliary proofs}\label{app:proofs}

\subsection{Proof of Lemma~\ref{lem:two-parameter}}
We first prove part~(1), then classify completely the solutions of one
prescribed-image equation, which gives part~(3).  Finally we impose two
prescribed images and prove part~(2).

\medskip
\noindent
\emph{A prescribed image in the upper half-plane.}
Fix $z,z'\in\Hh$, and write
\[
 w=\mathcal R_\beta(z)
   =\frac{S_\beta+Cz}{C-S_\beta z}
   =u+iv, \quad C = \cos\beta,\quad S_\beta = \sin\beta > 0, \quad v>0.
\]
The equation $F_{a,b}(z)=z'$ is
\[
 z'=b+\mathcal R_\beta(a+w).
\]
Since $b$ is real, it does not affect the imaginary part.  Hence
\[
 \operatorname{Im}z'
 =
 \frac{v}
 {(C-S_\beta(a+u))^2+S_\beta^2v^2}.
\]
Equivalently,
\[
 (C-S_\beta(a+u))^2+S_\beta^2v^2
 =
 \frac{v}{\operatorname{Im}z'}.
\]
The quadratic equation has leading coefficient
$S_\beta^2>0$, hence at most two real roots,
each determining $b$ uniquely through the
real part of the original equation:
\[
 b
 =
 \operatorname{Re}z'
 -
 \operatorname{Re}\mathcal R_\beta(a+w).
\]
This proves part~(1).

\medskip
\noindent
\emph{One prescribed image on the projective line.}
Fix $x,y\in M$, and put
\[
 r=\mathcal R_\beta(x).
\]
We have to solve
\[
 y=b+\mathcal R_\beta(a+r).
\]

Suppose first that $r,y\in\mathbb R$.  Then
\[
 \begin{aligned}
 b
 &=
 y-\mathcal R_\beta(a+r)
 =
 y-\frac{S_\beta+C(a+r)}{C-S_\beta(a+r)}\\
 &=
 \frac{y(C-S_\beta(a+r))-(S_\beta+C(a+r))}
      {C-S_\beta(a+r)}.
 \end{aligned}
\]
Thus
\[
 b=\psi(a)=\frac{Aa+B}{Da+E},
\]
where
\[
 A=-(yS_\beta+C),
 \qquad
 B=yC-yS_\beta r-S_\beta-Cr,
\]
and
\[
 D=-S_\beta,
 \qquad
 E=C-S_\beta r.
\]
Moreover,
\[
 \begin{aligned}
 AE-BD
 &=
 -(yS_\beta+C)(C-S_\beta r)
 +S_\beta(yC-yS_\beta r-S_\beta-Cr)\\
 &=
 -(C^2+S_\beta^2)
 =-1.
 \end{aligned}
\]
Hence $\psi$ is an orientation-reversing
real projective transformation.
Since $y$ is finite, the pole
$a=\cot\beta-r$ is excluded.
The solution set is therefore the graph
of $\psi$ with this point omitted.

Now suppose that $r=\infty$.  Since adding a finite real number leaves
$\infty$ unchanged,
\[
 a+r=\infty,
\]
and
\[
 \mathcal R_\beta(\infty)
 =
 -\frac{C}{S_\beta}
 =
 -\cot\beta.
\]
Therefore
\[
 F_{a,b}(x)=b-\cot\beta.
\]
If $y\in\mathbb R$, the equation is equivalent to
\[
 b=y+\cot\beta,
\]
while $a$ is arbitrary.  Thus the solution set is a horizontal line.
If $y=\infty$, there is no solution, since $b-\cot\beta$ is finite for
every real $b$.

For $r\in\mathbb R$ and $y=\infty$, the pole condition is
\[
 \mathcal R_\beta(a+r)=\infty,
 \qquad C-S_\beta(a+r)=0.
\]
The equation holds precisely when $a=\cot\beta-r$,
with arbitrary $b$, giving a vertical solution line.

This proves part~(3).

\medskip
\noindent
\emph{Two prescribed images on the projective line.}
Now fix distinct $x_1,x_2\in M$ and distinct $y_1,y_2\in M$.  Put
\[
 r_i=\mathcal R_\beta(x_i),
 \qquad i=1,2.
\]
Since $\mathcal R_\beta$ is one-to-one and $x_1\ne x_2$, we have
\[
 r_1\ne r_2.
\]
The two equations are
\[
 y_1=b+\mathcal R_\beta(a+r_1),
 \qquad
 y_2=b+\mathcal R_\beta(a+r_2).
\]

Suppose first that $r_1,r_2,y_1,y_2$ are all finite.  Subtracting the
two equations eliminates $b$.  A direct subtraction gives, for finite
$s,t$,
\[
 \mathcal R_\beta(s)-\mathcal R_\beta(t)
 =
 \frac{s-t}
 {(C-S_\beta s)(C-S_\beta t)},
\]
where we used $C^2+S_\beta^2=1$.  Taking
\[
 s=a+r_1,
 \qquad
 t=a+r_2,
\]
we obtain
\[
 y_1-y_2
 =
 \frac{r_1-r_2}
 {(C-S_\beta(a+r_1))(C-S_\beta(a+r_2))}.
\]
Since $y_1\ne y_2$, this is equivalent to
\[
 (C-S_\beta(a+r_1))(C-S_\beta(a+r_2))
 =
 \frac{r_1-r_2}{y_1-y_2}.
\]
The equation has at most two real roots in $a$,
since the quadratic coefficient is $S_\beta^2>0$,
and either original equation then determines
$b$ uniquely.

It remains to consider the cases involving $\infty$.

Suppose first that one of the $r_i$ is infinite.  After relabelling,
let
\[
 r_1=\infty.
\]
Then
\[
 F_{a,b}(x_1)=b-\cot\beta.
\]
If $y_1=\infty$, there is no solution.  If $y_1$ is finite, the first
equation fixes
\[
 b=y_1+\cot\beta.
\]
Since $r_1\ne r_2$, the value $r_2$ is finite.  With $b$ fixed, the
second equation becomes
\[
 \mathcal R_\beta(a+r_2)=y_2-b.
\]
If $y_2$ is finite, injectivity of $\mathcal R_\beta$ gives at most one
real value of $a$.  If $y_2=\infty$, the pole condition gives the only
possible value
\[
 a=\cot\beta-r_2.
\]
Thus in either case there is at most one pair $(a,b)$.

Finally, suppose that $r_1,r_2$ are finite and one of $y_1,y_2$ is
infinite.  Since $y_1\ne y_2$, exactly one of them is infinite; after
relabelling, let
\[
 y_1=\infty,
 \qquad
 y_2\in\mathbb R.
\]
The first equation forces
\[
 a=\cot\beta-r_1.
\]
Since $r_2\ne r_1$,
\[
 a+r_2
 =
 \cot\beta-r_1+r_2
 \ne
 \cot\beta,
\]
so $\mathcal R_\beta(a+r_2)$ is finite.  The second equation therefore
determines $b$ uniquely:
\[
 b
 =
 y_2-\mathcal R_\beta(a+r_2).
\]
Thus the cases involving $\infty$ give at most one pair.  Together with
the finite case, this proves part~(2).

\subsection{Proof of Lemma~\ref{lem:stationary-existence}}

We first describe a compact set of functionals
representing joint distributions with state distribution $F$.
We then average the iterates of the joint transition
and obtain an invariant distribution as a limit.

Let $\mathscr X=L^1(F,C(M))$, where $C(M)$ denotes
the real-valued continuous functions on $M$, with norm
\[
 \|H\|_{\mathscr X}
 =\int_S\|H(v,\cdot)\|_\infty\,dF(v).
\]
A measurable probability kernel $v\mapsto\mu_v$
defines the positive linear functional
\[
 \mathscr L_\mu(H)
 =\int_S\int_M H(v,\phi)\,d\mu_v(\phi)\,dF(v).
\]
Since each $\mu_v$ is a probability measure,
\[
 |\mathscr L_\mu(H)|
 \le\int_S\|H(v,\cdot)\|_\infty\,dF(v)
 =\|H\|_{\mathscr X}.
\]
Thus $\|\mathscr L_\mu\|\le1$, and
\begin{equation}\label{eq:kernel-functional-normalisation}
 \mathscr L_\mu(a(v)\mathbf1)
 =\int_S a\,dF,
 \qquad a\in L^1(F).
\end{equation}

Conversely, let $\mathscr L$ be a positive linear
functional on $\mathscr X$, with $\|\mathscr L\|\le1$,
satisfying \eqref{eq:kernel-functional-normalisation}.
We show that $\mathscr L$ has the same integral
representation for a measurable probability kernel.

Choose a countable rational vector subspace
$\mathcal D\subset C(M)$ containing $\mathbf1$
and dense in the uniform norm.
For each $h\in\mathcal D$, the map
\[
 a\longmapsto\mathscr L(a(v)h(\phi))
\]
is a bounded linear functional on $L^1(F)$.
By the duality between $L^1(F)$ and $L^\infty(F)$,
we can choose $\ell_v(h)\in L^\infty(F)$ such that
\[
 \mathscr L(ah)=\int_S a(v)\ell_v(h)\,dF(v),
 \qquad a\in L^1(F).
\]
For nonnegative $a\in L^1(F)$, positivity and
\eqref{eq:kernel-functional-normalisation} give
\[
 -\|h\|_\infty\int_S a\,dF
 \le\mathscr L(ah)
 \le\|h\|_\infty\int_S a\,dF.
\]
It follows that
\[
 |\ell_v(h)|\le\|h\|_\infty
 \quad\text{for $F$-almost every }v.
\]
Similarly, $\ell_v(h)\ge0$ almost everywhere
if $h\ge0$, and $\ell_v(\mathbf1)=1$
almost everywhere.
Uniqueness of the representing functions also gives
rational linearity in $h$, almost everywhere.

We chose $\mathcal D$ to be countable so that
these properties and all rational linear relations
hold outside one common $F$-null set.
For every remaining $v$, the map
$h\mapsto\ell_v(h)$ therefore extends uniquely
to a bounded real linear functional on $C(M)$.
This extension is positive: if $h\ge0$, choose
$h_j\in\mathcal D$ and positive rational numbers
$\delta_j\to0$ such that
$\|h_j-h\|_\infty<\delta_j$.
Then $h_j+\delta_j\mathbf1\ge0$, and passage
to the limit gives $\ell_v(h)\ge0$.
The extension also satisfies $\ell_v(\mathbf1)=1$.
By the Riesz representation theorem, there is
a probability measure $\mu_v$ on $M$ such that
\[
 \ell_v(h)=\int_M h\,d\mu_v,
 \qquad h\in C(M).
\]
On the exceptional null set, choose a fixed
probability measure for $\mu_v$.

For every $h\in C(M)$, uniform approximation
from $\mathcal D$ shows that
$v\mapsto\int_M h\,d\mu_v$ is measurable.
To obtain measurability of $v\mapsto\mu_v(U)$
for an open set $U\subset M$, approximate
$\mathbf1_U$ increasingly by continuous functions.
A monotone class argument then gives
measurability for every Borel set.
Thus $v\mapsto\mu_v$ is a measurable probability kernel.
The identity
\[
 \mathscr L(H)
 =\int_S\int_M H(v,\phi)\,d\mu_v(\phi)\,dF(v)
\]
holds first for finite sums $a(v)h(\phi)$.
Such sums are dense in $\mathscr X$, so the
identity holds for every $H\in\mathscr X$.

Let $\mathcal C$ be the set of positive functionals
in the unit ball of $\mathscr X^*$ satisfying
\eqref{eq:kernel-functional-normalisation}.
We have proved that $\mathcal C$ represents precisely
the joint distributions with state distribution $F$.
The positivity conditions and the normalisation
\eqref{eq:kernel-functional-normalisation}
are weak-$*$ closed, so $\mathcal C$ is a
weak-$*$ closed subset of the dual unit ball,
hence compact by the Banach--Alaoglu theorem.

We next define the operator corresponding to
the joint transition \eqref{eq:skew-transition}:
\[
 (\mathscr QH)(v,\phi)
 =\int_S H(w,T_{\lambda,v}\phi)\,
                         K^\leftarrow(v,dw).
\]
We verify that $\mathscr Q$ maps $\mathscr X$
into itself.
For $H\in\mathscr X$, put
\[
 g(v)=\int_S\|H(w,\cdot)\|_\infty\,
                         K^\leftarrow(v,dw).
\]
Stationarity of $F$ gives
\[
 \int_S g(v)\,dF(v)
 =\int_S\|H(w,\cdot)\|_\infty\,dF(w)
 =\|H\|_{\mathscr X}<\infty.
\]
In particular, $g(v)<\infty$ for almost every $v$.

Fix such a state $v$ and let $\phi_j\to\phi$.
Continuity of $T_{\lambda,v}$ and of $H(w,\cdot)$
gives
\[
 H(w,T_{\lambda,v}\phi_j)
 \longrightarrow H(w,T_{\lambda,v}\phi).
\]
The absolute values are bounded by
$\|H(w,\cdot)\|_\infty$, which is integrable
with respect to $K^\leftarrow(v,dw)$.
By the dominated convergence theorem,
$(\mathscr QH)(v,\cdot)$ is continuous.

For each fixed $\phi$, the function
$v\mapsto(\mathscr QH)(v,\phi)$ is measurable,
by measurability of the transition kernel
and of $(v,\phi)\mapsto T_{\lambda,v}\phi$.
The Borel sigma-algebra of $C(M)$ is generated
by evaluations on a countable dense subset of $M$.
Therefore
$v\mapsto(\mathscr QH)(v,\cdot)$ is measurable
as a $C(M)$-valued function.
Since $C(M)$ is separable, it is strongly measurable.
Moreover,
\[
 \|\mathscr QH\|_{\mathscr X}
 \le\int_S g(v)\,dF(v)
 =\|H\|_{\mathscr X}.
\]
This estimate also shows that $\mathscr Q$
does not depend on the representative of $H$
in $\mathscr X$.
Thus $\mathscr Q$ is a positive contraction
on $\mathscr X$.

Its adjoint $\mathscr Q^*$ is weak-$*$ continuous.
It preserves positivity and the unit ball.
For $a\in L^1(F)$, we also have
\[
 \mathscr Q(a(v)\mathbf1)
 =(P^\leftarrow a)(v)\mathbf1.
\]
Hence, for $\mathscr L\in\mathcal C$,
\[
 \begin{aligned}
 (\mathscr Q^*\mathscr L)(a(v)\mathbf1)
 &=\mathscr L((P^\leftarrow a)(v)\mathbf1)
 =\int_S P^\leftarrow a\,dF
 =\int_S a\,dF.
 \end{aligned}
\]
Therefore $\mathscr Q^*\mathcal C\subseteq\mathcal C$.

Choose $\mathscr L_0\in\mathcal C$ and form
the averages
\[
 \mathscr L_N
 =\frac1N\sum_{n=0}^{N-1}
                  (\mathscr Q^*)^n\mathscr L_0.
\]
Each $\mathscr L_N$ belongs to $\mathcal C$.
Subtracting the two sums gives
\[
 \mathscr Q^*\mathscr L_N-\mathscr L_N
 =\frac{(\mathscr Q^*)^N\mathscr L_0
                     -\mathscr L_0}{N}.
\]
Since $\mathscr Q^*$ is a contraction and
$\|\mathscr L_0\|\le1$, we get
\[
 \|\mathscr Q^*\mathscr L_N-\mathscr L_N\|
    \le\frac{\|(\mathscr Q^*)^N\mathscr L_0\|
                 +\|\mathscr L_0\|}{N}
    \le\frac2N.
\]

Since $S$ is standard Borel and $C(M)$ is
separable, $\mathscr X=L^1(F,C(M))$ is separable.
Thus $\mathcal C$ is weak-$*$ compact and
metrisable, and we can choose $N_j\to\infty$
and $\mathscr L\in\mathcal C$ such that
\[
 \mathscr L_{N_j}\xrightarrow{w^*}\mathscr L.
\]
For every $H\in\mathscr X$, we have
$\mathscr QH\in\mathscr X$, so
\[
 \begin{aligned}
 \mathscr L(\mathscr QH)-\mathscr L(H)
 &=\lim_{j\to\infty}
   (\mathscr Q^*\mathscr L_{N_j}
                    -\mathscr L_{N_j})(H),\\
 \left|(\mathscr Q^*\mathscr L_{N_j}
                    -\mathscr L_{N_j})(H)\right|
 &\le\frac{2}{N_j}\|H\|_{\mathscr X}
 \longrightarrow0.
 \end{aligned}
\]
Thus $\mathscr L(\mathscr QH)=\mathscr L(H)$
for every $H\in\mathscr X$.

By the representation of positive functionals,
$\mathscr L$ defines a joint probability
\[
 \mathcal P(d\phi,dv)
 =\mu_v(d\phi)\,dF(v).
\]
Testing $\mathscr L(\mathscr QH)=\mathscr L(H)$
with $H(v,\phi)=\mathbf1_D(v)h(\phi)$,
where $D\subseteq S$ is Borel and $h\in C(M)$,
shows that $\mathcal P$ is invariant under
\eqref{eq:skew-transition}.
Its state distribution is $F$ by construction.

Finally, let $\nu_{n,\lambda}$ be the direction
distributions obtained from the initial joint
distribution represented by $\mathscr L_0$.
For $h\in C(M)$,
\[
 \mathscr L_N(h(\phi))
 =\frac1N\sum_{n=0}^{N-1}
                  \int_M h\,d\nu_{n,\lambda}.
\]
Taking $N=N_j$ proves that the corresponding
averages of direction distributions converge
weakly to the direction distribution of
$\mathcal P$.

\end{document}